\documentclass[rmp,aps,nofootinbib,floatfix]{revtex4-1} 
\usepackage{dcolumn}
\usepackage{graphicx,color}
\usepackage{subfigure}
\usepackage{epsf}
\usepackage{amsmath}
\usepackage{bm}
\usepackage{xspace}	
\usepackage[utf8]{inputenc}
\usepackage{multirow}
\usepackage{ctable}
\usepackage{rotating}
\usepackage{hyperref}

\def\fm#1{\ifmmode #1 \else $#1$\fi}
\def\ket#1{{%
  \ifmmode |\,#1\,\rangle \else $|\,#1\,\rangle$\fi}}
\def\bra#1{{%
  \ifmmode \langle\,#1\,| \else $\langle\,#1\,|$\fi}}
\def\braket#1#2{{%
  \ifmmode \langle\,#1\,|\,#2\,\rangle \else $\langle\,#1\,|\,#2\,\rangle$\fi}}
\def\expect#1{{%
  \ifmmode \langle\,#1\,\rangle \else $\langle\,#1\,\rangle$\fi}}
\def\upket{\fm{\ket{\!\uparrow}}}
\def\downket{\fm{\ket{\!\downarrow}}}

\def\downN{\fm{\ket{\!\downarrow_1\downarrow_2\dots\downarrow_N}}}
\def\upN{\fm{\ket{\!\uparrow_1\uparrow_2\dots\uparrow_N}}}

\def\Tl{\fm{\mathrm{Tl}^{+}}\xspace}

\def\Al{\fm{\mathrm{Al}^{+}}\xspace}

\def\Alo{NIST-Al-1\xspace}
\def\Alt{NIST-Al-2\xspace}
\def\Ba{\fm{\mathrm{Ba}^{+}}\xspace}
\def\Hg{\fm{\mathrm{Hg}^{+}}\xspace}
\def\Hgonn{\fm{^{199}\Hg}\xspace}
\def\Sr{\fm{\mathrm{Sr}^{+}}\xspace}
\def\Ca{\fm{\mathrm{Ca}^{+}}\xspace}
\def\Yb{\fm{\mathrm{Yb}^{+}}\xspace}
\def\Yboso{\fm{^{171}\Yb}\xspace}
\def\In{\fm{\mathrm{In}^{+}}\xspace}
\def\B{\fm{\mathrm{B}^{+}}\xspace}
\def\Be{\fm{\mathrm{Be}^{+}}\xspace}
\def\Mg{\fm{\mathrm{Mg}^{+}}\xspace}
\def\tpz{\fm{{}^3\mathrm{P}_{0}}\xspace}
\def\tpo{\fm{{}^3\mathrm{P}_{1}}\xspace}

\def\ssz{\fm{{}^1\mathrm{S}_{0}}\xspace}
\def\spo{\fm{{}^1\mathrm{P}_{1}}\xspace}
\def\ssztpz{\ssz\fm{\leftrightarrow}\tpz\xspace}
\def\ssztpo{\ssz\fm{\leftrightarrow}\tpo\xspace}
\def\dsoh{\fm{{}^2\mathrm{S}_{1/2}}\xspace}
\def\dpoh{\fm{{}^2\mathrm{P}_{1/2}}\xspace}
\def\dpth{\fm{{}^2\mathrm{P}_{3/2}}\xspace}
\def\ddth{\fm{{}^2\mathrm{D}_{3/2}}\xspace}
\def\ddfh{\fm{{}^2\mathrm{D}_{5/2}}\xspace}

\def\dfsh{\fm{{}^2\mathrm{F}_{7/2}}\xspace}
\def\dsohdpoh{\dsoh\fm{\leftrightarrow}\dpoh\xspace}
\def\ddthdpoh{\ddth\fm{\rightarrow}\dpoh\xspace}

\def\Orf{\fm{{\Omega_\mathrm{rf}}}\xspace}
\def\Vrf{\fm{{V_\mathrm{rf}}}\xspace}
\def\Rrf{\fm{{R_\mathrm{rf}}}\xspace}
\def\VDC{\fm{{V_\mathrm{DC}}}\xspace}
\def\RDC{\fm{{R_\mathrm{DC}}}\xspace}

\begin{document}
\preprint{}
\date{\today}

\title{OPTICAL ATOMIC CLOCKS}

\author{Andrew D. Ludlow$^{1,2}$, Martin M. Boyd$^{1,3}$, Jun Ye$^1$}
\affiliation{$^1$ JILA, National Institute of Standards and Technology and University of Colorado, Boulder, CO 80309, USA \\ $^2$ National Institute of Standards and Technology (NIST), 325 Broadway, Boulder, CO 80305, USA \\ $^3$ AOSense, 767 N. Mary Ave, Sunnyvale, CA 94085, USA}

\author{E. Peik$^4$, P.O. Schmidt$^{4,5}$}
\affiliation{$^4$ Physikalisch-Technische Bundesanstalt, Bundesallee 100, 38116 Braunschweig, Germany \\ $^5$ Institut für Quantenoptik, Leibniz Universität Hannover, Welfengarten 1, 30167 Hannover, Germany}

\begin{abstract}
Optical atomic clocks represent the state-of-the-art in the frontier of modern measurement science. In this article we provide a detailed review on the development of optical atomic clocks that are based on trapped single ions and many neutral atoms. We discuss important technical ingredients for optical clocks, and we present measurement precision and systematic uncertainty associated with some of the best clocks to date. We conclude with an outlook on the exciting prospect for clock applications.
\end{abstract}
\maketitle
\tableofcontents

\section{INTRODUCTION }
\label{sec:intro}
An 1879 text written by Thomson (Lord Kelvin) and Tait \cite{thomson79,kelvin02,snyder73} included the following:
\vspace{.15in}

``The recent discoveries due to the Kinetic theory of gases and to Spectrum analysis (especially when it is applied to the light of the heavenly bodies) indicate to us {\em natural standard} pieces of matter such as atoms of hydrogen or sodium, ready made in infinite numbers, all absolutely alike in every physical property.  The time of vibration of a sodium particle corresponding to any one of its modes of vibration is known to be absolutely independent of its position in the universe, and it will probably remain the same so long as the particle itself exists."
\vspace{.15in}

\noindent Although it took a while to realize, this idea attributed to Maxwell \cite{thomson79,kelvin02}, is the basic idea behind atomic frequency standards and clocks.  In this review, we focus on frequency standards that are based on \emph{optical} transitions, which seems to be implicit in the text above.  Optical frequency references have certain advantages over their predecessors at microwave frequencies; these advantages are now starting to be realized.

The need for more accurate and precise frequency standards and clocks has continued unabated for centuries.  Whenever improvements are made, the performance of existing applications is enhanced, or new applications are developed.  Historically, the prime application for clocks has been in navigation \cite{major_quantum_2007, grewal_global_2013}, and today we take for granted the benefits of global navigation satellite systems (GNSS), such as the global positioning system (GPS) \cite{kaplan_understanding_2006, rao_global_2010, grewal_global_2013}.  With GPS, we can easily navigate well enough to safely find our way from one location to another.  We look forward to navigation systems that will be precise enough to, for example, measure small strains of the earth's crust for use in such applications as earthquake prediction.  In addition, frequency standards provide the base unit of time, the second, which is by definition derived from the electronic ground state hyperfine transition frequency in caesium.  Eventually the definition of the second might be based on an optical transition \cite{gill_when_2011}, but even now, accurate optical frequency standards are becoming de facto secondary standards \cite{cipm_recommendation_2013}.

Aside from the benefits of these practical applications, for scientists there is the additional attraction of being able to precisely control a simple quantum system so that its dynamics evolve in its most elemental form.  One exciting possibility is that the evolution may not be as originally expected.  For example, an area of current interest explores the idea that the relative strengths of the fundamental forces may change in time; this would indicate new physics~\cite{Fischer04a, Bize04a, rosenband_frequency_2008, Blatt08a}.  Comparing clocks based on different atoms or molecules may someday make such effects observable. Another example is the application of clock precision to the study of many-body quantum systems~\cite{Martin2013a, Rey2014a}.

\subsection{Ingredients for an atomic frequency standard and clock}

All precise clocks work on the same basic principle.  First, we require a system that exhibits a regular periodic event; that is, its cycles occur at a constant frequency, thereby providing a stable frequency reference and a basic unit of time.  Counting cycles of this frequency generator produces time intervals; if we can agree on an origin of time then the device subsequently generates a corresponding time scale.  For centuries, frequency standards were based on celestial observations, for example, the earth's rotation rate or the duration of one orbit of the earth about the sun \cite{jespersen_sundials_1999}.  For shorter time scales other frequency standards are desirable; classic examples include macroscopic mechanical resonators such as pendulum clocks, John Harrison's famous spring based clocks for maritime navigation, and starting in the early 20th century, quartz crystal resonators \cite{walls_fundamental_1995, vig_quartz_1999}.  However, each of these frequency standards had its limitations; for example, the earth's rotation frequency varies in time, and the frequency stability of macroscopic mechanical resonators are limited by environmental effects such as changes in temperature.

As Maxwell realized, an atom can be an ideal frequency standard because, as far as we know, one atom is exactly identical to another atom of the same species.  Therefore, if we build a device that registers the frequency of a natural oscillation of an atom, say the mechanical oscillations of an electron about the atom's core, all such devices will run at exactly the same frequency (except for relativistic effects discussed below), independent of comparison.  Therefore, the requirement for making an atomic frequency standard is relatively easy to state: we take a sample of atoms (or molecules) and build an apparatus that produces an oscillatory signal that is in resonance with the atoms' natural oscillations.  Then, to make a clock, we simply count cycles of the oscillatory signal.

Frequency standards have been realized from masers or lasers; in the context of clocks perhaps the most important example is the atomic hydrogen maser \cite{goldenberg60,kleppner62} which is still a workhorse device in many standards laboratories.  However, the more common method for achieving synchronization, and the primary one discussed here, is based on observing the atoms' absorption.  Typically, we first prepare the atom in one of the two quantum states ($|1\rangle$ = lower-energy state, $|2\rangle$ = upper state) associated with one of its natural oscillations.  We then use a ``local oscillator" that produces radiation around this oscillation frequency and direct the radiation towards the atoms.  The device will be constructed so that we can detect when the atoms change state; when these state changes occur with maximum probability, then we know that the oscillator frequency is synchronous with the atoms' natural oscillation.  The details of this process are discussed below.

\subsection{Characterization of frequency standards}

The degree to which we can synchronize a local oscillator's frequency to the atoms' natural oscillations is always limited by noise in the measurement protocol we use to establish this synchronization.  In addition, although isolated atoms are in a sense perfect, their natural frequencies can be shifted from their unperturbed values by external environmental effects, typically electric and magnetic fields.  Therefore, we must find a way to calibrate and correct for these ``systematic" frequency shifts.  Even then, there will always be errors in this correction process that we must characterize.  It is therefore convenient to divide the errors into two types: statistical errors that arise from measurement fluctuations and errors in the systematic-effect corrections that are applied to the measured frequencies.  We typically characterize these errors in terms of the fractional frequency errors, $\Delta f/f_0$ where $f_0$ is the reference transition frequency and $\Delta f$ is the frequency error.

For statistical errors, let us first suppose we have a perfect local oscillator whose frequency $f_s$ is near $f_c$, the frequency of the clock atoms under test ($f_c$ may be shifted from $f_0$ due to systematic effects).  We assume we can measure the fractional frequency difference $y \equiv (f_c - f_s)/f_0$ and average this quantity over various probe durations $\tau$.
A commonly used measure of the noise performance of clocks is the Allan variance \cite{allan66,riehle_frequency_2004, riley08}
\begin{equation}
\sigma_y^2(\tau) = \frac{1}{2(M-1)} \sum_{i=1}^{M-1} [\langle y(\tau) \rangle_{i+1}
- \langle y(\tau) \rangle_i ]^2,\label{allan_variance}
\end{equation}
where $\langle y(\tau) \rangle_i$ is the $i$th measurement of the average fractional frequency difference over duration $\tau$ and where we ideally assume there is no dead time between successive measurements $i$ and $i+1$.  The quantity $\sigma_y(\tau)$ is commonly called the stability (but is really proportional to the instability).  More efficient use of data uses overlapping samples of shorter-duration measurements resulting in the ``overlapping" Allan variance.  This and more sophisticated measures, which can reveal the spectrum of the noise are discussed in \citet{riehle_frequency_2004, riley08}, but the essence of the measure is contained in Eq.~(\ref{allan_variance}).  Many sources of noise are well-behaved (stationary) in the sense that if we average the output frequency of our standard for longer times, our precision on the measured frequency also improves ($\sigma_y(\tau)$ decreases).  However, other sources of noise, such as systematic shifts that drift over long durations, will cause $\sigma_y(\tau)$ to level off or increase with increased $\tau$.  Of course, we don't have perfect standards to compare to, so we always observe $\sigma_y(\tau)$ for comparison between two imperfect clocks.  Nevertheless, if we can compare three or more clocks it is possible to extract the noise performance of each separately \cite{riley08}.

Systematic errors are more challenging to document, in part because we may not always know their orgin, or even be aware of them!  If the measured frequency stability does not improve or becomes worse as $\tau$ increases, this indicates some systematic effect that we are not properly controlling.
Even worse is that stability may improve with $\tau$ but we have not accounted for a (constant) systematic offset.  Eventually such effects will likely show up when comparing different versions of the same clock; in the meantime, we must be as careful as possible to account for systematic shifts.

\subsection{Scope of paper}
In this paper we will be primarily interested in the physics of optical clocks, the performance  and limitations of existing devices, and prospects for improvements.  The status of the field has been summarized in various reviews and conference proceedings \cite{luiten_single-ion_2001,gill05,hollberg_optical_2005,maleki08,margolis09,gill_when_2011,derevianko_colloquium:_2011, poli_optical_2013}, so that we will not discuss the details of all experiments.  Rather, we will focus on aspects of a few high-performance clocks to illustrate the problems and issues that must be faced, as well as prospects for further advances in the state-of-the-art.
Our review covers optical atomic clocks based on both trapped single ions and many atoms. For simplicity, we will use the term ``atomic" clocks but of course a molecular or even a nuclear transition might be an equally viable candidate for a frequency reference.

\section{DESIDERATA FOR CLOCKS: QUANTUM SYSTEMS WITH HIGH-FREQUENCY, NARROWLINE RESONANCES}
\label{sec:systems}
\subsection{Stability}

Following the basic idea outlined above, to stabilize the frequency of a local oscillator to an atomic transition, we need to extract a sensitive discriminator signal $d S/d f$ where $S$ is the signal obtained from the atoms and $f$ is the frequency of applied radiation.  This signal can then be used to feed back and stabilize the oscillator's frequency.  There will be fluctuations $\delta S$ on the measured signal $S$ so that assuming no additional noise is injected during the protocol, the corresponding fractional frequency errors of the stabilized local oscillator during one feedback cycle can be expressed as
\begin{equation}
\delta y_1 = \biggl(\frac{\delta f}{f_0}\biggr)_1 = \frac{\delta S}{f_0 (d S/d f)}\ .\label{delta_y_1}
\end{equation}
From this expression we see that we want $f_0$ and $d S/d f$  as large as possible and $\delta S$ as small as possible.  If we denote the frequency width of the atomic absorption feature by $\Delta f$ and the signal strength on resonance as $S_0$, we can re-express Eq.~(\ref{delta_y_1}) as $\delta y_1 = \delta S/(S_0 Q \kappa_S)$, where $Q \equiv f_0/\Delta f$ is the $Q$-factor of the transition and $\kappa_S \equiv (dS/df) \Delta f/S_0$ is a parameter on the order of 1 that depends on the line shape.  From this expression for $\delta y_1$, it appears that the key parameters are signal-to-noise ratio and $Q$.  However, we must remember that this is for a single feedback cycle, which, for a given $Q$, requires a measurement duration $T_m$ proportional to $1/\Delta f$. If $\delta S$ is dominated by white frequency noise we then have for repeated measurements
\begin{equation}\label{sigma_y_2}
\sigma_y(\tau) = \biggl(\frac{\delta f}{f_0}\biggr)_1 \sqrt{\frac{1}{M}} = \frac{\delta S}{f_0 (d S/d f)} \sqrt{\frac{T_m}{\tau}}=\frac{\delta S}{S_0 Q \kappa_S} \sqrt{\frac{T_m}{\tau}} ,
\end{equation}
where $\tau$ is the total measurement duration and $M=\tau/T_m$ is the number of successive measurements.

To stabilize the local oscillator to the atomic transition, we will typically first prepare the atoms in one of the two clock states, here the lower-energy state $|1\rangle$.  We will then excite the clock transition resonance at a frequency near that which gives the maximum value of $(dS/df)/\delta S$, which is usually near or at the half-intensity points of the absorption feature.  In the absence of relaxation this leaves the atom in a superposition state $\alpha |1\rangle + \beta |2\rangle$ with $|\alpha|^2 \simeq 1/2$ and $|\alpha|^2 + |\beta|^2 = 1$.

In most cases discussed in this review, the observed signal is derived by use of what Hans Dehmelt termed the ``electron-shelving" technique \cite{dehmelt_mono-ion_1982}.  Here, one of the two states of the clock transition, say the lower-energy state $|1\rangle$, is excited to a third level by a strongly-allowed electric-dipole ``cycling" transition where this third level can only decay back to $|1\rangle$.  (We assume $|2\rangle$ is not excited by the cycling transition radiation).  By collecting even a relatively small number of fluorescent photons from this cycling transition, we can discriminate which clock state the atom is projected into upon measurement: if the atom is found in the state $|1\rangle$ it scatters many photons, if its optically-active electron is ``shelved" into the upper clock state $|2\rangle$, fluorescence is negligible.  With this method, we can detect the projected clock state with nearly 100 $\%$ efficiency.  When applied to $N$ atoms simultaneously, the atomic signal and its derivative will increase by a factor of $N$.   Upon repeated measurements of the state $\alpha |1\rangle + \beta |2\rangle$, there will be quantum fluctuations in which state the atom is projected into for each atom.  These quantum fluctuations contribute noise $\delta S = \sqrt{N p(1-p)} = \sqrt{N}|\alpha \beta|$ where $p = |\beta|^2$ is the transition probability \cite{itano_quantum_1993}. This ``projection" noise is the standard quantum noise limit in the measurements.  Added noise, for example phase noise from the probe local oscillator, will increase $\sigma_y(\tau)$.

In principle, to stabilize the oscillator to the atomic reference we would only need to probe one side of the absorption line, but in practice it is often necessary to alternately probe both sides of the line and derive an error signal based on the two different values of $p$. Doing so reduces influence of technical noise to the signal.  The feed-back servo is arranged to drive this difference to zero, in which case the mean of the two probe frequencies is equal to the atomic resonance frequency.  Equation~(\ref{sigma_y_2}) still holds, but since the absorption feature will be symmetric to a high degree, probing on both sides of the line makes the stabilization insensitive to slow variations in probe intensity, resonance linewidth, and detection efficiency.

A particularly simple expression for $\sigma_y(\tau)$ holds if we probe the resonance using the Ramsey method of separated fields \cite{ramsey_molecular_1985} with free-precession time $T_m\sim 1/(2\pi \Delta f)$ and assume (1) $\pi/2$ pulse durations are short compared to $T_m$, (2) unity state-detection efficiency, (3) relaxation rates are negligible compared to $1/T_m$, (4) the duration required for state preparation and measurement (dead time) is negligible compared to $T_m$, and (5) noise is dominated by quantum projection noise. In this case \cite{itano_quantum_1993},
\begin{equation}
\sigma_y(\tau) = \frac{1}{2 \pi f_0 \sqrt{N T_m \tau}}\ .
\label{eq:sql}
\end{equation}
This expression clearly shows the desirability of high frequency, large atom numbers, long probe times (with corresponding narrow line-widths), and of course long averaging times $\tau$.  If $N$, $T_m$, and $\tau$ can somehow be preserved, we see that the improvement in $\sigma_y(\tau)$ is proportional to $f_0$. Stated another way, if $N$ and $T_m$ are preserved, the time it takes to reach a certain measurement precision is proportional to $f_0^{-2}$, emphasizing the importance of high-frequency transitions.

\subsection{High-frequency clock candidates}

The advantage of high-frequency transitions had been appreciated for decades during which clock transitions based on microwave transitions (typically hyperfine transitions) prevailed.
Given the importance of high $f_0$ and narrow linewidths, one can ask why we don't make the jump to very high frequencies such as those observed in M\"{o}ssbauer spectroscopy.  For example, a M\"{o}ssbauer transition in $^{109}$Ag has $f_0 \simeq 2.1\times10^{19}$ Hz and a radiative decay time $\tau_\mathrm{decay} \simeq 60$ s corresponding to a natural $Q$ value $\simeq 1.3\times10^{22}$ \cite{alpatov_first_2007, bayukov_observation_2009}.    Even with practical limitations, the performance of actual M\"{o}ssbauer systems is still quite impressive.  For example, consider the 93 keV M\"{o}ssbauer transition in $^{67}$Zn \cite{potzel92}.  Here, $Q$'s of $5.8 \times 10^{14}$ were observed (\cite{potzel92}, Fig. 5) and a statistical precision of $10^{-18}$ was obtained in 5 days.  As is typical in M\"{o}ssbauer spectroscopy, a convenient local oscillator is obtained by using a M\"{o}ssbauer emitter of the same species whose frequency is swept via the first-order Doppler shift when this source is moved at fixed velocity relative to the absorber.  However, systematic effects in \cite{potzel92} were at a level of around $2 \times 10^{-17}$ due primarily to pressure effects in the host material and dispersive lineshape effects.  More importantly in the context of clocks, there is not a way to observe coherence of the local oscillator; that is, there is currently no means to count cycles of the local oscillator or compare clocks based on different transitions.  Moreover, comparison of M\"{o}ssbauer sources over large distances ($\gg 1$ m) is intractable due to the lack of collimation of the local oscillator radiation. On the other hand, if further development of extreme ultraviolet frequency combs~\cite{Jones05a, gohle_frequency_2005, cingoz_direct_2012} does produce spectrally narrow radiation sources in the keV region, it will be attractive to revisit the idea of M\"{o}ssbauer spectroscopy for clock applications.

In the optical region of the spectrum, suitable narrow-linewidth transitions were known to exist in many atoms; however, the missing ingredients until relatively recently were (1) the availability of lasers with sufficiently narrow spectra that could take advantage of these narrow transitions and (2) a convenient method to count cycles of the stabilized (laser) local oscillators.  These requirements have now been met with improved methods to lock lasers to stable reference cavities \cite{Young99a, Ludlow07a, dube_narrow_2009, millo_ultrastable_2009,  jiang_making_2011, Swallows12, kessler_sub-40-mhz-linewidth_2012,mcferran_laser_2012, Bishof2013a}  and the development of optical combs that provide the counters and convenient means for optical frequency comparisons \cite{udem_accurate_1999, diddams_direct_2000,stenger_ultraprecise_2002, Cundiff03a, hollberg_measurement_2005, ye_femtosecond_2005, Schibli08a, grosche_optical_2008, hall_nobel_2006, hansch_nobel_2006}.  These advances mark the beginning of high-precision clocks based on optical transitions.

\subsection{Systematic effects}
\label{sec:systematics}

To a high degree, the systematic frequency shifts encountered in optical atomic clocks are the same as for all atomic clocks.  We can divide the shifts into those caused by environmental perturbations (e.g., electric or magnetic fields) and those which we might call observational shifts.  The latter include instrumental effects such as servo offsets and frequency chirping in optical switches; these are apparatus-specific and best examined in each experimental realization.  More fundamental and universal observational shifts are those due to relativity, which we discuss below.

 \subsubsection{Environmental perturbations}
In simple terms, we need to examine all the forces of nature and consider how each might affect the atomic transition frequencies.  As far as we know, we can rule out the effects of \emph{external} strong and weak forces primarily because of their short range.  Gravitational effects are important but we include them below when discussing relativistic shifts.  The most important effects are due to electromagnetic fields; it is useful to break these into various categories, illustrated by some simple examples.  Details will follow in the discussions of the various clocks.

\paragraph{Magnetic fields}
Static magnetic fields $\vec{B} = B \hat{n}_B$ are often applied purposely to define a quantization axis for the atoms.  Here we implicitly assume the field is uniform, but inhomogeneties must be accounted for in the case of spread atomic samples. Shifts from these fields often cause the largest shifts that must be corrected for but these corrections can often be implemented with high accuracy.  We write
\begin{equation}
f - f_0 = \Delta f_M = C_{M1} B + C_{M2} B^2 + C_{M3} B^3 + \cdots,\label{B_dependence}
\end{equation}
where, for small $B$,the first two terms are usually sufficient.  The energies of clock states will depend on the atom's magnetic moment; for example, the electron spin Zeeman effect in the $^2S_{1/2} \rightarrow {}^2D_{5/2}$ transitions of $^{88}$Sr$^+$ gives a relatively large $C_{M1}$ coefficient on the order of $\mu_B/h \simeq 1.4 \times 10^{10}$ Hz/T where $\mu_B$ is the Bohr magneton and $h$ is Planck's constant.  Nevertheless, if the quantizing magnetic field is sufficiently stable, by measuring pairs of transitions that occur symmetrically around the unshifted resonance we can compensate for this shift \cite{bernard_laser_1998}.  As another example, $^1S_0 \rightarrow {}^3P_0$ transitions in $^{87}$Sr and \Al have a much smaller value of $C_{M1} \sim \mu_N/h$ where $\mu_N$ is the nuclear magneton, thereby reducing the shifts substantially.

For atoms with non-zero nuclear and electron spin, hyperfine structure will be present and both  $C_{M1}$ and $C_{M2}$ can be significant.   In this case we can often use the traditional ``clock" transitions between lower states $|F, m_F = 0\rangle$ and upper states  $|F',m_{F'} = 0\rangle$ where $F, F'$ and $m_F, m_{F'}$ are the total angular momenta and the projections of the angular momenta on the (magnetic field) quantization axis.  For these transitions, $C_{M1} = 0$ and for $B \rightarrow 0, \Delta f_M = C_{M2} B^2$ can be very small.  We can usually determine $B$ to sufficient accuracy by measuring a suitable field-dependent Zeeman transition.  Departures of $B$ from its nominal value $B_0$ might also vary in time.  If these variations are slow enough it might be feasible to intermittantly measure field sensitive transitions, or even the clock transition itself, to correct for or servo-compensate the slow variations~\cite{rosenband_observation_2007}.

Some isotopes of interest do not possess $m=0$ Zeeman sublevels because of their half-integer total angular momentum. An example are alkali-like ions without nuclear spin, where the absence of hyperfine structure facilitates laser cooling. In this case the linear Zeeman shift of the reference transition can be compensated by interrogating
two Zeeman components that are symmetrically shifted like $m \rightarrow m'$ and $-m \rightarrow -m'$ and determining the average of both transition frequencies. Consequently, the number of interrogations required for a frequency determination is doubled. However, this does not compromise the stability of the standard if magnetic field fluctuations are negligible during the time between interrogations. For the operation of $^{87}$Sr lattice clock, alternately interrogating opposite nuclear spin stretched states of $\pm$9/2 and taking their averages greatly suppresses the first order Zeeman shift.  The second-order Zeeman shifts can be determined by fast modulation of the bias magnetic field between high and low values. By using clock transition to directly sample and stabilize the magnetic field, the combined magnetic field related frequency shift can be measured below 1 $\times 10^{-18}$~\cite{Bloom2014,Nicholson2014}.

\paragraph{Electric fields}

Static electric fields at the site of the atoms can arise from potential differences in surrounding surfaces caused by, for example, differences in applied potentials on surrounding conductors, surface contact potential variations, or charge build up on surrounding insulators. Typically, clock states have well defined parity so that first-order perturbations vanish and shifts can often be calculated with sufficient precision in second-order perturbation theory. For the case of trapped ions, the static component of the electric field and corresponding Stark shifts vanish at the equilibrium position of the ions; since they don't move, the static field at their location must be zero. For neutral atom clocks the static electric field effects are usually small, however at the highest levels of accuracy they must be characterized~\cite{Lodewyck11a,Bloom2014} or even stabilized~\cite{Nicholson2014}.

Treating the quadratic Stark shift as a small perturbation of the linear Zeeman splitting, the shift of the state $\vert\gamma JFm\rangle$ is given by~\cite{angel_hyperfine_1968,itano00}
\begin{eqnarray}
h\Delta f_S(\gamma,J,F,m,\bm{E}) &=&-\left(2\alpha_S(\gamma,J)+\alpha_T(\gamma,J,F)\,g(F,m,\beta)\right)\frac{\left\vert\bm{E}\right\vert^2}{4}\label{sshift2}\\
g(F,m,\beta) &=& \frac{3m^2-F(F+1)}{F(2F-1)}\left(3\cos^2\beta-1\right)\,, \nonumber
\end{eqnarray}
where $\beta$ is the angle between the electric field vector and the orientation of the static magnetic field defining the quantization axis.
In general, the Stark shift is composed of a scalar contribution described by polarizability  $\alpha_S$ and, for levels with $J>1/2$ and $F>1/2$, by a tensor part that is proportional to $\alpha_T$.

In addition to a static electric field, AC electric fields can be present from several sources.  Important shifts for both neutrals and ions can arise from laser beams and background blackbody radiation.  For neutral atoms trapped by laser fields, the frequency and polarization of light can be chosen~\cite{Katori03a, Ye08a} so that the AC Stark shifts are the same for both clock levels to a high degree and the clock frequency is nearly unshifted (see Sec.~\ref{section:magic}).  For sympathetically cooled ions as in the $^{27}$Al$^+$ ``logic clock", the cooling light can impinge on the clock ion(s) causing Stark shifts that must be accounted for \cite{rosenband_frequency_2008,chou_frequency_2010}.  Ambient blackbody radiation shifts can be important for both neutrals and ions.  The uncertainty in the shift can be caused by uncertainty in the effective temperature $T$ at the position of the atoms and by uncertainties in the atomic polarizabilities.
In most cases the wavelengths of electric dipole transitions originating  from one of the levels of the reference transition are significantly shorter than the peak wavelength of the blackbody radiation spectrum of $9.7~\mu$m at room temperature. Consequently, a static approximation can be used and the shift is proportional to the differential static scalar polarizability $\Delta\alpha_s$
of the two levels constituting the reference transition and to the fourth power in temperature. This follows from the integration of Planck's radiation law, yielding the mean-squared electric field $\langle  E^2(T)\rangle =(831.9~$V/m$)^2(T({\rm K})/300)^4$.
The dependence of the shift on the specific transition wavelengths and matrix elements may be accounted for in a $T^2$-dependent dynamic correction factor $\eta$ \cite{porsev_multipolar_2006}. With these approximations, the Stark shift due to blackbody radiation is given by:
\begin{equation}
h\Delta f_\mathrm{BBR}=-\frac{\Delta\alpha_s  \langle  E^2(T)\rangle}{2}(1+\eta(T^2)),
\label{bbrshift}
\end{equation}
Since blackbody shifts scale as $T^4$, operation at low temperatures can be advantageous; by operating near liquid Helium temperatures, the shifts are highly suppressed \cite{itano00}.  Tables \ref{tab:ionatomicparams} and \ref{tab:systematics} lists blackbody shifts for some atoms/ions currently considered for optical clocks.

For ions confined in Paul traps, the trapping rf electric fields can produce quadratic Stark shifts.  These can be significant if ambient static electric fields push the ions away from the rf electric field null point in the trap; in this case the ions experience excess ``rf-micromotion", oscillatory motion at the rf trap drive frequency \cite{berkeland_minimization_1998}. The strength of the fields can be determined by observing the strength of rf micromotion induced FM sidebands of  an appropriately chosen optical transition (which need not be the clock transition).
As with the case of AC magnetic fields, the danger for both neutral atoms and ions is that AC electric fields may be present at the site of the atoms that otherwise go undetected.

If one or both of the clock states has a quadrupole moment, shifts can arise due to ambient electric fields gradients which can be strong in ion traps.  In several cases of interest,  one of the clock states is an atomic D level which will have such an atomic quadrupole moment that can give rise to significant shifts.
In the case of atomic ions, atomic quadrupoles can couple to gradients from the Coulomb field of simultaneously trapped ions.  In strongly binding traps where the ion separations are on the order of a few $\mu$m, shifts can be as large as 1 kHz \cite{wineland85}.

Shifts from collisions are typically dominated by electric field effects.  Since a precise theoretical description of these shifts is extremely complicated, experimentalists must typically calibrate them through measurements.  This can be particularly important in neutral atom clocks where multiple atoms might be held in a common location and the shift is dominated by collisions between clock atoms. In this situation scattering cross sections will strongly differ between fermionic and bosonic clock atom species. This is not an issue for ions, which are well separated in the trap.

Collision shifts from hot background gas atoms in vacuum can be even more difficult to characterize. At high vacuum, collisions with background gas atoms occur infrequently and it may be possible to establish a useful upper limit on collisional frequency shifts simply from the observed collision rate. Even tighter bounds can be established by a detailed analysis of the collision process using model potentials of the involved species~\cite{gibble_scattering_2013}. For example, the largest residual gas in the ultrahigh vacuum chamber for the Sr clock is hydrogen. An estimate of the Sr-H$_2$ van der Waals coefficients can be estimated to provide an upper bound of the background collision shift~\cite{Bloom2014}.

\subsubsection{Relativistic shifts}

In addition to environmental effects that perturb an atom's internal states and clock frequency, there can be errors in our determination of the clock atoms' frequency, even when atoms are perturbation free.  The most fundamental of these effects are relativistic shifts, due to the different frames of reference of the atoms, probing lasers, and other atomic clocks.

\paragraph{Doppler shifts}\label{sec:Doppler}

Basically, we want to relate an atom's transition frequency in its frame of reference to the frequency of the probe laser in the ``lab frame," which we assume is locked to the atomic transition \cite{chou_optical_2010}.  The frequency $f$ of the probe laser in  the lab frame has a frequency $f'$ when observed in a moving frame
\begin{equation}
f' = f \gamma \left(1 - \frac{v_{\parallel}}{c} \right),
\end{equation}
where $\gamma = (1 - (v/c)^2)^{-1/2}$, $v$ is atom's velocity relative to the lab frame, $v_{\parallel}$ is the atom's velocity along the probe laser beam direction, and $c$ is the speed of light.  The clock servo ensures that the frequency of the laser in the atom's frame equals the proper atomic resonance frequency $f_0$; that is $\langle f' \rangle = f_0$, where the angle brackets denotes the appropriate average over the laser probe duration.  If we can assume that f is constant over this duration, then $\langle f \rangle = f$, and we have
\begin{equation}
\frac{\delta f}{f_0} = \frac{f - f_0}{f_0} = \frac{1}{\langle \gamma (1 - v_{\parallel}/c) \rangle} - 1
\end{equation}
or
\begin{equation}
\frac{f - f_0}{f_0}= \frac{\langle v_{\parallel} \rangle}{c}
- \frac{\langle v^2 \rangle}{2 c^2}
+ \frac{\langle v_{\parallel} \rangle^2}{c^2} + O(v/c)^3.\label{rel_shift}
\end{equation}
The first term in Eq.~(\ref{rel_shift}), the first-order Doppler shift, can easily be the largest for clocks based on single photon transitions.  Historically, the relatively large size of the first-order Doppler shift was one of the motivations for probing confined atoms as opposed to atoms in an atomic beam.  Early work on the hydrogen maser \cite{goldenberg60} and high resolution hyperfine spectra of trapped $^3$He$^+$ ions \cite{fortson66} showed the advantages of confinement.  Trapping for long durations would seem to guarantee $\langle v_{\parallel} \rangle = 0$.  However, the distance between the mean position of the atoms and the location of the probe laser may be slowly drifting due to for example, thermal expansion, or any change in optical path, such as that due to a change in index of refraction in a transmitting fiber.  For example, to reach $\delta f/f_0 < 10^{-17}$, we must ensure $\langle v_{\parallel} \rangle <$ 3 nm/s.  More generally, any effect that leads to a phase change of laser beam field experienced by the atoms can be included in this category.  Fortunately, many of these effects can be compensated with Doppler cancellation schemes discussed in more detail in Sec.~\ref{sec:distribution}.
However, even with these measures, we must be cautious.  For example, during the laser probe and feedback cycle, there might be periods where the atom's position is correlated with the laser probe period and first-order Doppler shifts might occur.  To detect and compensate for this possibility, one can probe in multiple directions \cite{rosenband_frequency_2008}.

The next two terms in Eq.~(\ref{rel_shift}),  so called second-order Doppler shifts, are a form of time dilation.  Although they are fairly small for room temperature atoms, they may be difficult to characterize since the trapped atoms' velocity distribution may not be simple.  Of course, this was one of the early motivations for laser cooling and now various forms of cooling are used in nearly all high-accuracy clocks.  Even with laser cooling, in the case of ion optical clocks, the uncertainty in the second-order Doppler shift can be the largest systematic uncertainty due to limitations on characterizing the ions' thermal and rf micromotion \cite{rosenband_frequency_2008,chou_frequency_2010}. For neutral atoms laser cooled to near the motional ground state in an optical lattice trap, the primary concern is to reference the local oscillator and lattice laser beams to a common lab frame.

\paragraph{Gravitational red shift}

As predicted by relativity and the equivalence principle, if a gravitational potential difference exists between a source (one clock) and an observer (another clock, otherwise identical), the two clocks run at different rates \cite{vessot80}.  On the surface of the earth a clock that is higher by $\Delta h$ than another clock runs faster by $\delta f/f_0 = g \Delta h/c^2$ where g is the local acceleration of gravity. This phenomenon is regularly observed and taken into account when comparing various optical and microwave standards \cite{wolf_relativistic_1995, petit_computation_1997, blanchet_relativistic_2001, petit_relativistic_2005}.  For $\Delta h = $10 cm, $\delta f/f_0 \simeq 10^{-17}$ and this shift must be accounted for even when making measurements between nearby clocks. However, when clocks are separated by large distances, the differences in gravitational potential are not always easy to determine and may be uncertain by as much as an equivalent height uncertainty of 30 cm ($3 \times 10^{-17}$) \cite{pavlis_relativistic_2003}.  This can be important when comparing the best clocks over long distances \cite{kleppner06}, but might be turned to advantage as a tool in geodesy \cite{vermeer_chronometric_1983, bjerhammar_relativistic_1985, margolis09,chou_optical_2010}, as discussed in more detail in Sec.~\ref{sec:geodesy}.  The very high measurement precision afforded by optical standards forms the basis for proposals of space optical clocks as the most sensitive measurements of this relativistic effect~\cite{Schiller07a,wolf_quantum_2009,Schiller09a} and are described in   Sec.~\ref{sec:clocksinspace}.


\section{SPECTRALLY PURE \& STABLE OPTICAL OSCILLATORS }
\label{sec:oscillators}
As we have seen in the previous Sections, a key ingredient of the optical atomic clock is an optical resonance with a high quality factor.  Since the resonance results from light-atom interaction, both the light used to drive the atomic transition and the atomic states being driven must be highly coherent to achieve a high-$Q$ transition.  Lasers are traditionally viewed as exceptionally coherent sources of optical radiation.  However, relative to the optical coherence afforded by the exceedingly narrow electronic transitions between metastable states of an optical clock, most lasers are far too incoherent.  For this reason, a critical component of optical clock development is laser stabilization for generating highly phase coherent and frequency stable optical sources.

\subsection{Laser stabilization technique}

A simple laser consists merely of an optical gain medium located inside a resonant optical cavity.  The frequency of the laser is derived from the cavity resonance frequency where the laser gain is high.  The output frequency is susceptible to a variety of noise processes involving the gain medium, optical path length changes, other intracavity elements, and amplified spontaneous emission.  Such noise processes limit the temporal coherence of the laser, typically well below the needed coherence time required for high resolution spectroscopy of the optical clock transition.  In practice, a much more well-defined resonant frequency can be realized with a properly designed passive optical cavity, typically a simple two-mirror Fabry-Pérot interferometer.  A laser's frequency can be stabilized to such an optical resonance, yielding highly coherent optical radiation (e.g.~\cite{Young99a, Webster04a, Notcutt05a, Stoehr06a, Oates07a, Ludlow07a, Alnis08a, Dube09a, millo_ultrastable_2009, Zhao09a, dube_narrow_2009, jiang_making_2011, Leibrandt11a, Kessler11a, kessler_sub-40-mhz-linewidth_2012, Bishof2013a}).  To do so successfully, two important criteria must be met.

First, the laser output must be tightly stabilized to the cavity resonance.  This requires the ability to detect the cavity resonance with a large signal-to-noise ratio, together with the ability to adjust the laser frequency sufficiently fast to cancel the laser noise processes as they are detected with the optical cavity.  High bandwidth phase and frequency actuation is achieved using electro- and acousto-optic devices, intra-laser piezoelectric-transducers, diode laser current control, and more.  Many detection schemes exist, but the most widely utilized for high performance laser stabilization is the Pound-Drever-Hall (PDH) technique.  The interested reader is referred to the literature \cite{Drever83a,Black01a,Day92a,Hall92a,Zhu93a} for details on this popular scheme.  Here we simply point out that PDH stabilization utilizes the laser field reflected from the optical reference cavity to detect resonance.  The detection is performed at rf frequencies, by frequency modulating the incident laser field, and detecting the heterodyne beat between the optical carrier, in resonance with the cavity, and the FM sidebands, which are off-resonance and reflected by the cavity.  This rf signal can then be demodulated to yield a signal well-suited for feedback control of the laser frequency to track the cavity resonance.  The modulation frequency can be chosen at sufficiently high frequencies where technical laser amplitude noise is below the photon shot noise.  The modulation scheme, frequently employing an electro-optic phase modulator, can be designed to minimize unwanted residual amplitude modulation that contaminates the cavity resonance signal \cite{WongHall85,Zhang14}.  Intuitively, an optical cavity with a narrower resonance can more sensitively detect laser frequency excursions.  For this reason, high performance laser stabilization typically employs mirrors with very high reflectivity, achieving a cavity finesse approaching $10^6$.

Since PDH stabilization can be used to tightly lock a laser's frequency to the resonant frequency of an optical cavity, the second important criterium for achieving a highly coherent laser source is to ensure that the cavity resonant frequency is stable and immune or isolated from noise sources which cause resonance frequency changes.  Since cavity resonance is achieved for mirror spacing at half-integer multiples of the laser wavelength, the essential detail is to maintain exceptionally stable mirror spacing.  The mirrors are optically contacted to a mechanically-rigid spacer, whose primary function is to hold the mirror spacing constant.  Highly rigid spacer materials and mechanical isolation from ambient vibration sources help limit changes in the cavity length.  Properly chosen design of mechanical support of the cavity spacer and its shape can limit the effect of cavity length changes due to acceleration-driven deformation of the cavity spacer and mirrors \cite{Notcutt05a, Chen06a, Nazarova06a, Webster07a, Ludlow07a, millo_ultrastable_2009, Zhao09a, Webster11a, Leibrandt11a, Leibrandt11b}.  The spacer and mirrors are typically fabricated with materials (such as ultra-low-expansion (ULE) glass or low expansion glass-ceramics) to limit thermal drifts of the cavity length, and sometimes employ special design or material selection to further reduce thermally-driven drifts \cite{Alnis08a, Dube09a, Legero10a, jiang_making_2011}.  The cavity is held in a temperature-stabilized, shielded vacuum system, to thermally isolate the cavity from its environment and to reduce index of refraction fluctuations inside the cavity \cite{Saulson94a}.  Laser power incident on the cavity is typically limited and stabilized, in order to reduce heating noise from residual absorption by the mirrors \cite{Young99a,Ludlow07a}.  The most fundamental noise source stems from thermo-mechanical noise of the cavity spacer, the mirror substrates, and the optical coating~\cite{Numata04a,Notcutt06a,Kessler12a,kessler_sub-40-mhz-linewidth_2012}.  To reduce its influence, cavities sometimes employ special design considerations, including long spacers \cite{Young99a,jiang_making_2011,Nicholson12, amairi_reducing_2013, Bishof2013a}, mirror substrates made from high mechanical $Q$ materials \cite{Notcutt06a,millo_ultrastable_2009,jiang_making_2011}, or cryogenic cooling \cite{Kessler11a,Seel97a,Notcutt95a}. The more recent work has emphasized on the use of crystal materials to construct the cavity spacer and substrates~\cite{kessler_sub-40-mhz-linewidth_2012}, and even the optical coating \cite{cole_tenfold_2013}. An all-crystalline optical cavity has the prospect of stabilizing laser frequency to a small fraction of $10^{17}$, allowing further advances in clock stability and accuracy. Spectral analysis for these advanced stable lasers can be directly accomplished with clock atoms~\cite{Bishof2013a}.

Laser stabilization to optical cavities exploit narrow optical resonances detected with a high signal-to-noise ratio.  While cavities have historically been the most successful choice of optical resonance used for high-bandwidth laser stabilization, other systems can be used, including spectral hole burning in  \cite{Thorpe11a,Chen11a,Julsgaard07a,Strickland00a,thorpe_shifts_2013}, some atomic or molecular resonances (e.g.~\cite{Ye98a,Ye01a}), and optical-fiber delay lines \cite{Kefelian09a}.

\subsection{Remote distribution of stable optical sources}\label{sec:distribution}

Once a coherent optical wave is generated, it must be transmitted to the atomic system for spectroscopy, to an optical frequency comb for counting or linking to other optical or microwave frequency standards, or to other destinations in or outside the laboratory.  This can be done through free space or through optical fiber.  In either case, a variety of perturbing effects (e.g. thermal, acoustic, vibrational) can re-introduce frequency noise with deleterious effects on the laser coherence that has been so carefully realized.  For this reason, techniques for the transfer of coherent optical (or microwave) signals without the addition of noise are vital~\cite{Ma94a}. \cite{Foreman07a,newbury_coherent_2007} highlight optical techniques for the distribution of coherent signals, including microwave signals modulated on an optical carrier, coherent optical carrier transfer, and low-jitter transfer of the fs-pulses of an optical frequency comb.  A basic feature of these techniques is measurement of the additional noise introduced via transfer, followed by noise cancelation by writing the anti-noise onto the transmitted signal.  A popular technique for coherent optical carrier phase transfer exploits a heterodyne Michelson interferometer to measure the added noise and a fast-actuating acousto-optic modulator to cancel it~\cite{Bergquist92a,Ma94a}.  Noise-canceled transfer of a cw-laser plays a prominent role in optical clock measurements and comparisons.  First realized within the laboratory at the ten meter scale \cite{Bergquist92a,Ma94a}, it has now been extended to much longer distances, from many kilometers to hundreds of kilometers and beyond~\cite{Ye03a,Foreman07b,Williams08a,Pape10a,lopez_cascaded_2010,Kefelian09a,grosche_optical_2009,fujieda_all-optical_2011, predehl_920-kilometer_2012, droste_optical-frequency_2013}.  While transfer of an optical frequency signal through one kilometer of fiber would typically limit the transferred signal instability to worse than $10^{-14}$ at one second, proper implementation of noise cancelation techniques can preserve signal stability to below $10^{-17}$ at one second~\cite{Foreman07b,Williams08a}.  Transfer is conveniently achieved over fiber networks, although free-space propagation has been investigated~\cite{Djerroud10a,Sprenger09a,giorgetta_optical_2013} with promising potential.  Fiber network transfer has been used for high performance comparisons of optical frequency standards \cite{Ludlow08b,Pape10a,fujieda_all-optical_2011}, low-noise distribution of microwave signals or for high accuracy absolute frequency measurements \cite{Lopez08a,Narbonneau06a,Ye03a,Marra10a,Jiang08a,Campbell08a,Hong09a,Daussy05a}, and high performance remote timing synchronization \cite{Kim08a,Benedick12a,Holman05a,lopez_simultaneous_2013}.

\subsection{Spectral distribution of stable optical sources}

For many years, the benefits of atomic frequency standards operating at optical frequencies were outweighed by the difficulty of measuring the very high optical frequencies. Except for measurements between optical standards operating at very similar frequencies, comparison among and measurement of optical standards was difficult, as evidenced by the complexity of optical frequency chains (e.g.~\cite{Jennings86a,Schnatz96a}). Within the past 15 years, the development of the optical frequency comb has made optical frequency measurement relatively straightforward \cite{Reichert99a,diddams_direct_2000,Jones00a,Udem02a,Cundiff03a,Fortier03a}. With the pioneers of this technique rewarded by the 2005 Nobel Prize in Physics \cite{Hall06a,Hansch06a}, these optical measurements are now made regularly with amazing precision in laboratories around the world.  Furthermore, these optical combs have demonstrated the ability to phase coherently distribute an optical frequency throughout the optical spectrum, and even to the microwave domain.

\begin{figure}[t]
    \centering
    \includegraphics[width=3.5in]{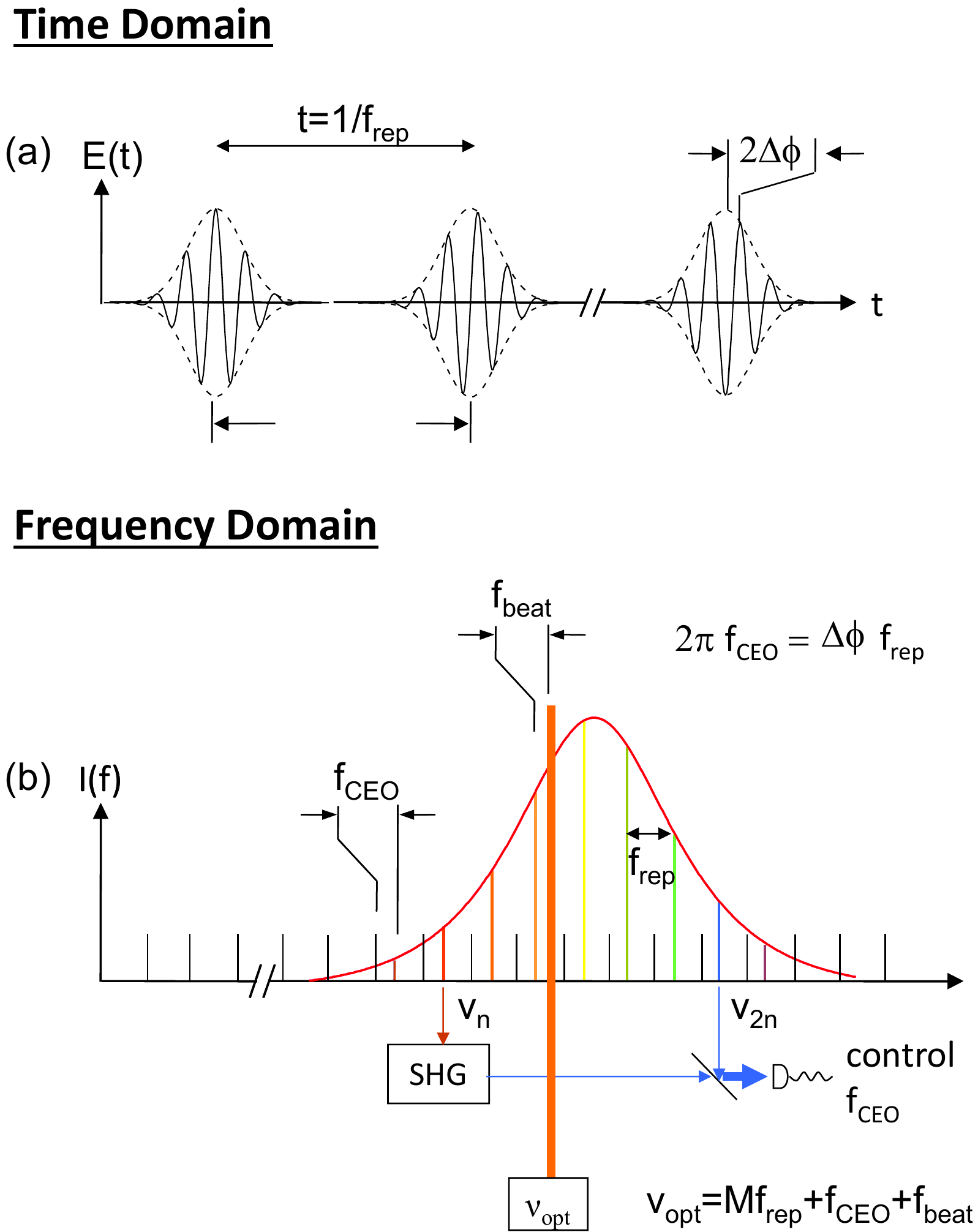}
    \caption[The frequency comb]{\label{FigComb} (a) In the time domain, the laser output generates fs pulse-width envelopes separated in time by $1/f_{rep}$.  Another important degree of freedom is the phase difference between the envelope maximum and the underlying electric field oscillating at the carrier optical frequency. (b) By Fourier transformation to the frequency domain, the corresponding frequency comb spectrum is revealed.  Each tooth in the comb, a particular single frequency mode, is separated from its neighbor by $f_{rep}$.  The relative carrier-envelope phase in the time domain is related to the offset frequency $f_{CEO}$ in the frequency domain.  $f_{CEO}$ is given by the frequency of one
    mode of the comb (e.\ g.\ $\nu_n$) modulo $f_{rep}$, and can be measured and stabilized with a f-2f interferometer.  In this interferometer, one comb mode, $\nu_n$, is frequency doubled and heterodyne beat with the comb mode at twice the frequency, $\nu_{2n}$.  Thus, by stabilizing $f_{CEO}$ and $f_{rep}$ to a well known frequency reference, each comb mode frequency is well known.  Measurement of the frequency of a poorly known optical frequency source (e.\ g.\ previously measured at the resolution of a wavemeter) can be determined by measuring the heterodyne beat between the frequency source and the frequency comb.}
\end{figure}

The optical frequency comb outputs laser pulses with temporal widths at the fs timescale and with a repetition rate of millions or billions of pulses per second. The advent of few-cycle lasers with a few femtosecond pulse width, where ultrafast Kerr-lens mode-locking mechanism ensures phase locking of all modes in the spectrum, along with the spectral broadening via microstructured fibers, have greatly facilitated the development of wide bandwidth optical frequency combs and their phase stabilization.  As in Fig.~\ref{FigComb}, the frequency and phase properties of this pulse train are given by two degrees of freedom: the relative phase between the carrier wave and the pulse envelope (known as the carrier envelope offset), and the pulse repetition rate. Applying a Fourier transform to this pulse train, the laser output consists of a comb of many single frequency modes.  The mode spacing is given by the laser repetition rate, and the spectral range covered by the frequency comb is related to the temporal width of each pulse. The frequency of each comb mode is given as a multiple of the mode spacing ($f_{rep}$) plus a frequency offset ($f_{CEO}$) which is related to the carrier envelope phase offset~\cite{telle_carrier-envelope_1999, udem_accurate_1999, Jones00a}. Control of these two rf frequencies yields control over the frequency of every comb mode \cite{udem_optical_2002, Ye00a, Ma04a}.  If these frequencies are stabilized to an accurate reference (caesium), the optical frequency of a cw-laser or optical frequency standard can be determined by measuring the heteroydne beat between the comb and optical standard.  A coarse, independent measurement of the unknown laser frequency using a commercially available wavelength meter allows one to determine which comb mode, $N$, makes the heterodyne beat with the laser.  The laser frequency is then determined straightforwardly by $\nu_{laser}= N f_{rep} + f_{CEO} \pm f_{beat}$, where $f_{beat}$ is the measured heterodyne beat frequency and the $\pm$ is determined by whether the comb mode or the unknown laser is at higher frequency. In this way, optical standards can be measured against caesium microwave standards.  Furthermore, by stabilizing the comb frequency directly to an optical standard, the comb allows direct comparison of optical standards at different frequencies within the spectral coverage of the comb~\cite{Schibli08a,nicolodi_spectral_2014}. These measurements can be made at the stability of the optical standards themselves, without being limited by the lower stability of most microwave standards.  The fs comb, using now standard laboratory techniques, thus enables microwave-to-optical, optical-to-microwave, and optical-to-optical phase-coherent measurement and distribution at the precision level slightly better than the current best atomic clocks~\cite{ma_optical_2004, ma_frequency_2007, kim_drift-free_2008, lipphardt_stability_2009, zhang_sub-100_2010, nakajima_multi-branch_2010, fortier_generation_2011, hagemann_providing_2013, fortier_photonic_2013, inaba_spectroscopy_2013}.

\section{MEASUREMENT TECHNIQUES OF AN OPTICAL STANDARD }
\label{sec:techniques}

All optical frequency standards that have been realized with cooled and trapped atoms are of the passive type, i.e. the oscillator of the standard is not the atomic reference itself, but a laser source whose output frequency is stabilized to the atomic signal. A further common feature of these standards is that the requirements of initial cooling and state preparation of the atoms lead to an operation in a cyclic sequence of interrogations and measurements.
This is in contrast to established atomic clocks like caesium clocks with a thermal atomic beam and hydrogen masers, that provide a continuous signal. In the optical frequency standard, the laser has to serve as a flywheel that bridges the intervals when no frequency or phase comparison with the atoms is possible. Its intrinsic frequency stability, the method for interrogating the atoms, and the use of the atomic signal for the frequency stabilization need to be considered together in the overall system design of the frequency standard. In this section we will discuss generic features of the methods and techniques that are applied for these purposes.

\subsection{Clock cycles and interrogation schemes}

The repetitive operation cycle of an optical frequency standard with cooled and trapped atoms consists of three distinct stages during which the following operations are performed:
(i) cooling and state preparation, (ii) interrogation, (iii) detection and signal processing.

For a clock with neutral atoms, the first phase comprises loading of a magneto optical trap or of an optical dipole trap from an atomic vapor or from a slow atomic beam. In the case of trapped ions, the same particles are used for many cycles, but some Doppler or sideband laser cooling is necessary to counteract heating from the interaction of the ion with fluctuating electric fields. The conditions that are applied during this trapping and cooling phase include inhomogenous magnetic fields and resonant laser radiation on dipole-allowed transitions. This leads to frequency shifts of the reference transition that can not be tolerated during the subsequent interrogation phase.
The first phase of the clock cycle is concluded with preparation of the initial lower-energy state of the clock transition by means of optical pumping into the selected hyperfine and magnetic sublevel.
Depending on the loading and cooling mechanism, this phase takes a time ranging from a few ms to a few hundred ms.

Before starting the interrogation, all auxiliary fields that would lead to a frequency shift of the reference transition need to be extinguished. Resonant lasers that are used for cooling or optical pumping are  usually blocked by mechanical shutters because the use of acousto-optic or electro-optic modulators alone does not provide the necessary extinction ratio. A time interval of a few ms is typically required to ensure the reliable closing of these shutters.

During the interrogation phase, radiation from the reference laser is applied to the atom. In an optimized system, the duration of this phase determines the Fourier-limited spectral resolution or line $Q$ of the frequency standard. Provided that the duration of the interrogation is not limited by properties of the atomic system, i.e. decay of the atomic population or coherence or heating of the atomic motion, it is set to the maximum value that is possible before frequency or phase fluctuations of the reference laser start to broaden the detected line shape. For a reference laser that is stabilized to a cavity with an instability $\sigma_y$ limited by thermal noise to about $5\times 10^{-16}$ around 1~s, a suitable duration of the interrogation interval is several 100~ms up to 1~s, resulting in a Fourier-limited linewidth of about 1~Hz.

Referring to pioneering work on molecular beams in the 1950s \cite{ramsey_molecular_1985}, one distinguishes between Rabi excitation with a single laser pulse and Ramsey excitation with two pulses that are separated by a dark interval. In Ramsey spectroscopy, the two levels connected by the reference transition are brought into a coherent superposition by the first excitation pulse and the atomic coherence is then allowed to evolve freely. After the second excitation pulse the population in one of the levels is detected, which shows the effect of the interference of the second pulse with the time-evolved superposition state. Assuming that the total pulse area is set to $\pi$ on resonance, Rabi excitation possesses the advantage of working with lower laser intensity, leading to less light shift during the excitation. Ramsey excitation, on the other hand, provides a narrower Fourier-limited linewidth for the same interrogation  time. If the duration of the excitation pulses is much shorter than the dark interval, Ramsey excitation keeps the atoms in a coherent superposition of ground and excited states that is most sensitive to laser phase fluctuations  -- with the Bloch vector precessing in the equatorial plane -- for a longer fraction of the interrogation time than Rabi excitation.

Generalizations of the Ramsey scheme with additional pulses permit one to reduce shifts and broadening due to inhomogeneous excitation conditions or shifts that are a result of the excitation itself. An ''echo'' $\pi$-pulse during the dark period may be used to rephase an ensemble of atoms that undergoes inhomogoenous dephasing \cite{warren_multiple_1983}. An example of such an excitation-related shift is the light shift and its influence may readily be observed in the spectrum obtained with Ramsey excitation~\cite{hollberg_measurement_1984}:
The position and shape of the envelope reflects the excitation spectrum resulting from one of the pulses, whereas the Ramsey fringes result from coherent excitation with both pulses and the intermediate dark period. The fringes are less shifted than the envelope, because their shift is determined by the time average of the intensity.  A sequence of three excitation pulses with suitably selected frequency- and phase steps can be used to cancel the light shift and to efficiently suppress the sensitivity of the spectroscopic signal to variations of the probe light intensity~\cite{Zanon06a, yudin_hyper-ramsey_2010,huntemann_generalized_2012}. While Rabi excitation is often used in optical frequency standards because of its experimental simplicity, these examples show that the greater flexibility of Ramsey excitation may provide specific benefits.

After the application of the reference laser pulses, the clock cycle is concluded by the detection phase. In most cases, the atomic population after an excitation attempt is determined by applying laser radiation to induce resonance fluorescence on a transition that shares the lower state with the reference transition. This scheme was proposed by Dehmelt and is sometimes called electron shelving \cite{dehmelt_mono-ion_1982}. In the single-ion case, the absence of fluorescence indicates population of the upper state and the presence of fluorescence population of the lower state. The method implies an efficient quantum amplification mechanism, where the absorption of a single photon can be read out as an absence of many fluorescence photons. It is therefore also advantageously used for large atomic ensembles. If the number of photons detected from each atom is significantly greater than $1$, photon shot-noise becomes negligible in comparison to the atomic projection noise.

A disadvantage of the scattering of multiple fluorescence photons is that it destroys the induced coherence on the reference transition and that it even expels trapped neutral atoms from an optical lattice. In a lattice clock this makes it necessary to reload the trap with atoms for each cycle. Since the loading and cooling phase takes a significant fraction of the total cycle time, reusing the same cold atoms would permit a faster sequence of interrogations, thereby improving the frequency stability. This can be realized in a
non-destructive measurement that detects the atomic state not via absorption but via dispersion as a phase shift induced on a weak
off-resonant laser beam~\cite{lodewyck_nondestructive_2009}.
If in addition to observing the same atoms, as it is the case with trapped ions, the internal coherence could also be maintained from one interrogation cycle to the next, a gain in stability can be obtained. If the atomic phase can be monitored over many cycles without destroying it, the frequency instability would average with $\sigma_y\propto \tau^{-1}$ like for white phase noise, instead of  $\sigma_y\propto \tau^{-1/2}$ as for white frequency noise in a conventional atomic clock. Such an atomic phase lock has been analyzed and an experimental realization proposed based on a measurement of Faraday rotation with trapped ions~\cite{shiga_locking_2012,Vanderbruggen2013} and for a dispersive interaction in a generic clock~\cite{borregaard_near-heisenberg-limited_2013}.

\subsection{Atomic noise processes} \label{atomicnoise}

In the atomic population measurement described above, noise may arise from fluctuations in the absolute atom number $N$ and in the atomic population distribution. For the frequency standards with cold trapped ions, $N$ is unity or a small number that is controlled in the beginning of each cycle, so that fluctuations are eliminated. If new atoms are loaded for each cycle from a reservoir, one may expect relative variations in the atom number $\delta N$. Since fluorescence detection permits to measure the atom number in each cycle, however, signals may be normalized to the atom number, so that the contribution from atom number fluctuations to the instability of the frequency standard scales as
$(\frac{1}{Nn_\mathrm{ph}}+\frac{2\delta N^2}{N^2})^{1/2}$, where the first term accounts for shot noise during detection of $n_\mathrm{ph}$ photons and the second term accounts for fluctuations in the atom number between cycles \cite{santarelli_quantum_1999}.

Sometimes the most severe noise contribution comes from quantum noise in the state measurement: physical measurement of a quantum system can be modeled by a Hermitian operator acting on the wave function of the system being measured, and the result of that measurement is an eigenvalue of the operator.  Thus, measurement of a superposition of eigenstates yields one of the corresponding eigenvalues, a statistical outcome given by the superposed weighting of the eigenstates.  This implies measurement fluctuation as the wavefunction collapses into a projection along a particular eigen-basis.  Let us consider the simple case of a single ion. The two levels that are connected by the reference transition
are denoted as $|1\rangle$ and $|2\rangle$ and it is assumed that the ion is initially prepared in the lower state $|1\rangle$.
After an excitation attempt the ion generally will be in a superposition state $\alpha |1\rangle + \beta |2\rangle$ and the measurement with the electron shelving scheme is equivalent to determining the eigenvalue $P$ of the projection operator $\hat P=|2\rangle\langle 2|$.
If no fluorescence is observed (the probability for this outcome being $p=|\beta|^2$) the previous excitation attempt is regarded  successful ($P=1$), whereas the observation of fluorescence indicates that the excited state was not populated ($P=0$). In one measurement cycle only one binary unit of spectroscopic information is obtained. Under conditions where the average excitation probability $p$ is $0.5$, the result of a sequence of cycles is a random sequence of zeros and ones and the uncertainty in a prediction on the outcome of the next cycle is always maximal.   These population fluctuations and their relevance in atomic frequency standards were first discussed by Itano et al., who named the phenomenon quantum projection noise (QPN) \cite{itano_quantum_1993}. A simple calculation shows that the variance of the projection operator is given by \cite{itano_quantum_1993}
\begin{equation}
(\Delta \hat P)^2=p(1-p).	
\end{equation}
For $N$ uncorrelated atoms, the variance is $N$-times bigger. For atoms with correlated state vectors, so-called spin squeezed states
\cite{wineland_spin_1992},  the variance can be smaller than this value, allowing for frequency measurements with improved stability
\cite{bollinger_optimal_1996} (see Sec.~\ref{sec:squeezing}).

In the servo-loop of an atomic clock, quantum projection appears as white frequency noise, leading to an instability as given in Eq.~\ref{eq:sql}, and decreasing with the averaging time like  $\sigma_y\propto \tau^{-1/2}$. It imposes the long-term quantum noise limit of the clock, that can be reached if an oscillator of sufficient short-term stability, i.e. below the quantum projection noise limit for up to a few cycle times, is stabilized to the atomic signal.

\subsection{Laser stabilization to the atomic resonance} \label{laserstabatom}

In an optical clock the frequency of the reference laser needs to be stabilized to
the atomic reference transition. In most cases, the error signal for the frequency lock
is derived by modulating the laser frequency around the atomic resonance and by measuring the resulting modulation of the frequency-dependent excitation probability $p$ to the upper atomic level. With a cyclic operation imposed already by the requirements of laser cooling and state preparation, the frequency modulation may be realized conveniently by interrogating the atoms with alternating detuning below and above resonance in subsequent cycles.
The value of the detuning will be chosen in order to obtain the maximum slope of the excitation spectrum, and is typically  close to the half linewidth of the atomic resonance.

Suppose the laser oscillates at a frequency $f$, close to the center of the reference line.
A sequence of $2z$ cycles is performed in which the atoms are interrogated
alternately at the frequency $f_+=f+\delta_m$ and at $f_-=f -\delta_m$.
The sum of the excited state populations is recorded as $P_+$ at $f_+$ and $P_-$ at $f_-$.
After an averaging interval of $2z$ cycles an error signal is calculated as
\begin{equation}
e= \delta_m \frac{P_+-P_-}{z},	
\label{errsig}
\end{equation}
and a frequency correction $g\cdot e$ is applied to the laser frequency before the next averaging interval is started:
\begin{equation}
f \rightarrow f+g\cdot e.	
\end{equation}
The factor $g$ determines the dynamical response of the servo system and can be regarded as the servo gain.
Since the frequency correction is added to the previous laser frequency, this scheme realizes an integrating servo loop \cite{bernard_laser_1998,barwood_development_2001,peik_laser_2006}

The time constant and the stability of the servo system are determined by the choice of the parameters $g$ and $z$.
If the laser frequency $f$ is initially one half linewidth below the atomic resonance and if $p_{\rm max}=1$,
the resulting
value of $(P_+-P_-)/z$ will also be close to one.
Consequently, with $g\approx 1$, the laser frequency will be corrected in a single step.
If $g\ll 1$, approximately $1/g$ averaging intervals will be required to bring the
frequency close to the atomic resonance and the demands on the short-term stability of the probe laser become more stringent.
For  $g\approx 1$ and a small value of $z$,
the short-term stability of the system may be unneccessarily degraded by strong fluctuations in the error signal because of quantum projection noise, especially if only a single ion is interrogated.
For $g\approx 2$, one expects unstable servo behaviour with the laser frequency jumping between $-\delta_m$ and $+\delta_m$.

A servo error may occur if the probe laser frequency is subject to drift, as it is commonly the case if  the short-term frequency stability is derived from a  Fabry-Pérot cavity which is made from material that shows aging or in the presence of slow temperature fluctuations. Laser frequency drift rates
$|df/dt|$ in the range from mHz/s up to  Hz/s are typically observed.
For a first-order integrating servo with time constant $t_{\rm servo}$, an average drift-induced error $\bar e =t_{\rm servo}\,df/dt$ is expected as the result of a constant linear drift.
Since the minimally achievable servo time constant has to exceed several cycle times  for stable operation, such a drift-induced error may not be tolerable.
An efficient reduction of this servo error is obtained with the use of a
second-order integrating servo algorithm \cite{peik_laser_2006} where
a drift correction $e_{\rm dr}$ is added to the laser frequency in regular time intervals $t_{\rm dr}$
\begin{equation}
f \stackrel{t_{\rm dr}}{\longrightarrow} f+e_{\rm dr}.	
\end{equation}
The drift correction is calculated from the integration
of the error signal Eq.~\ref{errsig} over a longer
time interval $T_{\rm dr}\gg t_{\rm dr}$
\begin{equation}
e_{\rm dr} \stackrel{T_{\rm dr}}{\longrightarrow} e_{\rm dr}+k \sum_{T_{\rm dr}} e,	
\end{equation}
where the two gain coefficients are related by $k\ll g$.

In the case of Ramsey excitation, an error signal may also be obtained by alternately applying phase steps of $\pm \pi/2$ to one of the excitation pulses while keeping the excitation frequency constant \cite{ramsey_molecular_1985,letchumanan_optical_2004,huntemann_generalized_2012}.  Whether a more precise lock is achieved with step-wise frequency- or phase-modulation depends on specific experimental conditions: While the former is more sensitive to asymmetry in the lineshape or a correlated power modulation, the latter requires precise control of the size of the applied phase steps.

Because of the time needed for preparation and read-out of the atoms, a dead time is introduced into each cycle during which the oscillator frequency or phase cannot be compared to the atoms. As first pointed out by Dick et al. \cite{Dick87a,dick_local_1990}, this dead time will lead to degraded
long-term stability of the standard because of down-conversion of frequency noise of the interrogation oscillator at Fourier frequencies near the harmonics of the inverse cycle time $1/t_c$.  The impact of the effect on clock stability depends on the fraction of dead time, the interrogation method (Rabi or Ramsey) and on the noise spectrum of the laser~\cite{Santarelli98a,Dick87a}:
\begin{equation}
    \label{eqn:stability4}
    \sigma _y \left(\tau\right) =
    \frac{1}{f_0}\sqrt{\frac{1}{\tau}\sum_{m=1}^{\infty}{\left(\frac{g_{c,m}^2}{g_0^2}+\frac{g_{s,m}^2}{g_0^2}\right)S_f\left(\frac{m}{T_c}\right)}}.
\end{equation}
Here $S_f(m/T_c)$ is the one-sided frequency noise power spectral density of the free running probe laser (local oscillator) at the Fourier frequency $m/T_c$, where $m$ is a positive integer.  The factors $g_{c,m}$ and $g_{s,m}$ correspond to the Fourier cosine and sine series coefficients giving the sensitivity spectral content at $f=m/T_c$ \cite{Santarelli98a}, and contain the physics of the atom laser interaction. For the case of Ramsey excitation one finds \cite{Santarelli98a}:
\begin{equation}
\sigma_{y\,{\rm lim}}(T)\approx 	 \frac{\sigma_{y\,{\rm osc}}}{\sqrt{2 \ln2}}\left| \frac{\sin(\pi t/t_c)}{\pi t/t_c}\right|\sqrt{\frac{t_c}{T}}
\end{equation}
where $\sigma_{y\,{\rm osc}}$ is the flicker floor instability of the oscillator. With achieved experimental parameters like $t/t_c>0.6$ and a flicker floor $\sigma_{y\,{\rm osc}}<5\cdot 10^{-16}$ \cite{kessler_sub-40-mhz-linewidth_2012}, it can be seen that the limitation from the Dick effect
$\sigma_{y\,{\rm lim}}\approx 2\cdot 10^{-16}\sqrt{t_c/T}$ is well below the quantum projection noise limited instability for single-ion clocks, but may impose a limit on the potentially much lower instability of neutral atom lattice clocks. For the frequency comparison between two atomic samples, the Dick effect may be suppressed by synchronous interrogation with the same laser \cite{takamoto_frequency_2011,chou_quantum_2011,Nicholson12}, whereas for improved stability of the clock frequency, a single oscillator may be locked to two atomic ensembles in an interleaved, dead-time free interrogation \cite{dick_local_1990,biedermann_zero-dead-time_2013, hinkley2013}.

\section{TRAPPED ION OPTICAL FREQUENCY STANDARDS}
\label{sec:ion_standards}
The invention of electro-magnetic traps for charged particles by Paul and Dehmelt in the 1950s marked an important step towards realizing the ideal environment for precision spectroscopy: an unperturbed system with long trapping times. Ion traps have played an important role in spectroscopy and precision measurements ever since, which was recognized by awarding the 1989 Nobel Prize in Physics to Dehmelt, Paul, and Ramsey \cite{dehmelt_experiments_1990,paul_electromagnetic_1990,ramsey_experiments_1990}. The absence of a magnetic field made Paul traps the preferred choice over Penning traps for frequency standards, thus avoiding undesired internal level shifts through the Zeeman effect. The basic principle behind the simplest form of a three-dimensional Paul trap is a time-varying electric quadrupole potential in which the balance between the Coulomb force and the inertia of the ions keeps the ions trapped \cite{straubel_zum_1955,paul_ionenkafig_1958,paul_electromagnetic_1990}. The traps typically provide several eV deep potentials, offering trap lifetimes that are limited by (photo-)chemical reactions with background gas atoms and range from several hours to months, depending on the atomic species and the background gas pressure. In spherical 3D Paul traps, only a single ion can be trapped at zero field. Linear Paul traps allow storage of strings of ions \cite{raizen_linear_1992}, potentially allowing an improvement in clock stability by interrogating several ions at once \cite{herschbach_linear_2012,pyka_high-precision_2013-1}. However, achieving the zero-field condition for many ions is a technological challenge. As a consequence, all optical single-ion frequency standards implemented up to now trade stability for accuracy and use a single ion.

The idea of using trapped ions as optical frequency references dates back to Dehmelt \cite{dehmelt_proposed_1973}, who proposed several species and experimental implementations \cite{dehmelt_proposed_1975,dehmelt_proposed_1975-1}, including the electron-shelving technique \cite{dehmelt_proposed_1975-2}. Doppler or sideband laser cooling \cite{hansch_cooling_1975, wineland_proposed_1975, neuhauser_optical-sideband_1978, neuhauser_visual_1978, wineland_radiation-pressure_1978,neuhauser_localized_1980,wineland_spectroscopy_1981}  localize the ion in a few ten nanometer large wavepacket around the zero point of the field. This strong localization in a nearly trapping-field-free environment allows spectroscopy in the recoil-free Lamb-Dicke regime \cite{dicke_effect_1953}. The experimental realization of Dehmelt's electron shelving state detection technique by observing quantum jumps in \Ba \cite{sauter_observation_1986,nagourney_shelved_1986} and \Hg \cite{bergquist_observation_1986} was an important prerequisite for high SNR spectroscopy of few particle systems. High resolution optical spectroscopy of trapped ions was first accomplished by optical two-photon excitation on the dipole-forbidden {S-D} transition  in a cloud of \Hg ions \cite{bergquist_energy_1985} and by direct laser excitation on a single \Hg ion \cite{bergquist_recoilless_1987}, laying the foundation for the development of optical ion clocks.


There are a number of excellent previous reviews on trapped ions and applications to microwave spectroscopy \cite{dehmelt_radiofrequency_1968, dehmelt_radiofrequency_1969, blatt_current_1992,fisk_trapped-ion_1997,wineland_high-resolution_1983}, early optical spectroscopy \cite{dehmelt_coherent_1981,dehmelt_mono-ion_1982}, and optical frequency standards  \cite{luiten_single-ion_2001,riehle_frequency_2004,hollberg_optical_2005,maleki08,margolis09,gill_when_2011,poli_optical_2013}.
In the following, we will discuss the principles and operation of trapped ion optical frequency standards and focus on the features and limitations of some of the most developed systems.


\subsection{Trapping Ions}
\label{sec:iontraps}
According to Earnshaw's theorem, stable trapping of charged particles in free space using only DC fields is not possible. This is a direct consequence of the Laplace equation for electro-static fields. Oscillating electro-magnetic fields provide a way around this fundamental limitation. The Paul trap \cite{paul_ionenkafig_1958, fischer_dreidimensionale_1959, paul_electromagnetic_1990} is a prime example for the realization of such a trap. It employs an oscillating quadrupole potential, resulting in stable confinement of a charged particle for certain operation parameters \cite{meixner_mathieusche_1954, mclachlan_theory_1947, ghosh_ion_1995}. In the most general case, we will consider a superposition of a static quadrupole and an AC electric quadrupole potential oscillating at an angular frequency $\Orf$. The electric fields are generated by two sets of electrodes with characteristic length scales $\RDC$ and $\Rrf$ and applied voltages $\VDC$, $\Vrf$, generating the DC and AC potential, respectively. The total potential can then be written as
\begin{equation}\label{eq:trappot}
\phi(\vec{r},t)= \VDC\frac{\alpha_x x^2+\alpha_y y^2+\alpha_z z^2}{2\RDC^2}+\Vrf\cos\Orf t\frac{\tilde{\alpha}_x x^2+\tilde{\alpha}_y y^2+\tilde{\alpha}_z z^2}{2 \Rrf^2},
\end{equation}with parameters $\kappa$, $\alpha_j$ and $\tilde{\alpha}_j$ that are determined by the electrode geometry. Laplace's equation $\Delta\phi(\vec{r},t)=0$ imposes the relations
\begin{equation}
\label{eq:laplace}
\sum_{j=x,y,z}\alpha_j=0 \;\;\; \mathrm{and}\;\;\; \sum_{j=x,y,z}\tilde{\alpha}_j=0
\end{equation}
between the geometrical factors. For spherical Paul traps, $\alpha_x=\alpha_y=-\frac{1}{2}\alpha_z$ and $\tilde{\alpha}_x=\tilde{\alpha}_y=-\frac{1}{2}\tilde{\alpha}_z$. Three-dimensional confinement of the charged particle is achieved solely through dynamical electric forces. Implementations of this type of traps are discussed in Sec.~\ref{sec:sphericaltraps}. Another popular choice of the geometry parameters is
$\alpha_x=\alpha_y=-\frac{1}{2}\alpha_z$ and $\tilde{\alpha}_x=-\tilde{\alpha}_y$, $\tilde{\alpha}_z=0$, corresponding to a linear Paul trap, discussed in Sec.~\ref{sec:lineartraps}. Radial ($x,y$) confinement is provided by the two-dimensional dynamical quadrupole potential, whereas axial ($z$) trapping is achieved through a three-dimensional static quadrupole potential. Deviations from the cylindrical symmetry can be described by $\alpha_x\neq \alpha_y$ and $\tilde{\alpha}_x\neq \tilde{\alpha}_y$, while maintaining Eqs.~\ref{eq:laplace}.

In this trapping potential, the ion performs slow harmonic oscillations with secular frequencies $\omega_j$ and a superimposed micromotion oscillation at the trap drive frequency. Micromotion is intrinsic to the trapping concept and thus unavoidable. Under stable trapping conditions, typically achieved by choosing $\Orf\gg\omega_j$, the micromotion amplitude is smaller by an order of magnitude compared to the amplitude of secular motion $u_j$. In this regime time scales separate and the effects of micromotion can be largely neglected. The dynamic and static trapping potentials can be approximated by a harmonic pseudo-potential of the form \citep{dehmelt_radiofrequency_1968}
\begin{equation}
\Phi_P(\vec{r})=\frac{1}{2}\sum_j m\omega_j^2 u_j^2.
\end{equation}
The trap frequencies are given by
\begin{equation}
\omega_j\approx\sqrt{\frac{Q\alpha_j\VDC}{mR^2_\mathrm{DC}}+\frac{Q^2\tilde{\alpha}_j^2V^2_\mathrm{AC}}{2\Orf^2m^2 \Rrf^4}},
\end{equation}
which are typically on the order of a few 100~kHz to a few MHz in traps for optical clocks.


In optical clocks, second order Doppler shifts from residual motion as further discussed in Sec.~\ref{sec:ionMotion} are significantly reduced through laser cooling of the trapped ion \cite{wineland_laser-cooling_1987}. Doppler cooling typically reduces the temperature of a single trapped ion to below mK temperatures, corresponding to a motional amplitude of $u_{j}\sim \sqrt{\frac{k_B T}{m\omega^2}}\sim 70$~nm for a single \Ca ion in a trap of 1~MHz trap frequency. This illustrates that the ion is very well localized and probes only the very bottom of the trap, where it is harmonic.

In the derivation of Eq.~\ref{eq:trappot}, we have implicitly assumed that the symmetry axis for the DC and AC electrodes coincide. In principle, this can be achieved through precision machining of the electrodes and by nulling any external DC or AC electric field that pushes the ion away from the trap center. In practice, however, machining tolerances and insufficiently compensated stray fields push the ion into the rf trapping field, causing so-called excess micromotion \cite{berkeland_minimization_1998}. Stray fields are believed to arise from patch charges on the electrodes and insulators, generated by contact potentials or charge buildup from ionization via electron bombardment during loading, or through the photo-effect from UV lasers \cite{harlander_trapped-ion_2010}. Patch fields can be nulled through additional compensation electrodes. However, they tend to fluctuate on slow time scales and need to be compensated from time to time \cite{tamm_stray-field-induced_2009}. There are several techniques for micromotion compensation. The simplest technique is based on the observation of the ion's position as the trapping conditions are changed: If patch fields are compensated, the ion does not move as the rf potential is lowered. In the non-resolved sideband regime, a phase-synchronous detection of fluorescence photons with the trap rf exhibits a modulation as a function of the relative phase in the presence of micromotion \cite{blumel_chaos_1989, berkeland_minimization_1998}. In the resolved sideband regime, the Rabi frequency of a micromotion sideband of the clock transition scales with the modulation index of the excess micromotion \cite{leibfried_quantum_2003}. A fourth technique uses parametric heating of the secular motion through amplitude modulation of the trap rf at the secular motional frequencies \cite{narayanan_electric_2011, ibaraki_detection_2011}. All techniques require probing the ion's micromotion along three non-coplanar directions. For two-ion crystals with ions having a different mass, the radial confinement of the two ions differs. Radial stray electric fields therefore result in a tilting of the ion crystals' symmetry axis with respect to the trap axis. As a consequence, radial and axial two-ion modes become coupled, which leads to additional motional sidebands when probing the excitation spectrum in the resolved sideband regime along the trap axis. Nulling of these sidebands provides another means for micromotion compensation \cite{barrett_sympathetic_2003}.

The choice of optimum trap electrode and support structure materials depends on a number of technical and practical considerations, as well as the dominant systematic shifts of the considered clock ion species. A careful calibration of the Black-body radiation shift requires the precise knowledge of the effective temperature seen by the ion. This is most easily achieved for ion traps at a homogeneous and well-defined temperature. Ohmic heating of the electrodes can be prevented by using a non-magnetic, high-conductivity metal, such as gold. Similarly, the mounting structure of the electrodes should be made from a dielectric material with a small rf loss tangent at the typical rf drive frequencies of $10\dots 100$~MHz. At the same time, the material should have a high thermal conductivity to ensure a homogeneous temperature. Materials fulfilling these conditions include diamond, sapphire (Al$_2$O$_3$ crystal), alumina (Al$_2$O$_3$ ceramics), aluminium nitride (AlN ceramics), and to a lesser extend, fused silica.
Another aspect concerns heating of the ions in the trap during interrogation, which increases the uncertainties in temperature-related shifts, such as the second order Doppler shift. Motional heating arises through electric field noise at the secular frequencies of the ion~\cite{wineland_experimental_1998, turchette_heating_2000}. Depending on the particular trap implementation there can be many origins for such heating, ranging from Johnson noise of drive and filter electronics to electrode surface contaminants, which are reviewed in detail in \citet{brownnutt_ion-trap_2014}.


\subsubsection{Paul traps}\label{sec:sphericaltraps}
\begin{figure}[t]
\includegraphics[width=\textwidth]{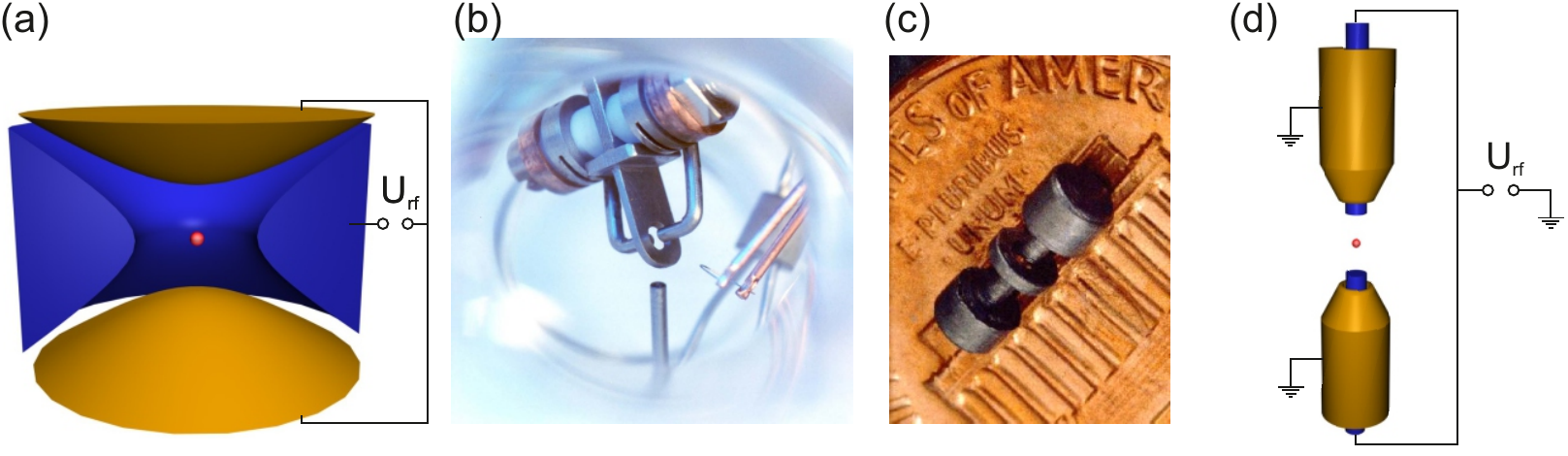}
\caption{Electrode configurations for spherical Paul traps. (a) Cut through the cylindrically symmetric electrode geometry used by Paul. The oscillating potential $U_\mathrm{rf}=\Vrf\cos\Orf t$ is applied between the ring (blue) and the endcap (yellow) electrodes. (b) Paul trap used for the \Yb frequency standard at PTB \cite{tamm_spectroscopy_2000}. (c) Paul trap used for the \Hg frequency standard at NIST \cite{oskay_measurement_2005}. (d) Endcap trap \cite{schrama_novel_1993}. The inner endcaps are 0.5~mm in diameter and are separated by 0.56~mm. The oscillating potential $U_{rf}$ is applied to the inner endcap electrodes (blue). The outer  electrodes (outer diameter 2~mm, yellow)  are normally grounded, however if required small potentials can be applied to compensate micromotion in the axial direction.}
\label{fig:spherical_traps}
\end{figure}

Figure~\ref{fig:spherical_traps}(a) shows the geometry studied by Paul and coworkers \cite{paul_ionenkafig_1958, fischer_dreidimensionale_1959, paul_electromagnetic_1990}. It consists of cylindrically symmetric ring and endcap electrodes between which a DC and AC voltage, $\VDC$ and $\Vrf$, respectively, is applied. The hyperbolic shape of the electrodes ensures a dominant quadrupole potential, even very close to the electrodes. However, the Paul trap only offers restricted optical access for laser cooling, clock interrogation, and fluorescence detection. \citet{beaty_simple_1987} introduced a quadrupole geometry with conical electrodes featuring larger optical access (see Fig.~\ref{fig:spherical_traps}(b,c)). This type of trap is used for the \Yb and \Hg ion clocks at PTB and NIST, respectively. A stronger variation of the original Paul design in which the endcap electrodes are pulled away from the ring was introduced by Straubel \cite{straubel_zum_1955}. It also offers larger optical access and can be implemented in several variations \cite{yu_miniature_1989, yu_demonstration_1991, schrama_novel_1993}. An even more open geometry with good approximation of quadrupole potential is obtained with the endcap trap \cite{schrama_novel_1993}, where the ring is replaced by two cylindrical shields that surround the rf-carrying  endcap electrodes (see Fig.~\ref{fig:spherical_traps} (d)). This geometry is employed in the \Sr ion clocks at NPL \cite{margolis_hertz-level_2004} and NRC \cite{dube_recent_2010, madej_<sup>88</sup>sr<sup>+</sup>_2012}. Typical trap frequencies in all ion clock experiments are on the order of a few MHz in all directions.

\subsubsection{Linear ion traps}\label{sec:lineartraps}
\begin{figure}[t]
\includegraphics[width=10cm]{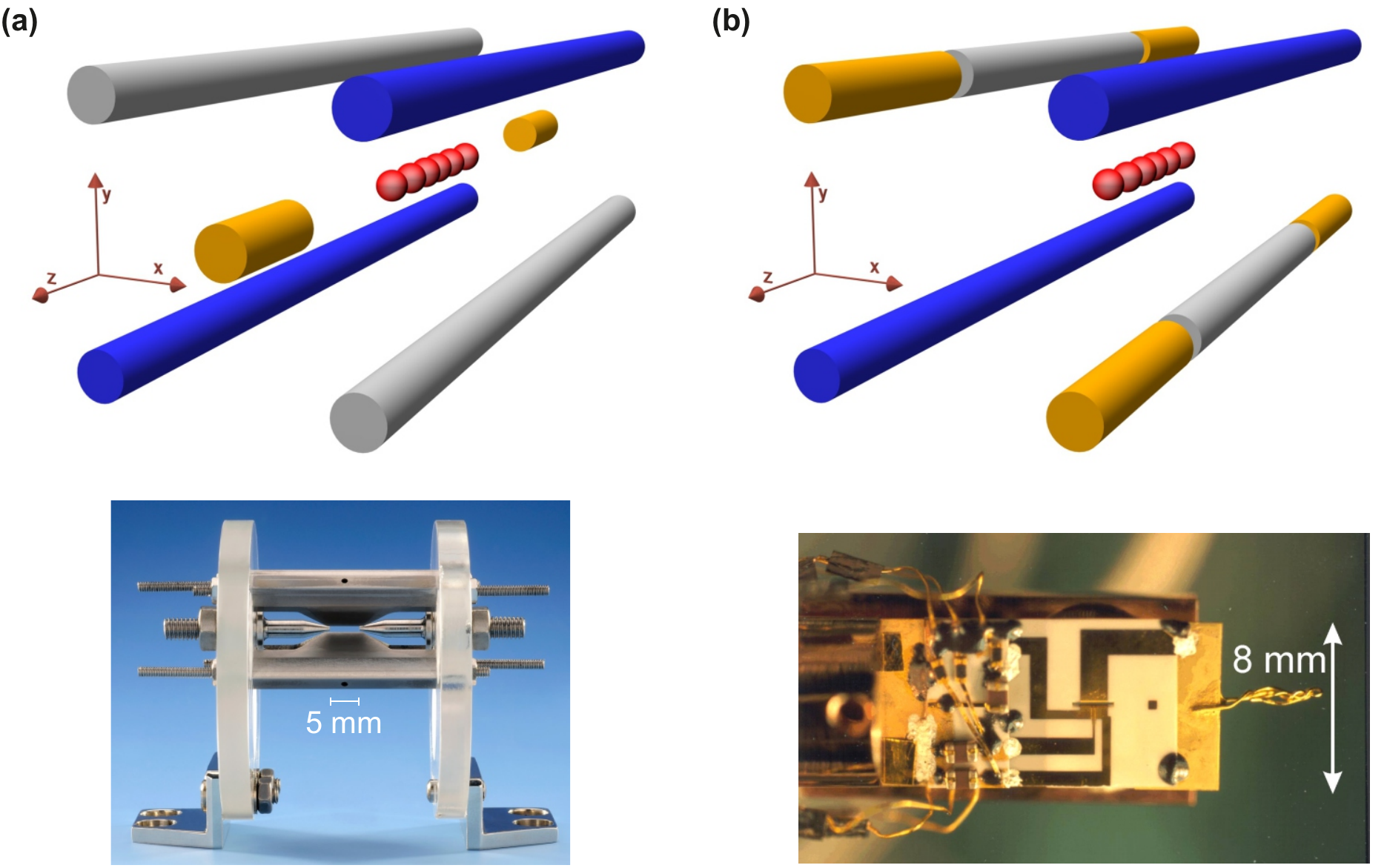}
\caption{Linear ion trap electrode geometries. (a) The Innsbruck trap geometry (upper panel, \cite{gulde_experimental_2003}) is implemented using elongated blades for the rf and two conical tips for the DC electrodes (lower panel). (b) The NIST trap geometry (upper panel, \cite{rowe_transport_2002}) is implemented using microstructured segmented electrodes (lower panel). This allows splitting the tip electrodes and moving them away from the axial symmetry line, enabling improved laser access. Blue electrodes are connected to rf potential, $\Vrf\cos\Orf t$, yellow electrodes are at a positive DC potential, $\VDC$, and grey electrodes are at ground.}.
\label{fig:linear_traps}
\end{figure}

In many applications it is desirable to trap more than one ion in a micromotion-free configuration \cite{raizen_linear_1992}. It is an important requirement for implementing quantum logic spectroscopy (see Sec.~\ref{sec:qls}) and scaling single ion to multi-ion optical clocks for improved stability~\cite{prestage_linear_1990, herschbach_linear_2012}.
Linear Paul traps provide such a micromotion-free environment along the zero line of the rf electric field if the radial confinement is much stronger than the axial. Their design is derived from the quadrupole mass filter \cite{paul_elektrische_1955, drees_beschleunigung_1964}, which provides radial confinement through an oscillating 2D quadrupole potential. Trapping in all three dimensions is accomplished by superimposing a 3D static quadrupole field, providing mostly axial confinement \cite{raizen_linear_1992}. This field configuration can be implemented through a variety of electrode geometries.
Figure~\ref{fig:linear_traps} shows two designs employed for the \Ca \cite{chwalla_absolute_2009} and \Al \cite{rosenband_observation_2007} optical frequency standards.
The Innsbruck design uses four symmetrically arranged blade electrodes with an electrode-electrode distance of 1.6~mm to which an rf voltage of  $\sim 1$~kV at a frequency of around 25~MHz is applied to create the 2D rf quadrupole. Two tip electrodes separated by 5~mm to which a positive DC voltage of around 2~kV is applied provide axial confinement~\cite{gulde_experimental_2003}. The electrodes are made from non-magnetic steel, whereas the ceramic support is made from Macor. When this trap is operated with two of the rf electrodes connected to ground (asymmetric driving), axial micromotion arises from a distortion of the 2D quadrupole, since the tip electrodes act as rf ground, thus removing the radial symmetry. This effect can be circumvented by either applying additional rf to the tip electrodes, or by applying rf voltages oscillating around rf ground, to all four rf electrodes (symmetric driving).

The first generation NIST trap is made from laser-structured and gold-coated alumina wafers, separated by 440~$\mu$m \cite{rowe_transport_2002}. This micro-structured approach allows for high accuracy in the electrode geometry and provides a path for scalable quantum information processing~\cite{kielpinski_architecture_2002}. An rf voltage of around 250~V and a DC voltage of up to 12~V results in secular frequencies of a single \Be ion of 8~MHz radially and 5~MHz axially.
The second NIST trap uses segmented gold-coated beryllium-copper electrodes, resembling the electrode geometry of the first generation NIST trap, but using conventional machining and larger dimensions (0.4~mm distance between ion and nearest electrode) with the goal of reducing micromotion and motional heating from fluctuating patch potentials \cite{chou_frequency_2010}. The blade-shaped electrodes are mounted onto and indexed to alumina rods that are mounted into a precision machined metal cage.

A linear trap geometry for multi-ion optical clocks has been designed that combines the precision of laser-machined wavers with large trap geometries for low motional heating rates and excellent laser access~\cite{herschbach_linear_2012}. High symmetry of the electrode geometry (e.g. by adding slots to the rf electrodes to match the gaps between DC segments) combined with integrated compensation electrodes allows storing tens of ions in a trap with small excess micromotion~\cite{herschbach_linear_2012, pyka_high-precision_2013-1}.

When two or more ions are stored in a linear Paul trap, their motion becomes strongly coupled and a normal mode description for the motion of the ions around their equilibrium position applies \cite{wineland_atomic-ion_1987, james_quantum_1998, kielpinski_sympathetic_2000,morigi_two-species_2001}. Each normal mode is associated with a mode frequency and motional amplitudes for the ions. For a two-ion crystal with a large mass ratio, the mode amplitudes differ significantly. The Doppler cooling rate scales with the motional amplitude. When cooling only on one of the ions, as is the case in the \Al clock, additional motional heating can thus result in an elevated temperature of weakly cooled modes \cite{wubbena_sympathetic_2012}.

\subsection{Cooling techniques and Lamb-Dicke regime}
\label{sec:lambdicke}
Techniques to suppress motion-induced frequency shifts have long played a central role in optical spectroscopy. Doppler laser cooling \cite{hansch_cooling_1975, wineland_proposed_1975, neuhauser_optical-sideband_1978, neuhauser_visual_1978, wineland_radiation-pressure_1978,neuhauser_localized_1980,wineland_spectroscopy_1981,stenholm_semiclassical_1986} on transitions with linewidth $\Gamma$ achieves temperatures of
\[
T_D=\kappa\frac{\hbar\Gamma}{2k_B},
\]
independent on the atomic mass and the trap frequency. The parameter $\kappa$ is of order unity and depends on the laser cooling geometry \cite{javanainen_light-pressure_1980}. For a few MHz broad transitions, this corresponds to temperatures in the few hundred $\mu$K regime, thus reducing second order Doppler shifts to well below $10^{-17}$ fractional frequency uncertainty for heavy clock ion species, such as \Yb or \Sr. In contrast to neutral atoms in free space, trapped ions require only a single cooling laser with $k$-vector components along all three trap axes \cite{wineland_laser_1979}. However, one has to ensure that all trap frequencies are different to spatially fix the normal mode axes to the geometry of the trap \cite{itano_laser_1982}. The MHz fast oscillations of the ion(s) in the trap allow efficient cooling when the ion(s) are moving towards the laser beam. This semi-classical picture is valid if the quantized mode structure of the ion's motion in the trap can be neglected, which is the case in the so-called \emph{weak binding regime} in which the trap frequency $\omega$ is much smaller than the linewidth of the cooling transition ($\omega\ll\Gamma$) \cite{wineland_laser_1979}. The situation changes when considering narrow transitions  ($\omega\gg\Gamma$). In this \emph{tight binding regime} motional sidebands are spectrally resolved from the carrier and can be individually addressed, resulting in a simultaneous change in the internal and motional state. In a simple picture, the spatial gradient of the laser's electric field along its propagation direction (characterized by the wavenumber $k$) interacts with the motional wavepacket of the ion in the trap (characterized by its ground state size $u_0=\sqrt{\hbar/2m\omega}$). The parameter describing the strength of the interaction is the so-called Lamb-Dicke parameter $\eta=k u_0$. Absorption and emission of photons by an unbound atom is associated with photon recoil, resulting in an energy shift $E_\mathrm{rec}=\frac{\hbar^2 k^2}{2 m}$ of the observed line. For trapped ions, this recoil is suppressed if $E_\mathrm{rec}/(\hbar\omega)<1$, which is equivalent to $\eta^2<1$, reminiscent of the Mößbauer-effect in nuclear physics. Optical clocks based on trapped ions are typically deep in this regime, thus eliminating recoil shifts. If we restrict ourselves to a 2-level system with ground ($\downket$) and excited ($\upket$) states coupled to a single motional mode (\ket{n}) with excitation $n$, the resulting system is described by a Jaynes-Cummings type model \cite{wineland_experimental_1998, leibfried_quantum_2003}. Particularly simple expressions for the transition strengths are obtained in the Lamb-Dicke regime for which the size of the motional wavefunction $\ket{\psi_m}$ is small compared to the wavelength: $\sqrt{\bra{\psi_m}{k^2\hat{u}^2}\ket{\psi_m}}\ll 1$. In this case three distinct transitions are dominant: (i) Carrier (CAR) transitions with Rabi frequency $\Omega$ change only the electronic state ($\downket\ket{n}\leftrightarrow \upket\ket{n}$); (ii) red sideband (RSB) transitions with Rabi frequency $\eta\Omega \sqrt{n}$ excite the electronic state and remove a quantum of motion ($\downket\ket{n}\leftrightarrow \upket\ket{n-1}$); (iii) blue sideband (BSB) transitions with Rabi frequency $\eta\Omega \sqrt{n+1}$ excite the electronic state and add a quantum of motion ($\downket\ket{n}\leftrightarrow \upket\ket{n+1}$). Outside the Lamb-Dicke limit, terms higher order in the Lamb-Dicke factor need to be considered, changing the Rabi frequencies of the transitions \cite{wineland_laser_1979, wineland_experimental_1998} and allowing higher-order motional transitions.
In the tight binding regime, the kinetic energy of the ion can be further reduced through resolved sideband cooling \cite{wineland_proposed_1975, dehmelt_entropy_1976}. By continuously driving the first red-sideband transition motional energy is removed and dissipated through spontaneous emission from the excited state. The latter step involves scattering of a photon, which provides the required dissipation and is recoil-free with a high probability in the Lamb-Dicke limit. Residual recoil from the dissipation step together with off-resonant excitation of CAR and BSB transitions determine the achievable average motional quantum number $\bar{n}$ of
\[
\bar{n}\approx C_s(\Gamma/\omega)^2,
\]
where $C_s$ is a numerical factor on the order of 1 depending on the selection rules of the atomic transition \cite{neuhauser_optical-sideband_1978, wineland_laser_1979, stenholm_semiclassical_1986, wineland_laser-cooling_1987}. Although up to now Doppler cooling for reducing second order Doppler shifts was sufficient, future ion clocks operating at or below the $10^{-18}$ fractional frequency uncertainty may require more involved cooling techniques. In typical ion trap experiments, the ground state can be populated with a probability reaching 99.9~\% \cite{roos_quantum_1999}, reducing this shift to its value in the ground state of the trap and its uncertainty to well below that. Different implementations of sideband cooling are further discussed in \citet{eschner_laser_2003}.
\subsection{Systematic frequency shifts for trapped ions}
\label{sec:importantsysteffects}

The most important systematic frequency shifts encountered in trapped-ion frequency standards are Doppler shifts resulting from the residual motion of the ion and shifts from the interaction with external electro-magnetic fields. For trapped ions, there is always a connection between the Doppler and the Stark shifts, because an ion with higher kinetic energy will also be exposed to higher field strength in the confining quadrupole potential of the trap. The sensitivity to field-induced shifts depends on the type of the reference transition and on properties of the specific ion. This has been an important criterion in the selection of suitable ions. In the following we expand on the main frequency shifting effects already introduced in Sec.~\ref{sec:intro} and discuss their specific properties in the context of trapped ions
\cite{itano00, luiten_single-ion_2001,madej_absolute_2004,Lea07a,dube_evaluation_2013}.

\subsubsection{Motion-induced shifts}\label{sec:ionMotion}
The oscillation frequencies of the ion in the trap are much higher than the linewidth of the optical reference transition.
Therefore, the linear Doppler effect (first term in in Eq.~\ref{rel_shift}) from secular and micromotion leads to sidebands in optical excitation or emission spectra, but does not shift the carrier. However, a shift can arise from a displacement of the ion in the direction of the probe laser beam if it is correlated with the interrogation cycle or continuous over an appreciable timescale.  Such an effect could be induced by electric fields correlated with the probe laser, or thermal effects changing the position between the ion trap and the reference phase of the probe laser. If the clock interrogation light is in the UV spectral regime (such as for the \Al clock), it can eject photo-electrons when hitting a surface \cite{harlander_trapped-ion_2010}. Depending on the geometry of the trap and laser direction, the created charges can alter the position of the ion, resulting in a linear Doppler shift. Some of these effects can be eliminated by probing the ion from two counter-propagating directions and averaging the observed transition frequencies \cite{rosenband_frequency_2008,chou_frequency_2010}.

Residual secular motion at the laser cooling limits determines the time dilation shift, which is mostly relevant for light ion species (second term in Eq.~\ref{rel_shift}). Moreover, secular motion results in an increased size of the ion's time-averaged wavepacket, which extends into the region of non-zero oscillating trap field. As a consequence, for typical trap operation parameters \cite{wineland_laser-cooling_1987, berkeland_laser-cooled_1998, wubbena_sympathetic_2012} the kinetic energy from secular motion is doubled through an equal contribution from micromotion.
The total kinetic energy is thus the sum of the secular kinetic energy $E_s=\frac{1}{2}\sum_\alpha\hbar\omega_\alpha (\bar{n}_\alpha+1/2)$ and the micromotion energy $E_{\mathrm{mm}}\approx E_\mathrm{emm}+E_s$, containing a term from excess micromotion and secular-motion induced micromotion. It is interesting to note that even for an ion in the ground state of the trap, the kinetic energy contribution from zero-point fluctuations result in a non-vanishing fractional time dilation shift of on the order of $-10^{-18}$ for \Al in a single mode with frequency 5~MHz.

\subsubsection{Zeeman effect}\label{sec:ionZeeman}
While a static magnetic field is not required for the operation of the Paul trap, a weak homogeneous field (typically in the range of 1 to 100~$\mu$T) is applied in order to separate the Zeeman components of the reference transition and to provide a controllable quantization axis for the interaction of the ion with the different laser fields. The methods for the control or compensation of resulting linear and quadratic Zeeman shifts are similar to those applied in other types of atomic clocks (see Section  II C).

\subsubsection{Quadrupole shift}\label{sec:ionQuadrupole}
In the case of an atomic state with $J>1/2$ (and $F>1/2$) the electronic charge distribution can have multipole moments that couple to an external electric field gradient, giving rise to the so called quadrupole shift of the energy level. A static electric field gradient is not required  for the operation of a Paul trap, but it turns out that because of the proximity of the ion to the trap electrodes and due to the presence of patch potentials on these, the ion may be exposed to an unintentionally applied field gradient as strong as
1~V/mm$^2$, that will lead to a level shift on the order of 1~Hz for a quadrupole moment of $ea_0^2$. While static electric stray fields can easily be diagnosed via the induced micromotion and can be nulled by compensation potentials on extra electrodes, the dynamics of the ion does not provide a similarly sensitive measure for residual field gradients and the strength and symmetry of these is initially unknown. Linear Paul traps require a static electric field gradient for closure along the trap axis. Since the gradient is related to the ion's axial trap frequency, it can be determined with high accuracy and allows a precision measurement of the electric quadrupole moment \cite{roos_designer_2006}.

The Hamiltonian describing the interaction of an external field gradient with the atomic quadrupole moment is \cite{itano00}
\begin{equation}
H_Q=\bm{\nabla E^{(2)}}\bm{.}\bm{\Theta^{(2)}}.
\label{qhamiltonian1}
\end{equation}
Here $\bm{\nabla E^{(2)}}$ is a symmetric traceless second-rank tensor describing the electric field gradient at the position of the ion and
$\bm{\Theta^{(2)}}$ is the electric-quadrupole operator for the atom. Transforming to principal axes, the electric potential creating the gradient can be written as
\begin{equation}
\Phi=A\left( (1+\epsilon)x^{\prime2}+(1-\epsilon)y^{\prime2}-2z^{\prime2}\right)
\label{qpot2}
\end{equation}
Treating the quadrupole shift as a small perturbation of the Zeeman shifts in the basis of states $|\gamma JFm_F\rangle$ and applying the Wigner-Eckart theorem to $\bm{\Theta^{(2)}}$,  the diagonal matrix elements of $H_Q$ can be written as
\begin{eqnarray}
H_Q&=&\langle \gamma JFm_F|H_Q|\gamma JFm_F\rangle\\
&=& \frac{-2(3m_F^2-F(F+1)) A \langle\gamma JF||\Theta^{(2)}||\gamma JF\rangle}{((2F+3)(2F+2)(2F+1)2F(2F-1))^{1/2}}
\times [(3\cos^2\beta -1)-\epsilon \sin^2\beta(\cos^2 \alpha - \sin^2\alpha)],
\label{qshift}
\end{eqnarray}
where $\alpha,\beta$ are the first two of the Euler angles that relate the principal axis frame to the laboratory frame where the $z$-axis is parallel to the magnetic field. The reduced matrix element of $\bm{\Theta^{(2)}}$ in the $(IJ)$ coupling scheme is
\begin{equation}
(\gamma IJF\vert\vert\Theta^{(2)}\vert\vert \gamma IJF)=(-1)^{I+J+F}(2F+1)
\left\{\begin{array}{ccc}J & 2 & J \\ F & I & F \end{array}\right\}
\left(\begin{array}{ccc}J & 2 & J \\ -J & 0 & J \end{array}\right)^{-1}\, \Theta(\gamma,J),
\label{redmatrixelement}
\end{equation}
where $\Theta(\gamma,J)$ is the quadrupole moment of the $(\gamma,J)$ state, which is defined as
\begin{equation}
\Theta(\gamma,J) = \langle \gamma JJ\vert \Theta_0^{(2)} \vert \gamma JJ\rangle \,.
\label{qmoment}
\end{equation}

Equation~\ref{qshift} possesses symmetry properties that can be used for a cancellation of the quadrupole shift without prior knowledge about strength or orientation of the electric field gradient.
\textcite{itano00} showed that  the sum of the angle-dependent factor in square brackets (a linear combination of spherical harmonics) vanishes for any three mutually perpendicular orientations of the quantization axis $z$. Therefore, the average of the transition frequency taken for three mutually perpendicular orientations of a magnetic field of the same magnitude does not contain the quadrupole shift. The method has been verified experimentally and has been used in frequency standards with $^{87}$Sr$^+$, $^{171}$Yb$^+$, and $^{199}$Hg$^+$
\cite{margolis_hertz-level_2004, oskay_measurement_2005, schneider_sub-hertz_2005, oskay_single-atom_2006}. In a mm-size Paul trap with $^{171}$Yb$^+$, stray field-induced, slowly variable quadrupole shifts of about 1~Hz, have been observed over a period of 74 days after loading the ion \cite{tamm_stray-field-induced_2009}.  The suppression of the quadrupole  shift that can be achieved depends on the precision to which the three magnetic field orientations are orthogonal. The uncertainty in the angles between field orientations has to be about $\pm1^\circ$ to get a suppression of the shift by a factor of 100. Such a precision and temporal stability of the magnetic field requires the use of magnetic shielding around the trap, with a set of field coils mounted inside the shield.

An alternative option for the cancellation of the quadrupole shift is based on the $m_F$ dependence in Eq.~\ref{qshift}: Because
\begin{equation}
\sum_{m=-F}^F (3m^2-F(F+1))=0,
\end{equation}
an average of the transition frequency over all Zeeman sublevels does not contain the quadrupole shift. For higher values of $F$ it will be more efficient to measure the transition frequencies  for two values of $|m|$ and to interpolate the linear dependence of the frequency on $m^2$ to the unperturbed value obtained at the ''virtual'' quantum number $m_0^2=F(F+1)/3$ \cite{dube_electric_2005}. Depending on the quantum numbers, different interrogation sequences may be used to simultaneously suppress combinations of $m$-dependent shifts, like for example the quadrupole shift and the linear Zeeman shift
\cite{margolis_hertz-level_2004,chwalla_absolute_2009,madej_<sup>88</sup>sr<sup>+</sup>_2012,dube_evaluation_2013}.
In all the reference transitions studied today, the ground state fulfills $J<1$ so that the quadrupole shift only needs to be considered for the excited state sublevels.
Unlike the method of averaging over three orientations of the quantization axis, averaging over the Zeeman components also eliminates higher orders of the quadrupole shift. In comparison to static patch fields, the oscillating quadrupole potential of the Paul trap generates much stronger field gradients on the order of 1~kV/mm$^2$. While the time-averaged first-order quadrupole shift produced by the oscillating field is zero, it could lead to a contribution from the second-order quadrupole shift \cite{cohen_quadrupole_1957}.

\subsubsection{Stark shift}\label{sec:ionStark}
An ion in a Paul trap will always be located at the point where DC electric fields vanish. However, it can be exposed to  oscillating electric fields arising from black-body radiation, laser fields, or motion around the field-free point in the oscillating trapping field.
Exposing the atom to a non-vanishing rms electric field displaces the energy levels via the quadratic Stark effect (c.f. Eq.~\ref{sshift2}).
Comparison of the tensor part of the Stark shift with the expression for the quadrupole shift (see Eq.~\ref{qshift}) shows that both effects  possess identical dependences on the orientation of the quantization axis and on the $m$-numbers. Therefore, averaging methods that suppress the quadrupole shift will also eliminate the tensorial Stark shift.

Since the oscillating trapping electric field drives a motion of the ion with $\left\vert\bm{E}\right\vert^2\propto \left\vert\bm{v}\right\vert^2$, there is a direct connection between the second order Stark and Doppler shifts. In cases such as \Ca and \Sr, where the scalar Stark shift from the differential polarizability $\Delta\alpha_S$ increases the transition frequency, a cancellation of both shifts is obtained for a specific value of the trap radiofrequency
$\Orf=(8e/mc)\sqrt{\hbar\omega_0/\Delta\alpha_S}$ \cite{itano00,dube_electric_2005,madej_<sup>88</sup>sr<sup>+</sup>_2012}.

A dynamic Stark effect will be produced by laser light impinging on the ion, and will be described by an expression like Eq. \ref{sshift2}, where the static polarizabilities are replaced by frequency-dependent dynamic polarizabilities and the rms electric field strength $\langle E^2\rangle=I_L/c\epsilon_0$ is proportional to the laser intensity $I_L$. The dynamic polarizability in general is composed of contributions from several dipole transitions coupling to the levels of the reference transition. In the case of a two-level system that is driven by near resonant light at Rabi frequency $\Omega_R$ and detuning $\delta$, the shift is $\Delta f_L=(2\delta \Omega_R^2)/(4\delta^2+\Gamma^2)$. Light from cooling and repumping lasers that couples resonantly to one of the levels of the reference transition is therefore usually blocked by mechanical shutters and care is taken to avoid the presence of stray light during the interrogation period. An exception is the quantum logic clock (see section \ref{sec:qls}), where cooling of the logic ion of a different species is continued during the clock interrogation. In the case of a strongly forbidden reference transition like the electric octupole transition in Yb$^+$, the  light shift induced by the reference laser itself through the coupling to other levels needs to be corrected for.

\subsubsection{Blackbody radiation shift}\label{sec:ionBBR}
The electric field associated with thermal radiation emitted by the trap structure and the inner surface of the vacuum chamber also gives rise to a quadratic Stark shift of the reference transition, the so-called blackbody radiation shift \cite{itano_shift_1982}.
If the thermal radiation field is isotropic, the tensor contribution to the Stark shift averages to zero.
Table \ref{tab:systematics} lists the expected shifts at $T=300$~K for the most important ion reference transitions.

At the present stage, the uncertainty from the blackbody radiation shift makes an important contribution to the systematic uncertainty budgets of many of the trapped ion optical frequency standards, resulting partly from uncertainty in the polarizabilities and partly from incomplete knowledge of the radiation field. The trap structure is subject to heating through the applied radiofrequency voltage, from ohmic losses in the conductors and from dieletric losses in the insulators. The employed materials possess very different emissivities for infrared radiation, ranging from $0.02$ for a polished metal surface to $0.9$ for ceramics.
In an experiment with thermistors attached to different parts of an ion trap, temperature differences up to about 25 K have been observed
\cite{dube_evaluation_2013}.  Attempts to analyze the temperature distribution in ion traps by finite-element modelling indicate that the use of materials with low electric losses and the provision of good thermal contact to a heat sink may constrain the rise of the effective radiation temperature seen by the ion due to applied rf voltages to below 1~K at room temperature.

\subsection{Ionic candidates and their electronic structure}
\label{sec:species}
Several different ion species have been considered for optical clocks. Each of them has advantages and disadvantages concerning systematic shifts and technical complexity. By definition, the most accurate frequency standard will be the one with the lowest uncertainties in the systematic shift evaluation. However, this does not necessarily imply that the shifts themselves are small. In fact they can be quite large if they are known with sufficient precision. This requires a thorough investigation of all shifts and associated uncertainties. The evaluation will depend on an accurate knowledge of the atomic shift coefficients and the fluctuations in the shift inducing effects, e.g. electric and magnetic fields. Without a priori knowledge of these parameters, it is impossible to judge which species will offer the most accurate reference transition. After a brief historical review, we will discuss in the following the order of magnitude of the systematic shifts introduced in Sec.~\ref{sec:importantsysteffects}, concentrating on ion species that have been brought to a sufficiently high level of accuracy to allow a comparison of each species' pros and cons. The discussion is complemented by Table~\ref{tab:ionatomicparams} which provides a detailed list of atomic coefficients for each species.

\begin{figure}[t]
\includegraphics[height=4cm]{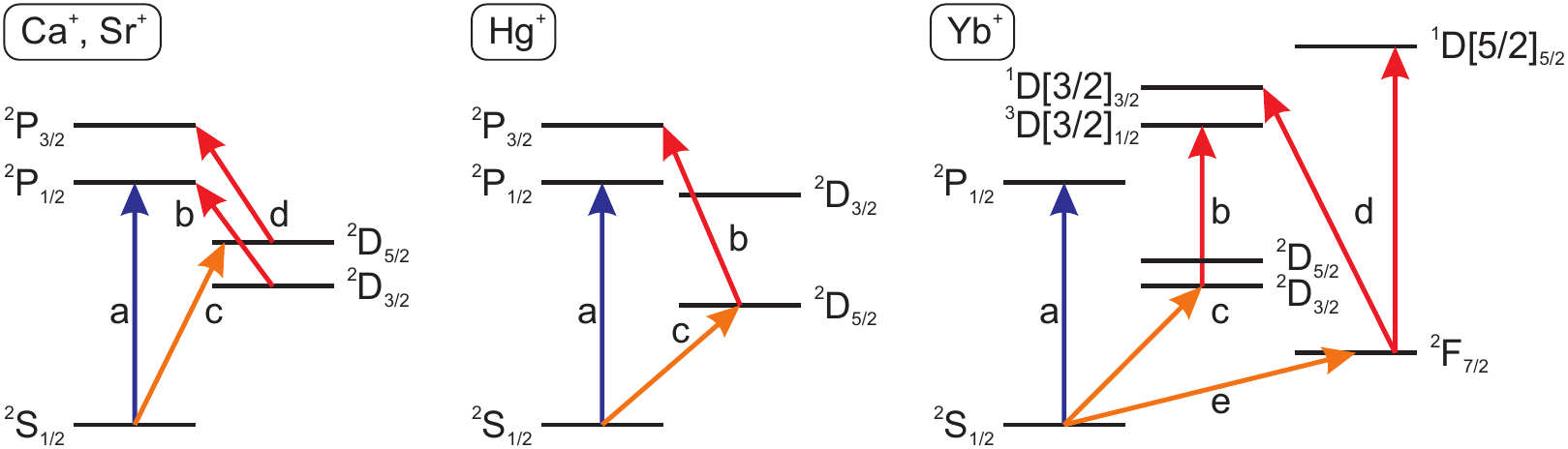}
\begin{tabular}{|l|l|l|l|l|}
\hline & \textit{\Ca} & \textit{\Sr} & \textit{\Hg} & \textit{\Yb}\\
\hline \textit{a: cooling \& detection} & 396.8 nm, 23.4 MHz & 421.7 nm, 20.2 MHz & 194.2 nm, 70 MHz & 369.5 nm, 19.6 MHz\\
\hline \textit{b: repump 1} & 866.2 nm, 1.69 MHz & 1091.5 nm, 1.5 MHz & 398.5 nm, 509 kHz & 935.2 nm\\
\hline \textit{c: clock (quadrupole)} & 729.1 nm, 0.207 mHz & 674.0 nm, 0.4 Hz & 281.6 nm, 1.8 Hz & 435.5 nm, 3.1 Hz\\
\hline \textit{d: repump 2} & 854.2 nm, 1.58 MHz & 1033.0 nm, 1.38 MHz & - & 760.1 nm/638.6 nm\\
\hline \textit{e: clock (octupole)} & - & - & - & 467 nm, 10$^{-9}$ Hz\\
\hline
\end{tabular}
\caption{Schematic energy level diagram of clock ions with a single valence electron and Yb$^+$, together with a table of the most relevant transition wavelengths and linewidths. All data are from NIST database \cite{Ralchenko_NIST_2012}, except where otherwise noted. Energy levels are not to scale and the term notation follows \cite{martin_atomic_1978}.}
\label{fig:groupII}
\end{figure}

The ideal optical clock ion species has a clock transition with a high $Q$ factor that is insensitive to external field perturbations and auxiliary transitions for laser cooling, state preparation through optical pumping and internal state detection. Historically, the first proposals for a single ion optical clock by Dehmelt and Wineland \cite{dehmelt_proposed_1973,dehmelt_proposed_1975-2,wineland_proposed_1975} were based on \Tl, since the comparatively short excited clock state lifetime of 50~ms seemed to allow fluorescence detection and laser cooling directly on the clock transition. However, a \Tl clock was never realized. The invention of the electron-shelving technique for internal state detection \cite{dehmelt_proposed_1975-2} and laser cooling on fast transitions \cite{hansch_cooling_1975, wineland_proposed_1975, neuhauser_optical-sideband_1978, wineland_radiation-pressure_1978}, allowed the investigation of other ion species with technologically more convenient laser cooling and clock transitions, featuring higher $Q$ factors and thus improved stability. Owing to their relatively simple electronic level structure, singly-charged ions with one and two valence electrons have been studied extensively. Figure~\ref{fig:groupII} shows the schematic energy level diagram of the Alkaline-like one-valence-electron systems \Ca, \Sr, \Hg, and \Yb. The clock transition is a quadrupole transition from the \dsoh ground to one of the \ddth, \ddfh excited states with linewidths ranging between 0.2~Hz and 3~Hz. All of these ions offer a fast, almost closed \dsohdpoh cycling transition for laser cooling and internal state discrimination. For \Ca and \Sr, the excited \dpoh state can also decay into the \ddth state, from which a repumper on the \ddthdpoh transition brings the electron back into the cycling transition. For \Hg, decay from the \dpoh to the \ddth and \ddfh is strongly suppressed \cite{bergquist_observation_1986}. After probing the clock transition, the \ddfh state can be repumped through the \dpth state for efficient initial state preparation. The ytterbium ion is an effective one-valence electron system belonging to the Lanthanoids. Besides the quadrupole clock transition, it also offers an octupole transition from the ground \dsoh to a low-lying \dfsh state with a particularly large $Q$ factor of $10^{23}$, corresponding to a natural linewidth of $10^{-9}$~Hz. Repumping back to the ground state after exciting the clock transitions can be performed in a number of ways as shown in Fig.~\ref{fig:groupII} \cite{huntemann_high-accuracy_2012}.

Some naturally occurring isotopes of these ions have non-vanishing half-integer nuclear spin and consequently hyperfine structure. This can be used to eliminate strong first order Zeeman shifts of a few 10~kHz/$\mu$T (1~MHz/G) on the clock transition by choosing $m_F=0\rightarrow m_F'=0$ transitions, at the expense of a more complex cooling laser system to address all hyperfine states. This has been implemented for the \Yboso and \Hgonn isotopes. Second-order Zeeman shifts arising from static and dynamic magnetic fields are on the order of a few 10~kHz/mT$^2$ and a few 10~Hz/mT$^2$ for isotopes with and without hyperfine structure, respectively (see Sec.~\ref{sec:ionZeeman}). In addition, the \ddth, \ddfh, and \dfsh-states have  $J>\frac{1}{2}$, thus exhibiting an electric quadrupole moment which couples to electric field gradients as outlined in Sec.~\ref{sec:ionQuadrupole}, producing shifts on the order of a few Hz in typical ion traps.
Linear Zeeman and electric quadrupole shifts can be simultaneously eliminated by averaging over appropriate Zeeman transitions.

\begin{figure}[t]
\includegraphics[height=4cm]{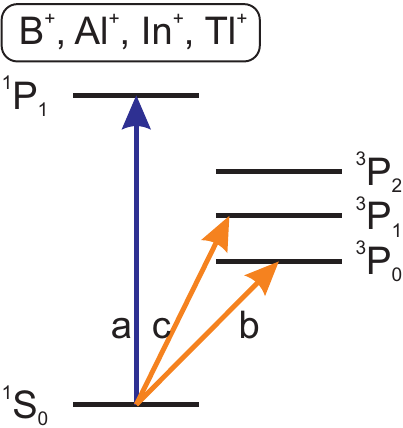}

\begin{tabular}{|l|l|l|l|}
\hline
& \textit{\B} & \textit{\Al} & \textit{\In} \\
\hline \textit{a: VUV cooling} & 136.2 nm, 191 MHz & 167.1 nm, 233 MHz & 159 nm, 1.5 GHz\\
\hline \textit{b: clock} & 267.8 nm, 37 $\mu$Hz & 267.4 nm, 7.7 mHz & 236.5 nm, 0.8 Hz\\
\hline \textit{c: narrow cooling/quantum logic} & 267.8 nm, 1.66 Hz & 266.9 nm, 520 Hz & 230.6 nm, 360 kHz\\
\hline
\end{tabular}
	\caption{Schematic level structure of ion clocks based on group~13 (formerly group~IIIA) singly charged ions, together with a table of the most relevant transition wavelengths and linewidths. All data from NIST database \cite{Ralchenko_NIST_2012}, except where otherwise noted. Energy levels are not to scale.}
	\label{fig:groupIII}
\end{figure}

It was realized early on that it is advantageous to have clock transitions between states with vanishing angular momentum, such as the \ssztpz clock transition in group 13 (formerly group IIIA) singly-charged ions 
\cite{dehmelt_coherent_1981,dehmelt_mono-ion_1982}. These transitions do not suffer from electric quadrupole shifts and offer smaller (nuclear) linear and quadratic Zeeman shifts of a few 10~kHz/mT and a few 10~Hz/mT$^2$, respectively. The common partial electronic level structure of the group 13 ions is shown in Fig.~\ref{fig:groupIII}. Single photon transitions between the pure states \ssz and \tpz ($J=0\rightarrow J'=0$) are rigorously forbidden by angular momentum selection rules. However, hyperfine interaction couples the \tpz state to the \tpo and \spo states with the same $F$ quantum number \cite{garstang_hyperfine_1962, marques_hyperfine_1993, peik_laser_1994, brage_hyperfine_1998, itano_optical_2007}.
As a consequence, what we label as the \tpz state actually contains admixtures of these other states, thus inheriting some of their properties, such as decay to the ground state, a modified $g$ factor, and a nonzero but very small electric quadrupole moment.
The ground state is a \ssz state, connected through a strong dipole transition to the \spo state, which could in principle be used for Doppler cooling and detection. However, for the considered ions the wavelength of this transition is in the VUV regime and not accessible by current laser technology. In the case of \In, laser cooling has been implemented on the narrow \ssztpo transition \cite{peik_laser_1994}. The corresponding transitions in \Al and \B are too narrow to allow efficient laser cooling. This limitation can be overcome by implementing quantum logic spectroscopy, described in the next section, where a co-trapped so-called logic ion provides laser cooling and internal state readout.

Black-body radiation shifts the energy of the two clock states by off-resonant coupling to other states. This effect is significant for most neutral and singly-charged ion species with typical shifts on the order of Hz at room temperature \cite{rosenband_blackbody_2006}. The large energy difference of the ground and excited clock states to other states connected by strong transitions results in a significantly reduced black-body radiation shift in group~13 ions \cite{rosenband_blackbody_2006,safronova_precision_2011,zuhrianda_anomalously_2012}.
If a single atom contains two clock transitions with different sensitivity to the black-body radiation shift (such as \Yb), a synthetic frequency can be established that eliminates the dominant $T^4$-dependent term of the shift \cite{yudin_atomic_2011}.

Table \ref{tab:ionatomicparams} summarizes the relevant atomic parameters for determining systematic shifts for the most developed ion clocks. Wherever available, we provide the experimentally determined coefficients, otherwise a theoretical prediction.

\ctable[
	sideways,
	notespar,
	pos=h!,
	cap = {Important atomic parameters of ion clock species},
	caption = {Important atomic parameters of ion clock species. Where available, experimentally measured quantities are given, otherwise the theoretical predictions either derived from measured quantities, or from  \textit{ab initio} calculations. The mass, nuclear spin and the Landée $g$-factors of the ground and excited clock states are labeled $m$, $I$, $g_g$ and $g_e$, respectively. The quadratic Zeeman shift $\Delta f_\mathrm{M2}$ is either given for the $m_F=0\rightarrow m_{F'}=0$ transition, or, where such a transition does not exist, for an average over Zeeman components that mimics such a transition. The static scalar differential polarizability $\Delta\alpha_S=\alpha_\mathrm{e}-\alpha_\mathrm{g}$ is the difference between the excited and ground state polarizability, similarly for the tensor polarizability $\Delta\alpha_T$.
	The dynamic correction factor $\eta$ accounts for the frequency-dependence of the polarizability~\cite{porsev_multipolar_2006} and corrects for the black-body spectrum around 300~K ($\Delta\alpha_\mathrm{300~K}=\Delta\alpha_S (1+\eta)$). The black-body radiation shift for 300~K is given by $\Delta f_\mathrm{300~K}$. The reduced electric quadrupole moment of the excited state are given by $\Theta$. The corresponding coefficients are defined in Sec.~\ref{sec:importantsysteffects}},
	label=tab:ionatomicparams,
	pos=h
]{|p{2.3cm}|c|c|c|c|c|c|c|}
{ 
	\tnote[a]{\textcite{tommaseo_g_j-factor_2003}}
	\tnote[b]{averaged over 6 transitions \cite{chwalla_absolute_2009}}
	\tnote[c]{only negligible contributions from mixing with \tpo and \spo states}
	\tnote[d]{\textcite{roos_designer_2006}}
	\tnote[e]{\textcite{barwood_characterization_2012}}
	\tnote[f]{\textcite{dube_evaluation_2013}}
	\tnote[g]{\textcite{jiang_blackbody-radiation_2009}}
	\tnote[h]{\textcite{barwood_measurement_2004}}
	\tnote[i]{\textcite{meggers_second_1967}}
	\tnote[j]{\textcite{tamm_cs-based_2014}} 
	\tnote[k]{\textcite{schneider_sub-hertz_2005}}
	\tnote[l]{\textcite{huntemann_unpublished_2014}}
	\tnote[m]{\textcite{huntemann_high-accuracy_2012}}
	\tnote[n]{\textcite{itano00}}
	\tnote[o]{\textcite{oskay_measurement_2005}}
	\tnote[p]{\textcite{rosenband_observation_2007}}
	\tnote[r]{\textcite{ting_nuclear_1953,becker_high-resolution_2001}}
	\tnote[s]{\textcite{herschbach_linear_2012}}
	\tnote[t]{\textcite{safronova_precision_2011}}
	\tnote[z]{averaged over all magnetic substates}
}{
\hline
& \Ca & \Sr & \Yb E2 & \Yb E3 & \Hg & \Al & \In \\ \hline
\centering{$m$ (u)} & 39.962 & 87.905  & 170.936 & 170.936 & 198.968 & 26.981 & 114.903\\ \hline
\centering{$I$ } & 0 & 0 & 1/2 & 1/2 & 1/2 & 5/2 & 9/2\\ \hline
\centering{transition}
& ${}^2$S$_{1/2} $
& ${}^2$S$_{1/2}$
& ${}^2$S$_{1/2}, F=0$
& ${}^2$S$_{1/2}, F=0$
& ${}^2$S$_{1/2}, F=0$
& ${}^1$S$_{0}, F=5/2$
& ${}^1$S$_{0}, F=9/2$ \\
& $\rightarrow {}^2$D$_{5/2}$
& $\rightarrow {}^2$D$_{5/2}$
& $\rightarrow {}^2$D$_{3/2}, F=2$
& $\rightarrow {}^2$F$_{7/2}, F=3$
& $\rightarrow {}^2$D$_{5/2}, F=2$
& $\rightarrow {}^3$P$_{0}, F=5/2$
& $\rightarrow {}^3$P$_{0}, F=9/2$
\\ \hline
\centering{$f_0$ (THz)}
& 411.042
& 444.779
& 688.358
& 642.121
& 1 064.72
& 1 121.02
& 1 267.40
\\ \hline
\centering{$g_g$ }
& 2.002 256 64(9)\tmark[a]
&  2.002\tmark[e]
& 1.998(2)\tmark[i]
& 1.998(2)\tmark[i]
& 2.003 174 5 (74)\tmark[n]
&  -0.00079248(14)\tmark[p]
&  -0.000 666 47\tmark[r] \\ \hline
\centering{$g_e$}
&  1.200 334 0(3)\tmark[a]
&  1.2\tmark[e]
&  0.802(2)\tmark[i]
& 1.145(2)\tmark[i]
& 1.1980(7)\tmark[n]
&  -0.00197686(21)\tmark[p]
&  -0.000 987(50)\tmark[r]\\ \hline
\centering{$\Delta f_\mathrm{M2}$ (Hz/mT$^2$) }
&  14.355(17)\tmark[b]
& 3.12225\tmark[f,z]
& 52 096(16)\tmark[j]
& -2030(20)\tmark[l]
& -18 900(2 800)\tmark[n]
&  -71.988(48)\tmark[p]
& 4.09\tmark[s]\\ \hline
\centering{$\Delta\alpha_S$ ($10^{-41}$~J m$^2$/V$^2$)}
& -73.0(1.0)\tmark[t]
&  -48.3(1.7)\tmark[g,f]
& 69(14)\tmark[k]
& 13(6)\tmark[m]
&  15\tmark[n]
& 0.82(8)\tmark[t]
&  3.3(3)\tmark[t]\\ \hline
\centering{$\Delta\alpha_T$ ($10^{-41}$~J m$^2$/V$^2$)}
& -24.51(29)\tmark[t]
& -78.6(5)\tmark[g] 
& -13.6(2.2)\tmark[k] 
& $\sim$ 1.3\tmark[m]
& -3\tmark[n]
& 0
& 0
\\\hline
%
%
\centering{$\eta$}
& -
&  -0.01\tmark[g]
& -
& -
& -
& $<10^{-4}$\tmark[t]
& $<10^{-4}$\tmark[t]\\ \hline
%
\centering{$\Delta f_\mathrm{300~K}$ (Hz)}
& 0.380(13) &  0.250(9) & -0.36(7) &  -0.067(31) &  -0.079 &  -0.0043(4)&  -0.017(2)\\ \hline
\centering{$\Theta$ (e a$_0^2$)}
& 1.83(0.01)\tmark[d]
& 2.6(3)\tmark[h]
&  2.08(11)\tmark[k]
&  -0.041(5)\tmark[m]
& -0.510(18)\tmark[o]
& 0.0\tmark[c]
& 0.0\tmark[c] \\ \hline
}

\subsection{Quantum logic spectroscopy of \Al}
\subsubsection{Quantum logic spectroscopy}
Efficient cooling of external motion and internal state discrimination of the clock atom(s) are an indispensable prerequisite for operating a clock. Typically, Doppler cooling and internal state detection are implemented on dipole-allowed cycling transitions. This puts severe restrictions to the level structure of the atomic species considered as clock references. In trapped ion systems, this restriction has been lifted by co-trapping a so-called logic ion together with the clock ion to provide sympathetic cooling \cite{larson_sympathetic_1986}. Furthermore, by employing techniques developed for quantum information processing \cite{wineland_experimental_1998, haffner_quantum_2008, blatt_entangled_2008}, the internal state information can be mapped through a series of laser pulses from the clock ion to the logic ion, where it is detected with high efficiency \cite{wineland_quantum_2002,  wineland_ion_2004}. It was first implemented for the \Al \ssztpo transition \cite{schmidt_spectroscopy_2005} and is the read-out scheme for the aluminum clock~\cite{rosenband_observation_2007}.

Quantum logic spectroscopy allows the selection of a clock ion species solely based on the features of the clock transition, since all other requirements are supplied by the logic ion. This enables spectroscopy of previously intractable ion species, such as group 13 (see Sec.~\ref{sec:systems}), highly charged \cite{dzuba_high-precision_2012, dzuba_ion_2012, derevianko_highly_2012, berengut_optical_2012, berengut_electron-hole_2011, berengut_enhanced_2010}, and molecular ions \cite{leibfried_quantum_2012, ding_quantum_2012, koelemeij_blackbody_2007, schmidt_spectroscopy_2006, vogelius_probabilistic_2006}.

\subsubsection{Clock operation}
\label{sec:qls}
\begin{figure}
\includegraphics[width=13cm]{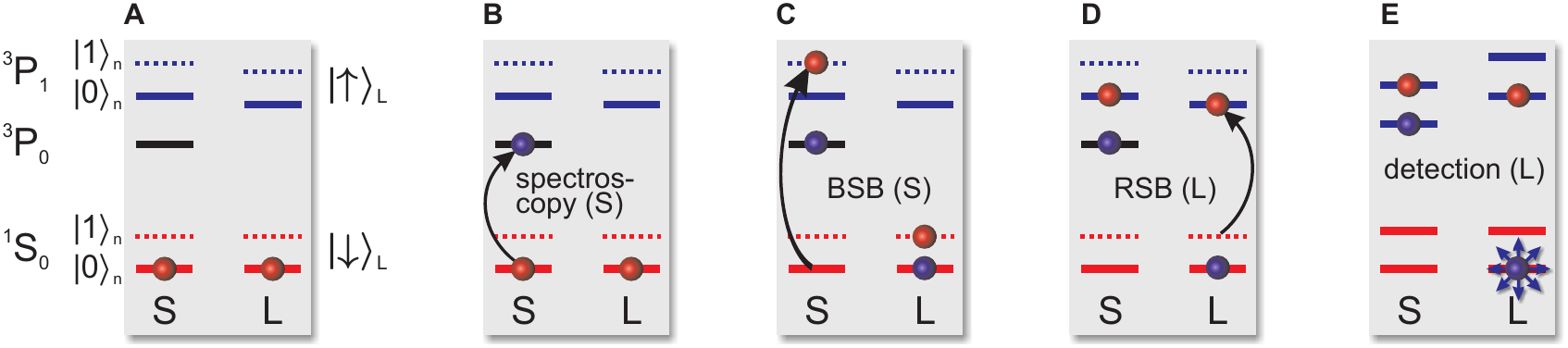}
\caption{Quantum logic spectroscopy sequence. Shown are the clock ground (\ssz) and excited (\tpz) clock states and an auxiliary metastable state (\tpo) together with the logic ion (qubit states $\ket{\downarrow}_L$, $\ket{\uparrow}_L$). In addition, two vibrational levels ($\ket{0}$, $\ket{1}$) of a common motional mode of the ions in the trap are shown. (A) Initially, both ions are prepared in the electronic and motional ground state. (B) After clock interrogation, the spectroscopy ion is in an equal superposition of the two clock states. (C) Resolved-sideband pulse on the blue sideband of the clock ion to the auxiliary state, mapping the ground state amplitude onto the first excited motional state. (D) Resolved-sideband pulse on the red sideband of the logic ion, mapping the first excited motional state amplitude to the electronically excited state of the logic ion. (E) Detection of the logic ion's internal state via the electron shelving technique \cite{dehmelt_proposed_1975-2}. Energy levels are not to scale.}
\label{fig:QLS_scheme}
\end{figure}
\begin{figure}[t]
\includegraphics[width=7cm]{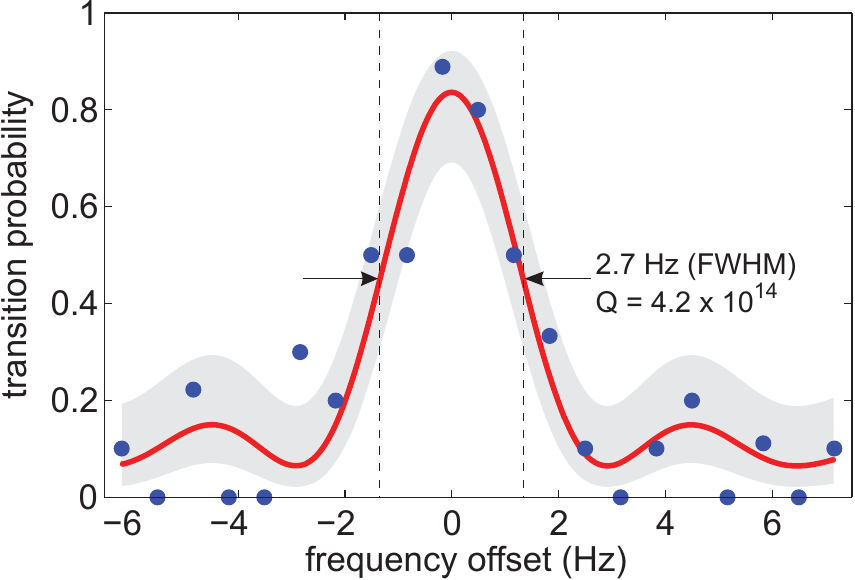}

\caption{Resonance of the \Al clock transition using quantum logic spectroscopy \cite{chou_optical_2010}. }
\label{fig:QLS_resonance}
\end{figure}

A simplified quantum logic spectroscopy scheme for interrogating the \Al clock is shown in Fig.~\ref{fig:QLS_scheme}.
The system is initialized in the electronic and motional ground state (we neglect motional heating for the moment) of a shared axial normal mode of the two ions (A). After interrogation of the clock transition \ssztpz (B), the internal state information is mapped through a pair of laser pulses onto the logic ion. The first pulse is implemented on the \ssztpo transition, allowing faster transfer compared to using the clock transition. When the ion is in the \ssz state, the pulse drives a blue sideband (BSB) transition changing the electronic state to \tpo, while adding a quantum of motion to the motional mode. A similar pulse tuned to the red sideband (RSB, removing a quantum of motion while changing the electronic state) is applied to the logic ion, reversing the mapping step. The internal state of the logic ion is then detected using the usual electron shelving technique on the logic ion.
If the clock ion was in the excited clock state at the beginning of the pulse sequence, none of the transitions can be excited since the state mapping laser is not resonant with any transition. The pulse sequence thus implements a faithful transfer of the clock ion's internal state after probing the clock transition to the logic ion. The term ``quantum logic spectroscopy'' is derived from the original proposal for quantum information processing with trapped ions by \textcite{cirac_quantum_1995} and many other quantum logic protocols, relying on internal-state dependent (de-)excitation of a motional state shared among several ions.
Figure~\ref{fig:QLS_resonance} shows a scan across the \Al clock resonance using this quantum logic spectroscopy technique.

\begin{figure}[t]
\includegraphics[width=15cm]{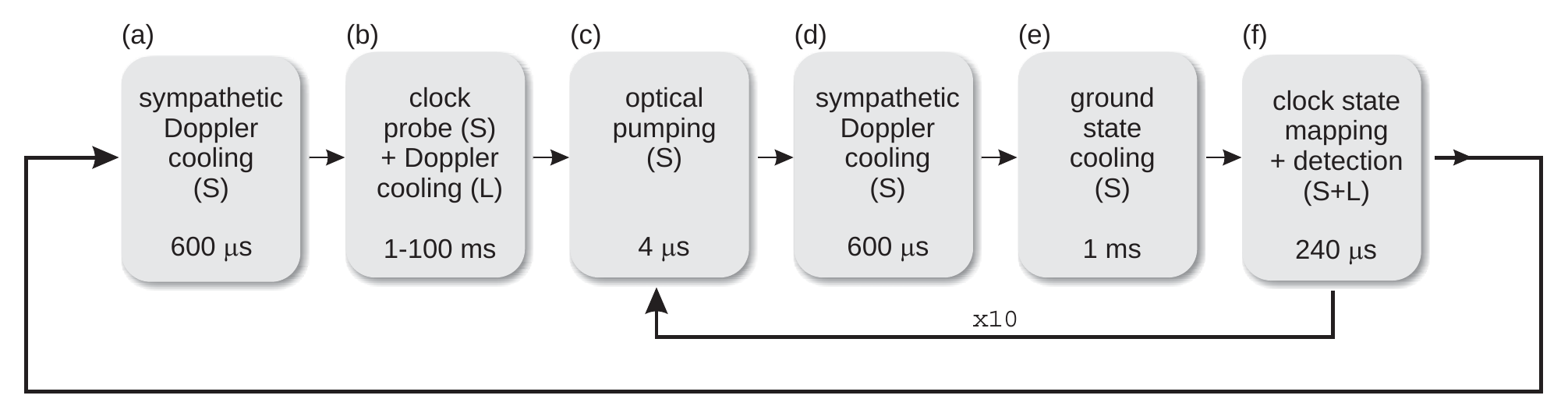}
\caption{Quantum logic clock interrogation cycle \cite{rosenband_observation_2007}. The following sequence describes interrogation of the \ssz $m_F=5/2\longrightarrow$ \tpz $m_F=5/2$ state. A similar protocol is used for the other ($m_F=-5/2$) stretched state interrogation.  (a) Sympathetic Doppler cooling, reaching $\bar{n}\approx 3$ in all modes with tilted ion crystal. (b) Probing of the \Al clock transition \ssztpz with simultaneous application of cooling light to maintain steady state motional occupation. (c) Optical pumping on the \ssztpo transition to \ssz, $m_F=5/2$ state. (d) Sympathetic Doppler cooling followed by ground state cooling of a selected axial mode to a mean motional excitation of $\bar{n}\approx0.05$ (e). (f) Quantum logic state detection (see Fig~\ref{fig:QLS_scheme} for details. Steps (c) through (f) are repeated 10 times for improved readout fidelity.}
\label{fig:QLS_cycle}
\end{figure}

In reality a few more steps are required to implement the full interrogation sequence.
A typical probe cycle is sketched in Fig.~\ref{fig:QLS_cycle}.

Imperfections in the transfer sequence and subsequent state detection on the logic ion results in reduced state detection fidelity. However, the state mapping (steps (c) through (f)) takes around 2~ms, which is sufficient for the \tpo state (lifetime 300~$\mu$s~\cite{johnson_transition_1986, trabert_measurement_1999}) to decay back to the ground state, whereas the excited clock state (lifetime 20.6~s~\cite{rosenband_observation_2007}) experiences negligible loss of population during this time.  Consequently, the mapping process can be repeated to improve state detection fidelity. State discrimination with up to 99.94~\% fidelity using Bayesian inference has been demonstrated for 10 detection repetitions~\cite{hume_high-fidelity_2007}. Every few seconds, the initial Zeeman state of the clock is changed via optical pumping using polarized light on the \ssztpo transition. Recording the center frequencies of both streched states ( $\ssz, m_F=\pm 5/2\leftrightarrow \tpz, m_F'=\pm 5/2$) allows the calculation of a linear-Zeeman shift free transition frequency from the sum of both frequencies \cite{bernard_laser_1998}. The difference frequency provides a direct measure of the magnetic field, which is then used to compute the DC component of the quadratic Zeeman shift \cite{rosenband_observation_2007}.
The total duration of a single interrogation cycle is approximately 120~ms of which 100~ms are used for probing~\cite{rosenband_frequency_2008}. This corresponds to a duty cycle of more than 80~\%, neglecting so-called service cycles during which slowly drifting parameters are recalibrated, such as micromotion compensation.
The additional overhead from calibration effectively reduces the duty cycle to between 45~\% and 65~\%, depending on the details of the implementation.

In principle, any ion that can be laser cooled and provides a pair of qubit states with internal state discrimination is a suitable candidate for a logic ion. However, the choice of logic ion influences the systematic effects of the clock. As discussed in Sec.~\ref{sec:lineartraps}, fluctuating electric fields lead to motional heating of the ions in the trap, and consequently a second order Doppler shift which is increasing during probe time. A steady state with a lower uncertainty in the second order Doppler shift can be achieved through laser cooling of the logic ion during interrogation. This imposes additional Stark shifts on the clock transition that depend on the cooling laser wavelength and have to be calibrated. The steady-state kinetic energy of the \Al ion in this situation depends on the achievable minimum laser cooling energy, usually determined by the linewidth of the cooling transition, and the mass ratio between clock and cooling ion. For realistic heating rates in typical ion traps for clocks, \Be,\Mg, and \Ca provide a similar residual second order Doppler shift of below $10^{-17}$ relative frequency uncertainty \cite{wubbena_sympathetic_2012}. In fact, for small external heating rates the narrow cooling transition linewidth of \Be and \Ca lets them outperform \Mg. Although the latter's mass is almost perfectly matched to Al+, allowing fast energy transfer and efficient sympathetic cooling, its Doppler cooling limit is hotter by nearly a factor of two.

\subsubsection{Experimental achievements of the \Al clocks}
As of the writing of this document, the performance of two \Al clocks has been reported. In the following, we will call them \Alo \cite{rosenband_frequency_2008} and \Alt \cite{chou_frequency_2010}, using \Be and \Mg as the cooling ion, respectively.
A number of impressive experimental results have been achieved with these, demonstrating the capabilities and future potential of optical clocks in terms of instability and inaccuracy.
An optical frequency ratio measurement between \Alo and a cryogenic single ion \Hg clock (see Sec.\ref{sec:mercury}) has been perfomed \cite{rosenband_frequency_2008}, resulting in a ratio of $\nu_{Al+}/\nu_{Hg+}= 1.052871833148990438(55)$ with a statistical uncertainty of $4.3\times 10^{-17}$. To date this is the most precise measurement of an optical frequency ratio of two different optical clock species. Combined with a previous measurement of the absolute frequency of the \Hg transition \cite{stalnaker_optical--microwave_2007} establishes an absolute frequency for the \Al clock transition of 1121015393207857.4(7)~Hz, limited by the uncertainty of the Cs fountain clock used for the calibration of the \Hg clock. The relative systematic uncertainty of the clocks was estimated to be $1.9\times 10^{-17} $ for \Hg and $2.3\times 10^{-17}$ for \Al. A comparison of the two clocks spanning almost a year yielded the currently lowest upper bound for a variation of the fine-structure constant from laboratory measurements (see Sec.\ref{sec:constants}).

The instability of single ion frequency standards is determined by the experimentally achievable $Q$-factor (Sec.~\ref{sec:techniques}), feedback strategy, and dead time.  Figure~\ref{fig:QLS_resonance} shows a Fourier-limited clock transition linewidth of 2.7~Hz with 80~\% contrast for 300~ms probe time, corresponding to a quality factor of $Q=4.2\times 10^{14}$ \cite{chou_optical_2010}.
Experimentally achieved instabilities are typically derived from frequency comparisons between two or more optical clocks.
The relative instability in a frequency comparison between \Alo and \Alt was  $2.8\times 10^{-15}/\sqrt{\tau/\mathrm{s}}$ \cite{chou_optical_2010} with probe times of 100~ms and 150~ms, respectively, and a duty cycle between 40~\% and 65~ \%. The frequencies of the two standards agreed to within $(-1.8\pm 0.7)\times 10^{-17}$, consistent with the evaluated inaccuracy of $2.3\times 10^{-17}$ and $8.0\times 10^{-18}$ for \Alo and \Alt, respectively (see next Section). Phase noise in the probe lasers, as discussed in Sec.~\ref{sec:oscillators}, limits the stability between two optical frequency standards. However, this noise source can be eliminated by correlating the phase noise seen by the atoms \cite{Bize00, lodewyck_frequency_2010}. Such a scheme has been implemented using two \Al ions trapped in the same trap and interrogated simultaneously by the same probe laser \cite{chou_quantum_2011}. The differential signal between the two ions is free of laser phase noise, since it is common mode suppressed. By adjusting the distance between the two ions before the second Ramsey pulse, the relative phase can be scanned. The relative coherence time can be extracted from the contrast of the observed fringes to be $T_C=9.7^{+6.9}_{-3.1}$~s, corresponding to a relative $Q$-factor of $Q=3.4^{+2.4}_{-1.1}\times 10^{16}$, and limited by the excited state lifetime of $T'=20.6\pm 1.4$~s \cite{rosenband_observation_2007}.
While such a suppression of laser phase noise conflicts with the operation of a clock, it may be useful for applications such as relativistic geodesy for which the local frequency of two frequency references operating in a different gravity potential are compared using length-stabilized optical fibers (see Sec.~\ref{sec:geodesy}).

\subsubsection{Systematic shifts of the \Al clocks}
\begin{table}
\setlength{\tabcolsep}{5pt}
\renewcommand{\arraystretch}{1.3}
\begin{tabular}{|l|c|c|c|c|l|}
\hline \multirow{2}{*}{\textbf{Shift}}& \multicolumn{2}{|c|}{\textbf{\Alo}} & \multicolumn{2}{|c|}{\textbf{\Alt}} & \multirow{2}{*}{\textbf{Limitation}}\\
\cline{2-5}  & \textbf{$\Delta f/f$} & \textbf{$\sigma$} & \textbf{$\Delta f/f$} & \textbf{$\sigma$} & \\
\hline Micromotion & -20 & 20 & -9 & 6 & static electric fields \\
\hline Secular motion & -16 & 8 & -16.3 & 5 & Doppler cooling \\
\hline Black-body radiation & -12 & 5 & -9 & 0.6 & DC polarizability \\
\hline Cooling laser Stark & -7 & 2 & -3.6 & 1.5 & Polarizability, intensity \\
\hline Clock laser Stark & - & - & 0 & 0.2 & Polarizability, intensity \\
\hline quadratic Zeeman & -453 & 1.1 & -1079.9 & 0.7 & B-field calibration \\
\hline First-order Doppler & 0 & 1 & 0 & 0.3 &  Statistical  imbalance\\
\hline Background gas collisions & 0 & 0.5 & 0 & 0.5 &  Collision model\\
\hline AOM phase chirp & 0 & 0.1 & 0 & 0.2 &  rf power\\
\hline\hline total & -513 & 22 & -1117.8 & 8.6 &\\
\hline
\end{tabular}
\caption{Systematic shifts and uncertainties for the \Alo \cite{rosenband_frequency_2008} and \Alt \cite{chou_frequency_2010} clocks. The fractional frequency shifts $\Delta f/f$ and the $1\sigma$ uncertainties are given in units of $10^{-18}$.}
\label{tab:aluminum-systematics}
\end{table}

The general physical principles of systematic frequency shifts in optical clocks have been outlined in Sec.~\ref{sec:systematics}, whereas the specifics of the shifts for trapped ions have been discussed in Sec.~\ref{sec:importantsysteffects}. In the following, we will discuss the mitigation of systematic shifts and their experimental evaluation in the two \Al clocks as prototypical systems of high-accuracy ion clocks. Table \ref{tab:aluminum-systematics} provides a summary of the shifts and their uncertainty.

The dominant uncertainty of both clocks arises from time dilation shifts caused by micromotion and residual secular motion of the ions. Micromotion compensation is limited by the measurement and control of static electric fields. The magnitude of this shift is bounded by the error in nulling it.
Micromotion compensation via mode-cross coupling as described in \citet{barrett_sympathetic_2003} and Sec.~\ref{sec:iontraps} was used in \Alo with a resolution of $(0\pm 10)$~V/m for the residual electric field. A field of 10~V/m in each of the radial directions corresponds to a fractional shift on the order of $10^{-17}$ \cite{wineland_laser-cooling_1987, berkeland_laser-cooled_1998}.  The more sensitive micromotion sideband technique \cite{berkeland_minimization_1998} was employed for \Alt, resulting in a reduced uncertainty in nulling this shift. The non-vanishing oscillating trap field in the presence of excess micromotion induces an additional AC Stark shift. However, in both standards this effect contributes less than 10\% to the total shift, and it can therefore be neglected~\cite{chou_frequency_2010}. Dynamic changes of excess micromotion, e.g. through charge buildup from the photo-electric effect, are highly sensitive to the duty cycle of the clock and need to be compensated while running the clock through interleaved calibration sequences.
Secular motion arises from insufficient cooling of the clock ion via the logic ion during the interrogation of 100~ms and 150~ms for \Alo and \Alt, respectively. As described in Sec.~\ref{sec:lineartraps}, in a dual-ion quantum logic clock, there is one less efficiently cooled mode along each trap axis. Additional heating from fluctuating electric fields raises the steady-state temperature above the Doppler cooling limit. For \Alo, the evaluation of the time dilation shift from secular motion is complicated by the fact that the initial temperature at the beginning of the probe time is below the steady-state temperature at the end of the probe time. The reason is that initial Doppler cooling is performed on a tilted ion crystal to provide better cooling through mode coupling. The tilt is induced by applying an additional static field of 300~V/m during cooling, which is adiabatically relaxed before probing the clock transition. During interrogation, the crystal is aligned with the rf zero line of the trap, resulting in a reduced cooling rate for two radial modes. The temperature rise in these modes during interrogation has been calibrated \cite[supplementary information]{rosenband_frequency_2008}. The expected uncertainty in determining the resulting shift of $-16\times 10^{-18}$ is 8 parts in $10^{18}$, arising from drifts in the experimental parameters and angular calibration errors. The influence of all the other modes to the time dilation shift is below $10^{-18}$ and has been neglected in table \ref{tab:aluminum-systematics}. The cooling rate for all modes is maximized through mass-matching the logic ion to the clock ion. This is the case for \Alt with a mass-mismatch of only 8\%, where the mean vibrational excitation matches the expected Doppler cooling limit \cite{chou_frequency_2010}. The uncertainty of this shift is given by the experimental error of 30\% in determining it.

Linear Doppler shifts that can potentially arise from charging of the trap electrodes by the clock laser (see Sec.~\ref{sec:ionMotion}) have been investigated using independent frequency servos for counterpropagating interrogation beams. For \Alo, they were found to be smaller than $1\times 10^{-18}$, whereas in \Alt a relative shift between the two probe directions of $(1.2\pm 0.7)\times 10^{-17}$ was observed. Imperfect frequency averaging of the two directions arising from slightly different gain settings in the servo loops leaves a residual uncertainty of $\pm 0.3\times 10^{-18}$ for this shift.
Another shift closely related to the linear Doppler shift arises from phase chirps of the interrogation laser during the probe time. These phase chirps can arise from instabilities in the optical setup that do not average to zero for long times \cite{falke_delivering_2012}. The major contribution originates from ringing and thermal expansion of acousto-optical modulators (AOMs) used for switching the probe beam \cite{degenhardt_influence_2005}.
It can be reduced by either combining the AOM with a mechanical shutter to keep the AOMs duty cycle close to 100~\%, or by applying only a very small rf power (e.g. 1~mW for \Alt) \cite{rosenband_frequency_2008-1}. This way, the relative frequency uncertainty could be reduced to $0.1\times 10^{-18}$ and $0.2\times 10^{-18}$ for \Alo and \Alt, respectively.

Several shifts arise from interaction of the clock states with external electric fields.
The most recent value for the differential polarizability between the two clock states has been estimated from \textit{ab initio} calculations to be $\Delta\alpha_S=\alpha(P)-\alpha(S)=0.82(8)\times 10^{-41}\mathrm{J}\mathrm{m}^2/\mathrm{V}^2$ \cite{safronova_precision_2011, mitroy_blackbody_2009}, resulting in a relative frequency shift of only $3.8(4)\times 10^{-18}$ at 300~K. This is the smallest BBR shift of an electronic transition in neutral or singly-charged atoms considered for optical clocks. The effective BBR environment seen by the ion would need to be known with an uncertainty of only 15~K to achieve $10^{-18}$ relative frequency uncertainty. The values given in Tab.~\ref{tab:aluminum-systematics} are derived from $\Delta\alpha_S=1.7(6)\times 10^{-41}\mathrm{J}\mathrm{m}^2/\mathrm{V}^2$, which was inferred from a measurement of the dynamic polarizability at a wavelength of 1126~nm and an extrapolation to zero frequency via experimental oscillator strengths \cite{rosenband_blackbody_2006}. This extrapolation becomes possible, since all contributing transitions are in the deep UV spectral regime, far away from the calibration wavelength at 1126~nm, and all strong transitions lie around $171$~nm, compensating each other to a large degree. Recently, the polarizability of \Al has been remeasured using a 976~nm Stark-shifting laser to be $\Delta\alpha_S=0.702(95)\times 10^{-41}\mathrm{J}\mathrm{m}^2/\mathrm{V}^2$ \cite{chou_unpublished_2014}, which is very close to the theoretical value given above and in Tab.~\ref{tab:systematics}.

Doppler laser cooling of the logic ion during interrogation with a laser beam illuminating both ions also causes a Stark shift of the clock ion. For \Alt this shift has been evaluated by calibrating the intensity of the cooling laser beam through off-resonant excitation of the \Mg dark state \ket{F=2, m_F=-2} and applying the model for the BBR shift extrapolated to the cooling laser wavelength of 280~nm. The model yields a shift of $(-3.5\pm 0.6)\times 10^{-17} s$, with saturation parameter $s=I/I_s$ and saturation intensity $I_s\approx 2470$~W/m$^2$. The measured saturation parameter of $s=0.103\pm 0.04$ results in a total shift of $(-3.6\pm 1.5)\times 10^{-18}$. This shift can be further reduced by focusing the laser beam onto the cooling ion, or by using a logic ion species with smaller saturation intensity and further wavelength detuning, such as \Ca.
Off-resonant coupling of the clock laser to other levels has been evaluated by significantly increasing the intensity of the interrogation pulse in \Alt and comparing to \Alo. No shift has been detected at a fractional frequency level of $2\times 10^{-15}$, corresponding to an uncertainty of $0.2\times 10^{-18}$ when scaled down to the normal operating power.

The largest shift of the clock transitions stems from the quadratic Zeeman effect. The shift is proportional to the average of the square of the magnetic induction $\expect{B^2}=\expect{B_\mathrm{DC}}^2+B_\mathrm{AC}^2$, consisting of a static and dynamic contribution, $B_\mathrm{DC}$ and $B_\mathrm{AC}$, respectively. The static component arises from a small applied quantization field of around $B_\mathrm{DC}\approx 0.1$~mT. Its slow drifts can directly be deduced from the difference of the \Al stretched state frequencies $f_1$ as described above, exhibiting a linear Zeeman shift of $\Delta f_\mathrm{M1}=-82884(5) B~$Hz/mT~\cite{rosenband_observation_2007}. The  corresponding quadratic shift of $\Delta f_\mathrm{M2}=-7.1988(48) \times 10^7$~Hz/T$^2$ has been calibrated by deliberately varying the static field and measuring the transition frequency against another frequency standard~\cite{rosenband_frequency_2008-1}. The dynamic contribution arises mostly from charge/discharge currents of the rf trap electrodes and can be calibrated from hyperfine spectroscopy on the logic ion by e.g. comparing the clock transition  $[(^2\mathrm{S}_{1/2}, F=2, m_F = 0) \longrightarrow (^2\mathrm{S}_{1/2}, F=1, m_F = 0)]$ frequency in \Be to the transition $(^2\mathrm{S}_{1/2}, F=2, m_F = -2) \longrightarrow (^2\mathrm{S}_{1/2}, F=1, m_F = -1)$ with large linear magnetic field sensitivity. Similar transitions were used in \Mg for \Alt. The measured magnetic fields were $B_\mathrm{AC}=5\times 10^{-8}$~T and $B_\mathrm{AC}=5.2\times 10^{-6}$~T for \Alo and \Alt, respectively.
The combined quadratic Zeeman shifts for \Alo and \Alt are $-453\pm 1.1\times 10^{-18}$ and $-1079.9\pm 0.7\times 10^{-18}$, respectively.

Although collision shifts between cold and localized trapped ions are absent, collisions with background gas can result in a differential shift between the ground and excited clock state. Two types of collisions are distinguished by comparing the impact parameter $b$ to the Langevin radius $r_L=\sqrt{\frac{4e^2\alpha}{4\pi\epsilon_0\mu v_r^2}}\approx 0.5~nm$, where $\alpha$ is the polarizability of the background gas atom (assumed to be a hydrogen molecule), $\mu$ is the reduced mass, and $v_r$ is the mean relative velocity.
In glancing collisions ($b>r_L$), the background gas particle flies by the clock ion at large distance. The charge of the ion induces a dipole moment in the background gas particle, resulting in a $C_4/r^4$  interaction. The fractional resulting shift can be estimated to be below $10^{-20}$ at a pressure fo $10^{-9}$~Pa for the \Al ion~\cite{rosenband_frequency_2008-1}. In Langevin collisions ($b\leq r_L$) with thermal background gas, significant phase shifts of the ion can occur. The short collision time ($1~\mu$s) allows to model the effect of the collision as an instantaneous phase shift of up to $2\pi$ at arbitrary times during the interrogation pulse of several ten to hundreds of ms. It has been shown in a numerical study, that a worst case phase shift of $\pi/2$ in the middle of a Rabi pulse causes a frequency shift of $0.15 R_\mathrm{coll}$, where $R_\mathrm{coll}$ is the collision rate~\cite{gioumousis_reactions_1958, rosenband_frequency_2008-1}. If  a collision can be detected, e.g. through a drop in fluorescence during laser cooling from the large energy transfer during the collision, such events can be discarded and no shift correction has to be applied. For the \Al clocks, the mean time between collisions has been estimated from ion crystal reordering to be on the order of a few hundred seconds. This results in a fractional shift of up to $0.5\times 10^{-18}$ \cite{rosenband_frequency_2008-1}.

\subsection{Other optical ion frequency standards}

Here we discuss specific properties of different atomic ions other than $^{27}$Al$^+$ and the main experimental achievements that have been obtained with these ions in the development of optical frequency standards. The energy level schemes, transition wavelengths and linewidths, and sensitivity factors for the most important systematic frequency shifts are given in Figs. \ref{fig:groupII}, \ref{fig:groupIII} and in Tab. \ref{tab:ionatomicparams}.
Table  \ref{tab:absfreqion} lists the results of the most precise absolute frequency measurements that are available for these ions.

\begin{table}
  \caption{Selected absolute frequency measurements of optical clocks with trapped ions.}
  \label{tab:absfreqion}
  \begin{tabular}{|l|l|l|c|}\hline
   Ion & Transition & Absolute frequency and uncertainty (Hz) & reference \\ \hline
		$^{27}$Al$^+$ & $^1S_0-^3P_0$ & 1 121 015 393 207 857.4(7) & \cite{rosenband_frequency_2008} \\
    $^{40}$Ca$^+$ & $^2S_{1/2}-^2D_{5/2}$ & 411 042 129 776 393.2(1.0) & \cite{chwalla_absolute_2009} \\
							& & 411 042 129 776 393.0(1.6) & \cite{huang_hertz-level_2012} \\
							& & 411 042 129 776 398.4(1.2) & \cite{matsubara_direct_2012}\\
		$^{88}$Sr$^+$ & $^2S_{1/2}-^2D_{5/2}$ & 444 779 044 095 484.6(1.5) & \cite{margolis_hertz-level_2004} \\
						& & 444 779 044 095 485.5(9) & \cite{madej_<sup>88</sup>sr<sup>+</sup>_2012} \\
$^{171}$Yb$^+$ & $^2S_{1/2}-^2D_{3/2}$ & 688 358 979 309 307.82(36) & \cite{tamm_cs-based_2014} \\
$^{171}$Yb$^+$ & $^2S_{1/2}-^2F_{7/2}$ & 642 121 496 772 645.15(52) & \cite{huntemann_high-accuracy_2012} \\
			& & 642 121 496 772 646.22(67) & \cite{king_absolute_2012} \\
$^{199}$Hg$^+$ & $^2S_{1/2}-^2D_{5/2}$ & 1 064 721 609 899 144.94(97) & \cite{oskay_single-atom_2006} \\
 \hline
  \end{tabular}
\end{table}

\subsubsection{Calcium}
$^{40}$Ca$^+$ is an isotope without hyperfine structure and therefore convenient for laser cooling. It has found many applications in experiments on quantum computing \cite{haffner_quantum_2008}. The electric quadrupole reference transition  $^2S_{1/2}-{^2D}_{5/2}$ has been investigated in single ions \cite{chwalla_absolute_2009, huang_hertz-level_2012,matsubara_direct_2012} and also in entangled states of two ions that can be designed to suppress selected frequency shifts like the linear Zeeman shift \cite{roos_designer_2006}. The same groups have reported absolute frequency measurements.
The isotope $^{43}$Ca$^+$ has been investigated because its half-integer nuclear spin $I=7/2$ leads to the existence of magnetic-field insensitive $m_F=0$ Zeeman sublevels \cite{champenois_evaluation_2004,kajita_prospect_2005,benhelm_measurement_2007}. The high value of $I$ and the resulting high number of sublevels, however, makes it difficult to obtain cyclic excitation for laser cooling and also to efficiently populate a selected $m_F=0$ state for interrogation of the reference transition.

\subsubsection{Strontium}
$^{88}$Sr$^+$ has advantages similar to those of $^{40}$Ca$^+$ in terms of simplicity of the level scheme and availability of reliable solid-state laser sources for cooling and interrogation \cite{barwood_observation_1993,marmet_precision_1997}. The methods of averaging the transition frequency over several Zeeman components for the elimination of the linear Zeeman, electric quadrupole and quadratic Stark shift have been developed and first applied here on the  electric quadrupole reference transition  $^2S_{1/2}-{^2D}_{5/2}$ \cite{bernard_laser_1998,dube_electric_2005,margolis_hertz-level_2004,madej_<sup>88</sup>sr<sup>+</sup>_2012,dube_evaluation_2013}. A recent evaluation has resulted in a systematic uncertainty of $2.3\times10^{-17}$, dominated by the contribution from the blackbody radiation shift \cite{dube_evaluation_2013}. As in Ca$^+$, the use of an odd isotope $^{87}$Sr$^+$ with half-integer nuclear spin $I=9/2$ has been discussed \cite{boshier_polarisation-dependent_2000}. Again, the high value of $I$ leads to the same difficulties as mentioned for $^{43}$Ca$^+$. The $^{88}$Sr$^+$ optical frequency standard is presently being investigated in two laboratories, NPL in Great Britain and NRC in Canada, and both groups have performed absolute frequency measurements that show good agreement of the results.

\subsubsection{Ytterbium}
The rare-earth ion Yb$^+$ presents an alkali-like level scheme with similarities to Ca$^+$ and Sr$^+$. Apart from even isotopes with $I=0$, an isotope
$^{171}$Yb$^+$ with $I=1/2$ exists, so that a magnetic field insensitive $F=0$ hyperfine sublevel of the ground state becomes available and the problem of state preparation is reduced to hyperfine pumping. Work on Yb$^+$ frequency standards therefore concentrates on this isotope. The relatively high atomic mass of Yb$^+$ leads to smaller Doppler shift at a given temperature. Experiments with trapped Yb$^+$ consistently observe the longest storage times \cite{tamm_stray-field-induced_2009} -- exceeding several months  --  of a single ion among the elements investigated as optical frequency standards, facilitating the long-term continuous operation of the standard. While in other ions chemical reactions with background gas seem to ultimately limit the storage time, this loss process is  prevented for Yb$^+$ by the near-coincidence of photodissociation resonances for YbH$^+$ with the 370~nm cooling laser light
\cite{sugiyama_production_1997}.   Several reference transitions in $^{171}$Yb$^+$ have been studied, including the 12.6~GHz microwave frequency standard based on the ground state hyperfine splitting \cite{fisk_trapped-ion_1997} and the $^2S_{1/2}-{^2D}_{5/2}$ electric quadrupole transition
\cite{taylor_investigation_1997}. Work has focussed on the
$^2S_{1/2}-{^2D}_{3/2}$ electric quadrupole transition \cite{tamm_spectroscopy_2000} and on the  $^2S_{1/2}-{^2F}_{7/2}$ electric octupole transition
\cite{roberts_observation_1997}. Both frequency standards are presently pursued at PTB in Germany and  NPL in Great Britain. The quadrupole transition has been used in a sub-hertz optical frequency comparison between two trapped ions that has also made it possible to measure the relevant polarisabilities and the quadrupole moment of the $^2D_{3/2}$ state \cite{schneider_sub-hertz_2005}.
The octupole transition between the $^2S_{1/2}$ ground state and the
lowest excited $^2F_{7/2}$ state is unusal because of its extremely small natural linewidth in the nanohertz range. While allowing for very high resolution, at the limit imposed by noise of the interrogation laser, an associated disadvantage is a significant light shift of the transition frequency \cite{webster_kilohertz-resolution_2002}. This shift is proportional to the laser intensity so that a $\pi$-pulse with Fourier-limited spectral width $\Delta f$ causes a shift proportional to $(\Delta f)^2$. The shift contains both scalar and tensorial contributions and scales like
$0.65(3)$~Hz$^{-1}(\Delta f)^2$ if the polarization and magnetic field orientation are chosen to maximize the excitation probability \cite{huntemann_high-accuracy_2012}.
Disregarding the light shift, the sensitivities of the Yb$^+$ octupole transition frequency to electric field induced shifts are significantly lower than those of the quadrupole transitions in the alkali-like ions, as has been pointed out in theoretical estimates \cite{Lea07a} and measured in the frequency standard \cite{huntemann_high-accuracy_2012}. Qualitatively, this can be explained by the electronic configuration $(4f^{13}6s^2)$ of the
$^2F_{7/2}$ level that consists of a hole in the $4f$ shell that is surrounded by the filled $6s$ shell, and therefore less polarizable than an  outer $d$-electron. PTB and NPL have both reported absolute frequency measurements of the octupole transition with respect to primary caesium fountain clocks, obtaining Cs-limited uncertainties below $1\times10^{-15}$ and excellent agreement of the values
\cite{huntemann_high-accuracy_2012,king_absolute_2012}. With the application of a generalized Ramsey interogation method that suppresses the uncertainty due to the light shift from the interrogation laser \cite{huntemann_generalized_2012} and improved control of the blackbody radiation shift, this systems offers prospects for a systematic uncertainty below $10^{-17}$.

\subsubsection{Mercury}\label{sec:mercury}
$^{199}$Hg$^+$, like $^{171}$Yb$^+$, has also been investigated as a frequency standard in the microwave
\cite{prestage_ultra-stable_1992,berkeland_laser-cooled_1998}, as well as in the optical frequency range, based on the $^2S_{1/2}-{^2D}_{5/2}$ electric quadrupole transition \cite{bergquist_recoilless_1987}. The Hg$^+$ optical frequency standard developed at NIST in the USA makes use of a cryogenic ion trap that reduces ion loss due to reactions with the background gas and the frequency shift induced by blackbody radiation \cite{poitzsch_cryogenic_1996}.
The suppression of the quadrupole shift through averaging over three orthogonal orientations of the quantization axis was first demonstrated in this system \cite{oskay_measurement_2005,oskay_single-atom_2006}. The total systematic uncertainty has been evaluated to $1.9\times10^{-17}$ fractional frequency uncertainty. A number of precise absolute frequency measurements of this transition have been performed at NIST over an extended time span, so that, together with data on transitions in $^{171}$Yb$^+$ and $^{27}$Al$^+$, it can be used to constrain a temporal drift of the fine structure constant
\cite{peik_limit_2004,fortier_precision_2007,rosenband_frequency_2008}.

\subsubsection{Barium}
$^{138}$Ba$^+$ was used in the pioneering experiments on laser-cooling of ions in Paul traps \cite{neuhauser_localized_1980} and kHz-resolution spectroscopy has been performed of the  $^2S_{1/2}-{^2D}_{5/2}$ electric quadrupole transition at $1.76~\mu$m wavelength
\cite{nagourney_high_1990,yu_stark_1994,appasamy_quantized_1995} and on the 24-THz fine structure transition between the two $D$-levels \cite{whitford_absolute-frequency_1994}. In the latter case, an  absolute frequency measurement has also been performed \cite{whitford_absolute-frequency_1994}. More recently, use of the $^2S_{1/2}-{^2D}_{3/2}$ electric quadrupole transition at $2.05~\mu$m wavelength in $^{137}$Ba$^+$ has been proposed
\cite{sherman_progress_2005}. With a nuclear spin $I=3/2$, this isotope possesess a hyperfine sublevel $F=0$ of the $^2D_{3/2}$ state so that the transition would be free from the linear quadrupole shift.

\subsubsection{Indium}
$^{115}$In$^+$ was the first ion where laser excitation of the hyperfine-induced $^1S_0 - {^3P}_0$ transition was demonstrated \cite{peik_double_1995}.
Unlike Al$^+$, the intercombination line $^1S_0 - {^3P}_1$ in In$^+$ is sufficiently fast to allow for laser sideband cooling that leads to a vibrational quantum number $\langle n \rangle <1$ in a one-stage cooling process \cite{peik_sideband_1999}.
Precision laser spectroscopy of the  $^1S_0 - {^3P}_0$ transition has led to measurements of the lifetime and g-factor of the excited state
\cite{becker_high-resolution_2001} and to early frequency measurements using a mode-locked femtosecond laser and a calibrated, methane-stabilized He-Ne laser as a reference \cite{zanthier_absolute_2000}. Similarly to Al$^+$, In$^+$ offers very low sensitivity to field-induced systematic shifts
\cite{becker_high-resolution_2001,safronova_precision_2011}. Among the singly charged ions of the third group of the periodic system, In$^+$ is most amenable for laser cooling, but the relatively small linewidth of the cooling transition results in a low photon count rate for fluorescence detection, whereas the reference transition with a natural linewidth of 0.8~Hz would limit the obtainable spectral resolution and thus the instability of a In$^+$ single-ion frequency standard. It has therefore been proposed to use larger numbers of laser-cooled In$^+$ ions in a linear Paul trap for a multi-ion optical frequency standard with improved stability \cite{herschbach_linear_2012}.

\newpage
\section{NEUTRAL ATOM ENSEMBLE OPTICAL FREQUENCY STANDARDS}

Optical atomic clocks based on neutral atoms possess the advantage of enhanced clock signals that offer improved clock stability. However, only recently with improved local oscillators are these benefits being exploited. At the present time, rapid advances are being made with these systems, and we foresee continued advances in both stability and accuracy for neutral atom optical clocks.

\subsection{Atomic candidates: Alkaline Earth(-like) Elements}

The choice of a quantum reference depends on a number of important factors. As already emphasized earlier, a good starting point is to find a clock transition that supports a superior line quality factor $Q$ and whose frequency is insensitive to external fields.  For the long coherence times demanded by state-of-the-art frequency standards, it is also crucial that the atoms have well-defined motion -- namely that they can be efficiently prepared by laser cooling and trapping.  Alkali atoms such as caesium and rubidium have played a prominent role in atomic clocks.  Notably, $^{133}$Cs has served as the primary standard of time and frequency since 1967.  At the same time, these alkali systems have played a pioneering role in laser cooling and quantum control.  Properties such as strong laser accessible transitions, ground state magnetic moments, and magnetic Feshbach resonances have made this control possible.  The ability to manipulate these atomic systems also became important for the development of the most accurate caesium and rubidium standards, relying on laser cooled samples in an atomic fountain.  Yet while the fractional accuracy of the most advanced Cs clocks is now approaching the 1 part per 10$^{16}$ level, improvement in its fractional stability, and ultimately the accuracy, is hindered by the relatively small hyperfine transition frequency of 9.2 GHz.

Among possible atomic candidates, alkaline earth (-like) atoms and ions (Mg, Ca, Sr, Yb, Hg, Al$^{+}$,In$^{+}$, \ldots) are turning into increasingly popular choices for frequency standards due to their narrow intercombination transitions and simple level structure~\cite{Hall89a}. A representative diagram of this level structure is shown in Figure~\ref{levels}.  With two valence $s$-shell electrons, the spin of each electron can add parallel or anti-parallel, yielding singlet and triplet states.  Strong transitions exist among the various singlet or triplet states, while weaker spin-forbidden transitions occur between them.  In many cases the strong $^1S_0$-$^1P_1$ cycling transition can be used for cooling, trapping, and sensitive state detection, while the spin-forbidden $^1S_0$-$^3P_1$ can be used for cooling to ultra low temperatures.  Transitions from the $^3P$ states to $^3S_1$ or $^3D$ are useful for repumping the $^1S_0$-$^1P_1$ cooling transition or for optical pumping used in state detection.  The doubly-forbidden $^1S_0$-$^3P_0$ transition in isotopes with nuclear spin have attracted the most attention.  The low lying metastable $^3P_0$ excited state has only very weak coupling to $^1S_0$, with a laser accessible energy interval.  The $^1S_0$-$^3P_0$ transition linewidth is very small (ranging from 1 Hz to well below 1 mHz) offering a line Q reaching 10$^{18}$, optimal for optical clock development.  Furthermore, the lack of electronic angular momentum in these clock states reduces the size of many potential systematic uncertainties in the system. For atomic confinement in an optical potential, these Group II species are ideal due to the existence of Stark cancelation wavelengths and because of the minimal dependence of the clock frequency on the light polarization. While each atomic specie has individual advantages as a frequency standard, currently Sr, Yb, and Hg are popular choices for standards under development.  Here we discuss optical lattice clocks with particular emphasis on Sr and Yb, recognizing that many features of these systems are shared by other alkaline earth(-like) systems.

\subsection{Laser cooling and trapping of alkaline earth(-like) atoms}

\begin{figure}[t!]
   \centering
     \includegraphics[width=\textwidth]{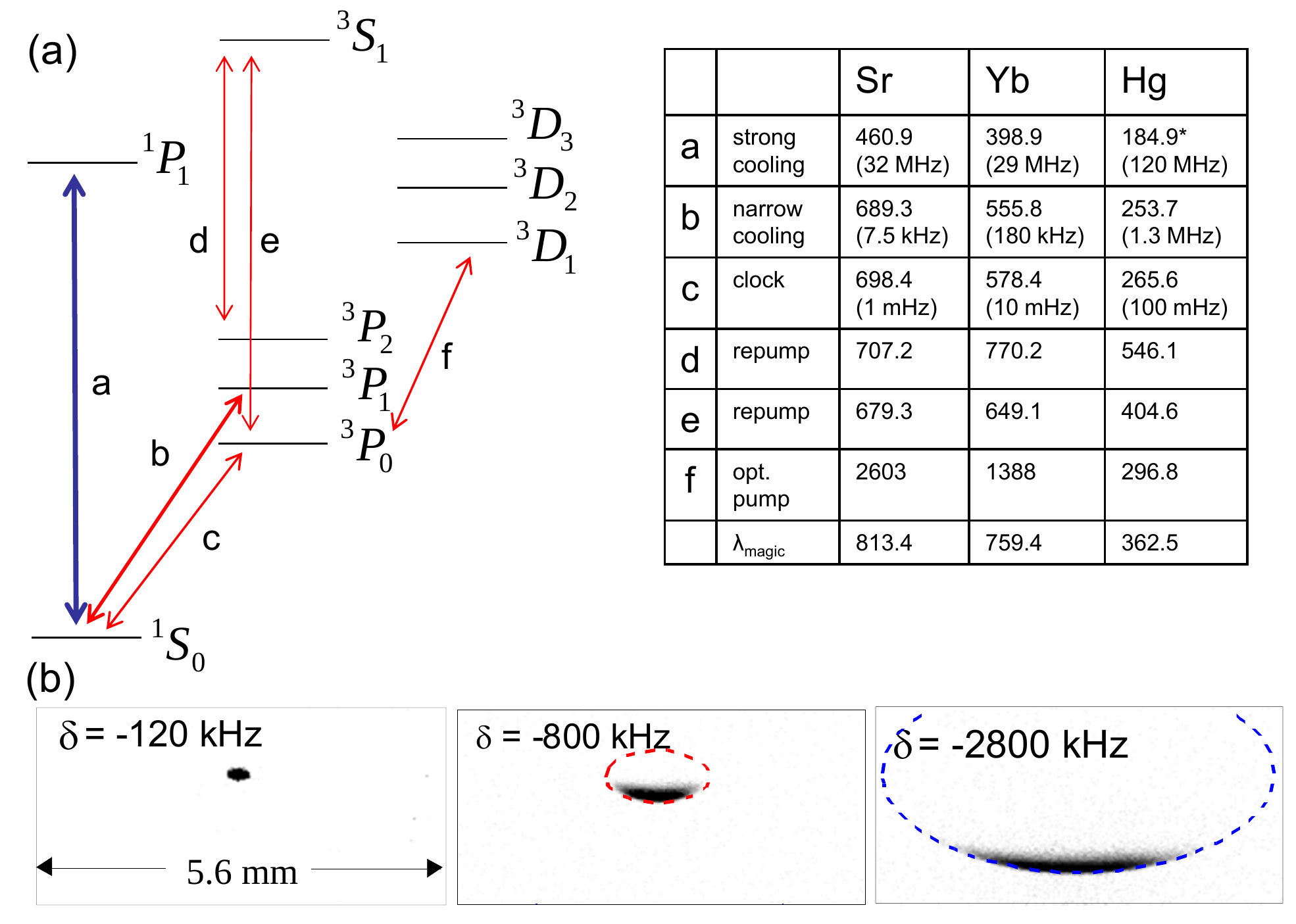}
    \caption[]{\label{levels} (a) A simplified energy level diagram representative of many Group 2 (-like) atoms.  Some of the most relevant transitions are indicated with arrows, and their wavelengths (in nm) and linewidths are specified in the adjacent table for the case of Sr, Yb, and Hg. (b) The importance of gravity on narrow line cooling dynamics is clearly seen from the \textit{in-situ} red MOT images as the laser detuning $\delta$ is varied in a narrow-line MOT for Sr. The dashed ovals represent the spatial position where the photon scattering rate is the highest as the laser frequency detuning matches the Zeeman level shift induced by the MOT magnetic field. In the absence of gravity, the dashed ovals would define the MOT region.}
\end{figure}

The $^1S_0$-$^1P_1$ transition (Figure~\ref{levels}) is well suited for laser cooling and trapping from a thermal source.  The transition typically has natural linewidths of several tens of MHz or more, allowing relatively fast photon scattering for efficient cooling.  It is also a nearly closed transition, enabling many photons to be cycled.  In most cases, the transition is not completely closed: excited $^1P_1$ population weakly decays into the triplet manifold and eventually makes its way to the lower lying metastable $^3P$ states.  This decay is particularly weak for Yb, but even for cases like Sr, it is sufficiently weak that MOT operation does not require a repumping laser.  Nevertheless, a repumping configuration can yield longer MOT lifetimes and more trapped atoms, and some possible repumping configurations are shown in Figure ~\ref{levels}.  Another benefit of the $^1S_0$-$^1P_1$ transition is that it links a $J=0$ to a $J=1$ state, making it a simple one for achieving MOT spatial confinement.

Some experimental complexity does exist with laser cooling alkaline earths on this transition, and this is one reason that these systems historically have been studied less extensively than their alkali metal counterparts.  The first is that the $^1S_0$-$^1P_1$ transition tends to lie in the bluer regions of the optical spectrum, where achieving high laser power has been more challenging.  The broad natural linewidth also dictates relatively large magnetic field gradients in a magneto-optic trap (MOT), requiring MOT coils that are driven with up to 100 A or more.  Finally, the $J=0$ ground state is a simple one for laser cooling, but with a small magnetic moment of a nuclear origin (for fermionic species), popular magnetic trapping of the ground-state alkaline earth atoms is essentially prohibited.  All of these experimental complexities can be addressed and have been overcome in their various applications. Over the recent decades the number of options for reliable blue laser sources continues to increase.  Work in laser cooling calcium, magnesium, and strontium began twenty years ago or more (e.g.\ \cite{Kurosu90a,Beverini89a,Witte92a,Sengstock93a,Fox93a}) and more than ten years ago in ytterbium (e.g.\ \cite{Honda99a,Loftus00a}).

The natural linewidth of the $^1S_0$-$^1P_1$ transition offers the potential for a strong cooling force, but at the expense of a relatively high Doppler cooling temperature limit, near the mK level.  A second stage of Doppler cooling on the narrow $^1S_0$-$^3P_1$ intercombination transition offers a much lower limit which, depending on the choice of element, can approach the $\mu$K level or below.  Second stage cooling \cite{Hall89a} using this intercombination transition was first carried out for strontium \cite{Vogel99a,Katori99a}, and shortly thereafter on calcium \cite{Binnewies01a,Curtis01a}.  In the case of calcium, the $^1S_0$-$^3P_1$ transition linewidth is so narrow (400 Hz), that quenching is required to generate an optical force to exceed that of gravity.  The operation of a narrow line MOT has been studied extensively for strontium~\cite{Loftus04a,Loftus04b,Mukaiyama03a}, providing unique insights into narrow line cooling dynamics.  For the case of Yb, where the intercombination transition linewidth is 180 kHz, it is possible to load atoms from a slowed atomic beam directly into a MOT operating on the $^1S_0$-$^3P_1$ transition~\cite{Kuwamoto99a}.  For the case of Hg, where the intercombination transition linewidth is 1.3 MHz and generation of the 185 nm light for the $^1S_0$-$^1P_1$ transition is difficult, atoms are cooled and loaded directly into a MOT utilizing the $^1S_0$-$^3P_1$ transition~\cite{Hachisu08a,Petersen08a}.  The convenience of doing so is traded for a higher Doppler cooling limit on this intercombination transition (31 $\mu$K).

To give more detail to the cooling and trapping as realized in an optical lattice clock, here we discuss one particular example, for Sr~\cite{Ludlow06a,Boyd07b}.  Sr atoms are first loaded from an slowed atomic beam into a MOT operating on the strong $^1S_0$-$^1P_1$ transition, which is used as a pre-cooling stage to reach mK temperatures.  During this cooling stage, the weak decay path from the $^1P_1$ state results in population buildup in the $^3P_2$ state~\cite{Xu03a}. Repumping lasers are used to drive the population back to the ground state though the $^3P_{2,0}$$\rightarrow$$^3S_1$$\rightarrow$$^3P_1$$\rightarrow$$^1S_0$ channel and typically enhance the trap population by more than an order of magnitude.  Atoms are then released from the blue trap and undergo a brief stage of broadband $^1S_0$-$^3P_1$ molasses cooling to reduce the temperature to about 10~$\mu$K. Next, the atoms are loaded into a single-frequency MOT operating on the 7.4 kHz $^3P_1$ line for direct laser cooling below 1~$\mu$K~\cite{Loftus04a,Loftus04b}.

The narrow-line cooling offers a rich system of mechanical and thermodynamic properties that have been explored extensively ~\cite{Loftus04a, Loftus04b}. Here we mention just a couple of interesting effects. For strong transitions, such as the singlet line, the maximum scattering force from the cooling beams is about five orders of magnitude larger than the force of gravity. Conversely, for the narrow $^1S_0$-$^3P_1$ transition in Sr, the maximum light scattering force is only about 16 times larger than
gravity.  Therefore, gravity, which can be safely ignored in traditional laser cooling/MOT experiments, becomes a significant effect for a narrow line MOT.  As noted above, for lighter alkaline earth atoms with weaker intercombination lines (e.g. Ca), the cooling force is sufficiently weak such that the force of gravity dominates, making it impossible to realize a MOT with direct Doppler cooling, and other cooling schemes are required~\cite{Curtis01a,Binnewies01a}. The effect of gravity on the dynamics of a Sr MOT can be easily observed in Fig.~\ref{levels}(b), where a $^1S_0$-$^3P_1$ $^{88}$Sr MOT is imaged \textit{in situ} for different frequency detunings under a fixed saturation parameter $s$ = 250 of the trapping laser. As the detuning is increased, the gravitational force becomes more important, the atomic cloud sags until it reaches a spatial location where the corresponding magnetic field results in the maximum scattering rate. This self-adjusting feature results in a constant scattering rate at the trap boundary that is independent of the laser detuning. In contrast to standard laser cooling, this effect leads to a detuning-independent atomic temperature in the MOT~\cite{Loftus04a,Loftus04b}. In this case the temperature is 2~$\mu$K and is unchanged over a range of detunings from 100-400 times the transition linewidth.

Another significant feature of narrow line cooling is the importance of the photon recoil on cooling dynamics. For broad transitions we have the situation that $\Gamma_E$/$\omega_R$ $>>$ 1, where $\Gamma_E$=$\Gamma\sqrt{1+s}$ is the power broadened transition linewidth and $\omega_R/(2\pi)=\frac{\hbar k^2}{4 \pi M}$ is the photon recoil frequency.  For the Sr intercombination line (ignoring saturation), the ratio $\Gamma/\omega_R=1.6$.  In this case the relevant energy scale is that of a single photon recoil. Consequently, quantum (not semi-classical) scattering governs trap dynamics. When operating the red MOT at low saturation we observe temperatures as low as 250(20) nK, in good agreement with the predicted half recoil limit in quantum cooling~\cite{Castin89a}.

The cooling mechanisms described here were studied systematically with $^{88}$Sr.  The Group II atoms offer an abundance of both bosonic and fermionic isotopes.  Generally speaking, due to nucleon spin pairing, the bosonic isotopes of these Group II (-like) atoms have even numbered atomic mass and a nuclear spin equal to zero.  The fermionic isotopes have odd numbered atomic mass, and non-zero nuclear spin.  The non-zero nuclear spin introduces hyperfine splittings into the level structure shown in Figure~\ref{levels}.  This additional complication usually makes bosonic isotopes somewhat simpler systems for manipulation such as laser cooling. But sometimes hyperfine structure can bring unexpected benefit, such as sub-Doppler cooling for fermionic isotopes~\cite{Xu03b}. For narrow line cooling, the difference for fermionic isotopes is highlighted by the additional laser requirements for operation of a $^1S_0$ ($F=9/2$)--$^3P_1$ ($F=11/2$) MOT with fermionic $^{87}$Sr~\cite{Mukaiyama03a}.  The complexity arises due to the significant difference in the Land\'{e} $g$-factors for the ground and excited states, which are determined by the nuclear spin and the electronic spin, respectively.  This issue is exacerbated by the small natural linewidth of the transition, which results in a scattering rate, and even the direction of the force, that depends strongly on a specific $m_F$ sublevel populated. To achieve stable trapping, a two-color scheme~\cite{Mukaiyama03a} can be used with additional MOT beams driving the $^1S_0$($F=9/2$)-$^3P_0$($F=9/2$) transition. The $F=9/2$ excited state has a smaller $g$-factor than that of the $F=11/2$ state and it enables sufficient optical pumping to keep the atomic population within the states that are trapped by the primary MOT beams.  For lattice clock experiments, an optical lattice is typically overlapped with the MOT cloud during the entire cooling sequence to allow loading. This typically results in the capture of between $10^3$ and $10^6$ atoms for clock spectroscopy.

\subsection{Free Space Standards}

For neutral atoms, lack of a net electric charge precludes a straightforward method for confining the atoms without altering the natural electronic structure. Nevertheless, with advances in laser stabilization, nonlinear spectroscopy, and other experimental techniques, interest in probing narrow electronic spectra in dilute thermal samples gained momentum as early as the 1970s, with particular interests in the intercombination transitions in alkaline earth atoms~\cite{Barger76a,Bergquist77a,Barger79a}.  Atomic motion led to significant first and second order Doppler shifts, and consequently laser cooling played an important role in unlocking the potential of this type of atomic frequency standard. Researchers explored frequency standards using untrapped calcium, magnesium \cite{Sengstock94a,Friebe08a,He09a}, and strontium~\cite{Ido05a}.  Of considerable note, significant effort spanning more than a decade explored laser-cooled, ballistically-expanding calcium~ \cite{Kurosu92a,Witte92a,Kisters94a,Oates99a,Binnewies01a,Curtis01a,Curtis03a,Wilpers03a,Udem01a,Degenhardt05a,Wilpers06a,Wilpers07a}.  These cold-calcium systems often employed four-pulse optical-Ramsey interrogation \cite{Borde1984} and the later implementations benefitted from multiple stages of laser-cooling to reduce Doppler effects, including quenched narrow-line cooling to reach 10 $\mu$K temperatures or below. Impressive experimental efforts with these optical frequency standards achieved total uncertainties of $\leq1\times10^{-14}$.  While residual Doppler effects did not dominate the final uncertainty, they were nevertheless significant.  It was anticipated that another round of improvements could perhaps push the accuracy of the free space calcium standard to the $10^{-15}$ level.  However, at the same time, the neutral atom optical lattice clock was proposed.  If the residual AC Stark shift from atomic confinement in an optical lattice could be canceled at the `magic' wavelength, motional effects could be reduced to far below the $10^{-15}$ level. Consequently, momentum in the neutral atom optical frequency standard community moved towards optical lattice clock systems based on the $^1S_0$-$^3P_0$ transition, in other alkaline earth (-like) elements such as strontium, ytterbium, and mercury.

\subsection{Strong atomic confinement in an optical lattice}
A common feature of the optical lattice and single-ion clocks is the tight atomic confinement provided by a trap. In both systems, this confinement accomplishes a critical goal: decoupling the external (motional) and internal (atomic state) degrees of freedom, so that a precise measurement of the internal degree of freedom can be made without troubling systematics arising from atomic motion.  To approach confinement capable of a pure internal state measurement (free of motional effects), several important criteria must be met and are described below.

\subsubsection{Spectroscopy in the well-resolved sideband and Lamb-Dicke regimes}

\begin{figure}[b!]
   \centering
     \includegraphics[width=5in]{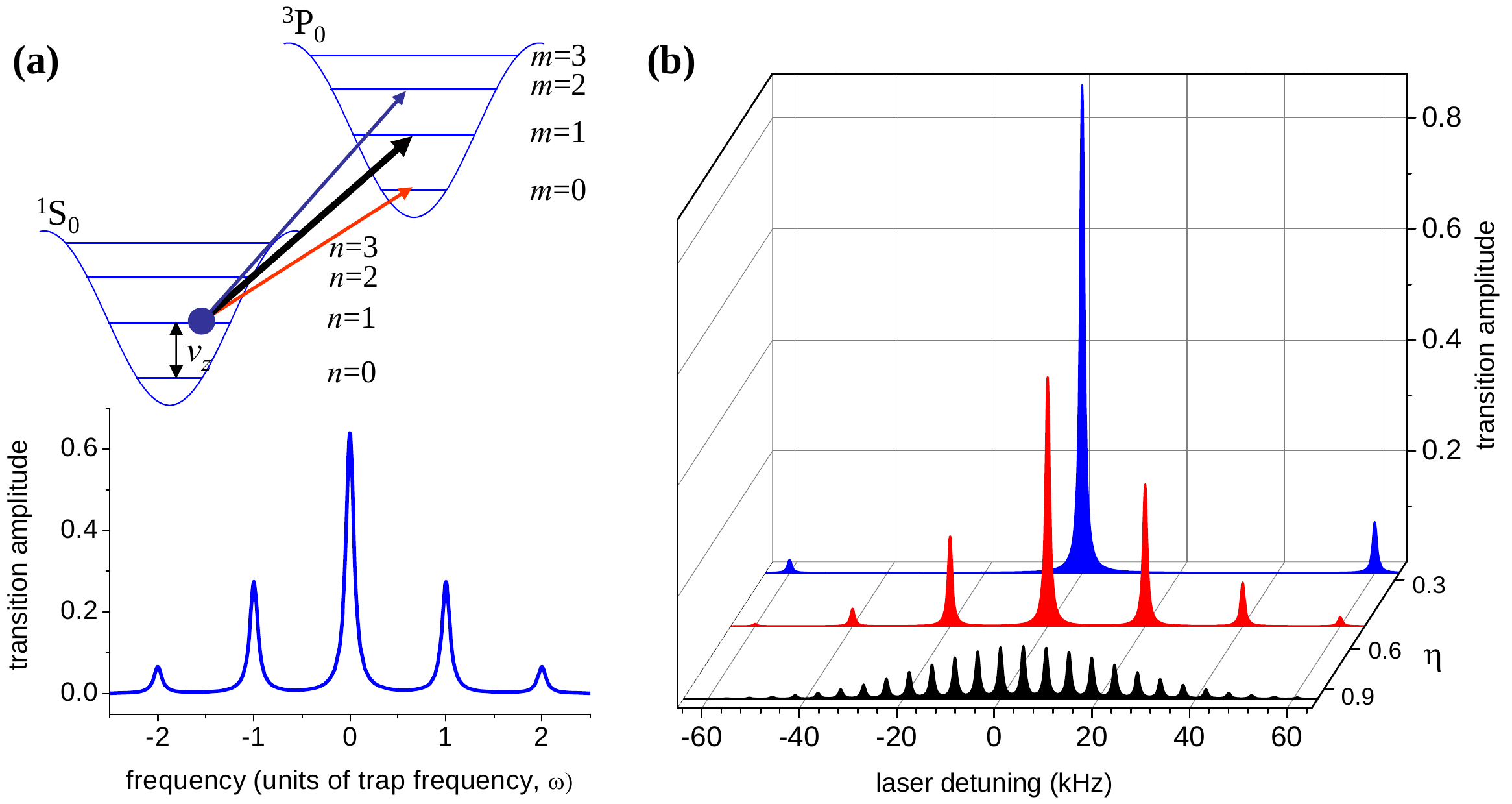}
    \caption[]{\label{FigConfine1} (a) A sketch of generic absorption spectrum in the well-resolved sideband limit. (b) Absorption spectrum of the $^1S_0$-$^3P_0$ clock transition in ultracold $^{87}$Sr under various levels of confinement.  Tighter confinement corresponds to higher trap frequency and smaller $\eta$.  The spectral lines are well-resolved, and the strong confinement curve (blue) falls in the Lamb-Dicke regime.}
\end{figure}

The evolution of a resonantly driven, two-level atom (at rest) is given by the Rabi flopping solution, with population exchange between the two levels at the Rabi frequency $\Omega$.  In the frequency domain, population is excited with the characteristic $sinc^2$ Rabi lineshape, or in the presence of sufficient decoherence, a Lorentzian lineshape whose width is the decoherence rate, $\Gamma$, divided by $2\pi$.  In the presence of atomic motion, this lineshape becomes inhomogeneously broadened from the Doppler shift across the atomic velocity distribution, yielding a Gaussian or Voigt lineshape.
However, for an atom confined in a harmonic potential, the atomic motion is not a continuous variable, but is restricted to the quantized motional states of the system.  The excitation of the two level atom with an initial motional state $|n\rangle$ and final motional state $|m\rangle$ is given by a modified Rabi rate:
\begin{equation}
    \label{eqn:confine21}
    \Omega_{mn}=\Omega
    e^{-\frac{\eta^2}{2}}\sqrt{\frac{n_<!}{n_>!}}\eta^{|m-n|}L_{n_<}^{|m-n|}(\eta^2),
\end{equation}
where $\Omega$ is the corresponding Rabi frequency for the atom at rest, $n_<$ ($n_>$) is the lesser (greater) of $n$ and $m$, and
$L_n^\alpha(x)$ is the generalized Laguerre polynomial.  $\eta$ is the so-called Lamb-Dicke parameter which is roughly the ratio of the spatial extent of the ground motional state and the wavelength of the probing radiation ($\eta=kx_0/\sqrt{2}$).  The resonant transition rate is given by the modified Rabi frequency $\Omega_{mn}$ and the transition frequency is determined by the energy difference between the initial and final states that include both electronic and motional degrees of freedom.  An overly-simplified illustration of the resulting excitation spectrum is shown in Fig.~\ref{FigConfine1}(a).  Here, the tall central feature corresponds to pure electronic excitation and to either side is the blue (red) sideband associated with both electronic excitation and excitation (de-excitation) of the atomic motion. The relative size of the decoherence rate, $\Gamma$, and the trap frequency, $\omega$, strongly influences the two level dynamics and the observed spectroscopic features illustrated in Fig.~\ref{FigConfine1}(a).  If $\Gamma > \omega$, the sideband structure in Fig.~\ref{FigConfine1} would be unresolvable, preventing a clean discrimination of the purely electronic excitation (the carrier transition) from a mixed electronic and motional excitation (sideband transitions). Indeed, in this limit, the various spectral features blend into each other, leaving spectroscopic measurements sensitive to motional shifts and broadening. This is in contrast to the case of $\Gamma \ll \omega$, where the Doppler effects are manifest at high modulation frequencies far from the carrier transition.  Consequently, the influence of motion on the purely electronic excitation is reduced to line-pulling effects from the motional sidebands. The ability to discriminate carrier and sidebands ($\Gamma \ll \omega$) is named the resolved-sideband or strong binding regime, and was first demonstrated in trapped-ion experiments~\cite{wineland_laser_1979,Stenholm86a}.

Atomic recoil also plays an important role in the observed spectra, and is influenced by the atomic confinement.  Figure~\ref{FigConfine1}(b) shows the absorption spectrum for ultracold $^{87}$Sr on the $^1S_0$-$^3P_0$ clock transition, for three cases of increasing atomic confinement (increasing $\eta$ from the bottom to top traces).  Here the effect of atomic recoil is included, and the spectrum is integrated over the Boltzmann distribution of motional states.  For the weakest confinement case, the recoil effect is clear: the transition with largest amplitude is not the pure electronic excitation at zero detuning, but rather the first blue-detuned motional sideband.  In the absence of any confinement, the continuous spectrum would be peaked at the blue-detuned recoil shift value.  As confinement becomes stronger (decreasing $\eta$), we move into the Lamb-Dicke regime where $\eta \ll 1$.  In this regime, the recoil effect on the line intensities is reduced: the carrier transition at zero detuning emerges as the dominant feature with maximum amplitude, and the sideband spectra, distributed at harmonics of the trap frequency on either side of the carrier, have amplitudes that are significantly suppressed relative to the carrier.  The Rabi rate for motional excitation (de-excitation) in Eq.~\ref{eqn:confine21} simplifies to $\Omega \eta \sqrt{n}$ ($\Omega \eta \sqrt{n+1}$) in this regime.  The Lamb-Dicke effect is equivalent to the suppression of sideband excitation \cite{dicke_effect_1953}, caused by the fact that the optical potential, not the atom, takes up the recoil momentum from an absorbed photon. This phenomenon is analogous to the much studied absorption of $\gamma$-rays in Mossbauer spectroscopy \cite{Mossbauer00a}.

Operating in both the resolved-sideband regime ($\Gamma \ll \omega$) and the Lamb-Dicke regime $\eta \ll 1$ provides maximum benefit to spectroscopically probe the transition virtually free of Doppler and recoil effects: In the resolved-sideband regime, motional effects are pushed to sideband frequencies far from the carrier, and in the Lamb-Dick regime, the motional sideband amplitudes are suppressed.  However, to realize the full separation between excitation of internal and external atomic degrees of freedom, one more critical condition must be met.  The confinement experienced by the atom must be the same regardless of which internal clock state is populated.  This is equivalent to saying that the carrier transitions shown in Figure \ref{FigConfine1} occur at a true zero detuning relative to the unperturbed atomic transition.  For large sample of cold neutral atoms, this is accomplished by confinement in an optical lattice operating at the `magic' wavelength.

\subsubsection{The magic wavelength}\label{section:magic}

The optical lattice confines the atoms by inducing a dipole moment in the atom, and exerting a force on this dipole through a laser field gradient. In general, the induced polarizabilities of the two clock states of the atom will differ such that the trapping field results in an ac Stark shift of the clock transition frequency, substantially deteriorating the clock accuracy. Furthermore, since the light field is inhomogeneous, atomic motion within the trap will couple the external and internal degrees of freedom, degrading the coherence in spectroscopic measurement. Although the dynamic polarizabilities (or equivalently the ac Stark shift) of the two clock states will have a
different form, they do have a dependence on the wavelength and polarization of the trapping light. In some special cases it is possible to tailor the trapping field so that the polarizabilities are equal and the clock states experience identical perturbations \cite{Ye08a}.  In this case the atoms can be measured in a pseudo Stark-shift-free environment, allowing Lamb-Dicke confinement and clean separation of the atomic motion from the internal degrees of freedom, similar to a trapped ion system but with many more atoms available for spectroscopy \cite{Katori03a}.  The idea that a magic wavelength lattice could be useful for a high accuracy optical frequency standard was first proposed by Katori et al.~in 2001 \cite{Katori03a}.

The two-electron level structure (see Fig.~\ref{levels}(a)) results in nearly independent series of singlet and triplet states such that
the Stark shift of the clock states can be tuned semi-independently. Consider the case of Sr: the ground state $(5s^2)\,^1S_0$, ignoring weak
intercombination transitions, is coupled predominantly to excited $(5snp)\,^1P_1$ states by an optical field. For all lattice wavelengths
longer than 461 nm (the lowest lying excited state transition wavelength), we have the situation of a red-detuned far-off-resonance optical dipole trap, in which the ac Stark shift will always be negative and the atoms will be trapped at the anti-nodes of the standing wave. The upper clock
states, $(5s5p)\,^3P_{1,0}$, are markedly different as three series of triplet states are coupled by the trapping laser, specifically
the $(5sns)\,^3S$ and $(5snd)\,^3D$ series, and the $(5p^2)\,^3P$ states. The Stark shift for the $S$ and $P$ state contributions will
be negative for all wavelengths above 700 nm. However, the low-lying $(5s4d)\,^3D$ state will contribute a positive shift for wavelengths
below 2600 nm.  In the wavelength range 700-2600 nm there exists a sign change in the polarizability and Stark shift of the
$(5s5p)\,^3P$ state. However, the $(5s^2)\,^1S_0$ polarizability changes very little in the same wavelength range. Additionally, the presence of
resonances in the $^3P$ polarizability provides sufficient amplitude swings to essentially guarantee a magic crossing point where the $^1S_0$ and $^3P$ polarizabilities match.

To find this magic wavelength, the dynamic ac Stark shifts can be calculated for the clock states of interest.  The Stark shift, $\Delta f$, of an energy level $i$ in the presence of an optical field with an electric field amplitude $E$ is given by $h\Delta f = - \frac{1}{2} \alpha_i |E|^2$. For a 1D optical lattice geometry the potential is described by a longitudinal standing wave with a Gaussian distribution in the radial dimension, given by \cite{Friebel98a}
\begin{equation} U(r,z)=4 U_{m} e^{-\frac{2r^2}{w(z)^2}}\cos^2(2\pi z/\lambda_L). \label{potential}
\end{equation}
Here $U_m=P\alpha_i /(\pi c \epsilon_0 w(z)^2)$, where $P$ is the average laser power of the incoming beam, $w(z)$ is the beam waist
(radius) at a longitudinal distance $z$ from the focus of the beam, $r$ is the radial distance from the beam center, and $\lambda_L$ is
the laser wavelength.  The trap depth can be characterized in terms of the harmonic oscillation frequency as $ U_T= \nu_z^2 \frac{M^2\lambda_L^4}{h^2}E_R$, where $E_R = \hbar \omega_R$, and $U_T/E_R$ characterizes the lattice intensity.
For a complete description of the trap properties, the polarizability of the relevant atomic states must be evaluated. In
the presence of a laser field of frequency $\omega_L$, the dynamic dipole polarizability of a state $i$ involves the sum over the
dipole interaction between state $i$ and excited states $k$,
\begin{equation}
\alpha_i(\omega_L, p)= 6 \pi  \epsilon_0 c^3
\sum_{k}\frac{A_{ik}(p)}{\omega_{ik}^2(\omega_{ik}^2-\omega_L^2)}\label{polarizability2},
\end{equation}
which depends only on the lattice frequency, the atomic transition
rates $A_{ik}(p)$ between states $i$ and $k$ for polarization $p$, and
the corresponding energy difference $\hbar\omega_{ik}$.

\begin{figure}[t!]
    \centering
    \includegraphics[width=4in]{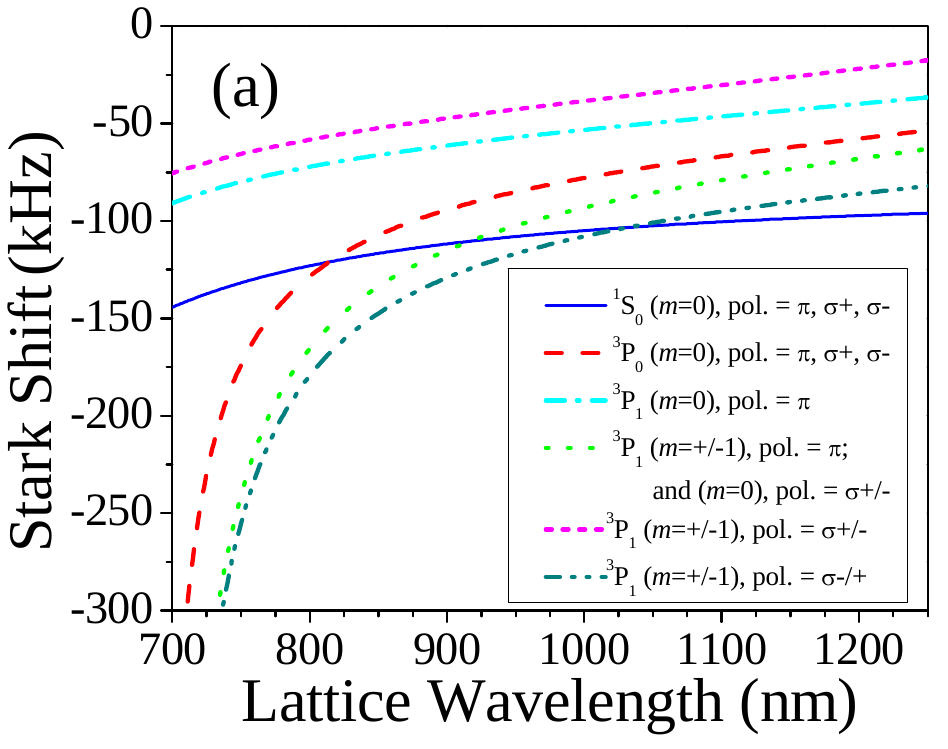}
    \caption[Magic wavelength calculation and measurements]{\label{5Hz} (a) Calculations of the wavelength-dependent ac Stark shift for the $^1S_0$, $^3P_0$, and $^3P_1$($m_J$=0,$\pm$1) states in $^{88}$Sr.  Values are given for linear ($\pi$) and circular ($\sigma^{\pm}$) polarizations.  The
    $^1S_0$ (solid line in blue) and $^3P_0$ (long-dashed line in red) states exhibit no polarization dependence and cross at a wavelength of 815 nm in good agreement with experimental results.  The $^3P_1$ state reveals a significant polarization- and $m_J$-dependence due to the tensor and vector nature of the light shifts.}
\end{figure}

Figure~\ref{5Hz} shows the calculated wavelength-dependent light
shifts for these states in Sr under various polarization configurations using Eq.~\ref{potential} and~\ref{polarizability2}, with $P$ = 150 mW and $w(z=0)$ = 65 $\mu$m. The light shift for the $^3P_1$ state shows a
significant dependence on the magnetic sublevel ($m_J$) and
polarization due to the tensor and vector light shift contributions.
An interesting region occurs at 917 nm, where the $^3P_1$($m_J=\pm1$)
states experience the same light shift as the $^1S_0$ state when
linearly polarized light is used.  This magic wavelength could be
used for development of a lattice clock based on the $^1S_0$-$^3P_1$
transition. However, the final accuracy of such a clock will likely be
limited by the polarization sensitivity. The $^1S_0$ and $^3P_0$ states
show no polarization dependence since $m_J$=0 are the only sublevels
present.  Therefore the polarization sensitivity is removed (aside
from the small corrections arising from nuclear spin in $^{87}$Sr) and the transition is more suitable for the highest accuracy spectroscopy. The calculated crossing point for the two clock states occurs just below 815 nm, convenient for developing high power stabilized laser systems.  For the parameters used here, the Stark shift of $U_0\sim h\times125$ kHz (or $U_0\sim 35 E_R$) corresponds to a longitudinal trap frequency of $\nu_z\sim$ 40 kHz such that $\eta$=0.33. Similarly, the sensitivity of the clock transition to deviations from the magic wavelength are calculated to be 10 (Hz/nm)/($U_T$/$E_R$), such that for this particular trap depth the lattice laser frequency can deviate by up to 500 kHz from the cancelation value without degrading the clock accuracy at the $10^{-18}$ level.

\begin{table}
\caption{\label{MagicTable}Some measured magic wavelength values for the $^1\!S_0$ - $^3\!P_0$ clock transition.  Bold text indicates the quantity (wavelength or optical frequency) reported directly in the given reference.}
\begin{tabular}{|l|c|c|c|}\hline
Atomic Species & Magic wavelength (nm, in vacuum) & Magic optical frequency (GHz) & Reference \\ \hline
 $^{87}$Sr & \textbf{813.420(7)} & 368,558(3) GHz & \cite{Takamoto05a} \\
& \textbf{813.418(10)} &  368,559(4.5) & \cite{Ludlow06a}\\
& \textbf{813.428(1)} &  368,554.4(0.5) & \cite{Brusch06a} \\
&  \textbf{813.4280(5)} & 368,554.4(0.2) & \cite{Boyd07a} \\
& 813.42735(40) & \textbf{368,554.68(18)} &  \cite{Ludlow08b} \\
&  813.427270(11) & \textbf{368,554.718(5)} & \cite{Westergaard11a}\\
&  813.427746(33) & \textbf{368,554.502(15)} & \cite{Falke11a} \\ \hline
$^{88}$Sr& 813.42757(62) & \textbf{368,554.58(28)} & \cite{Akatsuka10a} \\ \hline
$^{174}$Yb& \textbf{759.35(2)} & 394,800(10) & \cite{Barber06a}\\
&  759.353740(67) & \textbf{394,799.475(35)} &  \cite{Barber08a} \\ \hline
$^{171}$Yb & 759.355944(19) & \textbf{394,798.329(10)} & \cite{Lemke09a} \\
& \textbf{759.353(3)} & 394,800(1.6) & \cite{Kohno09a} \\
& 759.35565(15) & \textbf{394,798.48(79)} & \cite{Park2013} \\ \hline
\end{tabular}
\end{table}

An early step towards developing the lattice clock is the determination of the magic wavelength for the clock transition.  As a first indication, a number of theoretical and semi-empirical calculations of varying complexity have been made for different atomic species such as Sr, Yb, Hg, Cd, Zn, Mg, and Ca (see for example, \cite{Katori03a,Porsev04a,Degenhardt04a,Ovsiannikov07a,Hachisu08a,Ye08b,Dzuba10a} and references therein).  Ultimately, experimental measurement must be used to sufficiently constrain the value of the magic wavelength. To determine this experimentally, the transition frequency is measured for a variety of trap depths and wavelengths \cite{Ido03a}.  Table \ref{MagicTable} lists a number of such measurements for isotopes of Sr and Yb.  A measurement for Hg is reported in \cite{Yi2011}.

\begin{figure}[t!]
    \centering
    \includegraphics[width=4 in]{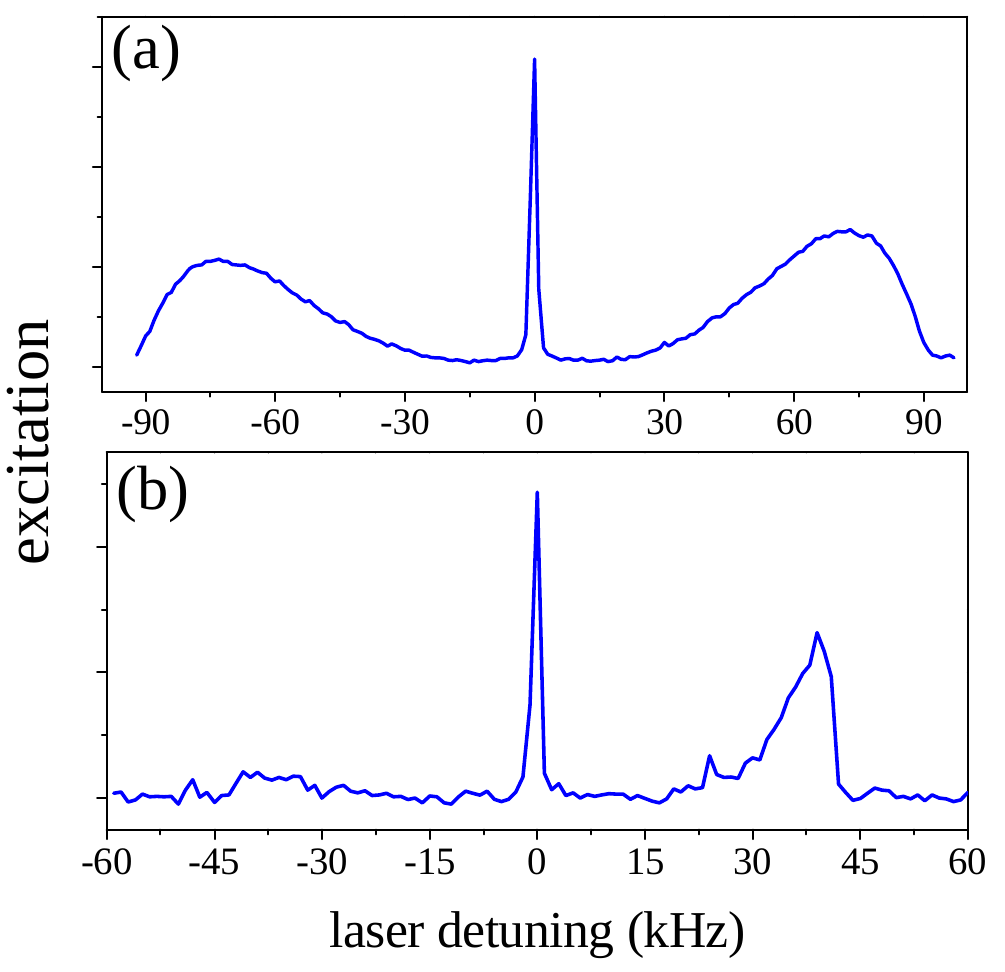}
    \caption[$^1S_0$-$^3P_0$ Spectroscopy in a 1D lattice: Spectroscopy of the Motional Sidebands]{\label{Trapspec}Spectroscopy of the clock transition in the optical lattice. When the clock transition is strongly driven into saturation, the motional sidebands can be more easily observed. From these spectra trap parameters such as motional frequency, trap depth, Lamb-Dicke parameter, as well as the atomic temperature can be extracted. Examples of sideband spectra for (a) Yb and (b) Sr are shown under trapping conditions described in the text.}
\end{figure}

\subsubsection{Spectroscopy of lattice confined atoms}

Even well into the Lamb-Dicke regime and the well-resolved sideband regime, the excitation spectrum shown in Fig.~\ref{FigConfine1} can be altered by details of the confinement.  This is particularly true for a 1-D optical lattice, presently a common choice of confinement for the lattice clock systems.  Figure~\ref{Trapspec} shows longitudinal sideband spectra of the clock transition for (a) Yb and (b) Sr for diverse trapping conditions in a 1-D optical lattice.  To make clear observations of the sideband, the carrier transition was driven strongly into saturation. Notably the red-detuned and blue-detuned sidebands are smeared out over a broad range of frequencies, unlike the motional sidebands observed in trapped single ion experiments.  Since the atoms are only tightly confined along the longitudinal axis of the 1-D optical lattice, weak transverse confinement means that the atomic wave-function extends into the Gaussian intensity profile of the lattice laser beam, especially for atoms occupying the higher transverse motional states.  At the lower intensity regions, the corresponding longitudinal trap frequency is smaller, and thus the sideband features bleed into lower frequencies~\cite{Blatt09}.  Furthermore, as the lattice trap depth is usually only a few tens of $\mu$K for these systems, higher longitudinal motional states sample the trap anharmonicity, which also results in lower trap frequencies for higher motional states.

The sharp edge of the blue sideband gives a good estimate of the longitudinal trap frequency. If the probe beam is aligned along the tight trap axis then the amplitude of the radial trap sidebands is significantly suppressed. Trap frequencies are most commonly measured with direct spectroscopy of the motional sidebands, but can also be measured using parametric excitation to induce trap heating and loss \cite{Friebel98a}.

The longitudinal temperature of the atom sample can be estimated from the relative areas under the blue and red sidebands.  Strong suppression of the red sideband indicates low atomic temperature as the $n=0$ atoms have no lower motional state to transfer to.  For example, in Figure~\ref{Trapspec}(b),  the relative strengths of the two sidebands are about 5:1, which for the relatively low trap frequencies yields $\langle n\rangle$ = 0.25, or a temperature of $\sim$ 1.5 $\mu$K.

While the motional sideband spectra, corresponding to both electronic and motional transitions, are strongly modified by the atomic confinement, the pure electronic transition for the central carrier maintains only a weak and indirect dependence. The Rabi excitation frequency for a given atom depends on the motional quantum numbers \cite{wineland_laser_1979}.  This dependence leads to excitation dephasing between atoms in different motional states of the Boltzman-distributed ensemble~\cite{Blatt09}. As a result of this dephasing, Rabi flopping contrast collapses and eventually revives.  For atoms in a 1-D optical lattice, the effect could be strong enough to reduce excitation from a mean $\pi$-pulse to 90$\%$ or less.

\subsubsection{Ultrahigh resolution spectroscopy} \label{highres}
The narrow central feature in Fig.~\ref{Trapspec} is the primary interest for clock development.  This carrier transition ($\Delta n=0$) provides a narrow atomic resonance, minimally affected by atomic motion in the Lamb-Dicke and resolved-sideband limits.  For saturation intensities below unity, the longitudinal sidebands' amplitudes are found to be at the percent level, while the radial sideband are estimated to be at least a factor of ten smaller.  In this case our absorption spectrum is a single strong feature at the clock transition frequency, with its width determined by the Fourier limit of probe laser pulse, when other broadening mechanisms are negligibly small and the laser is sufficiently coherent.

\begin{figure}[t!]
    \centering
    \includegraphics[width=\textwidth]{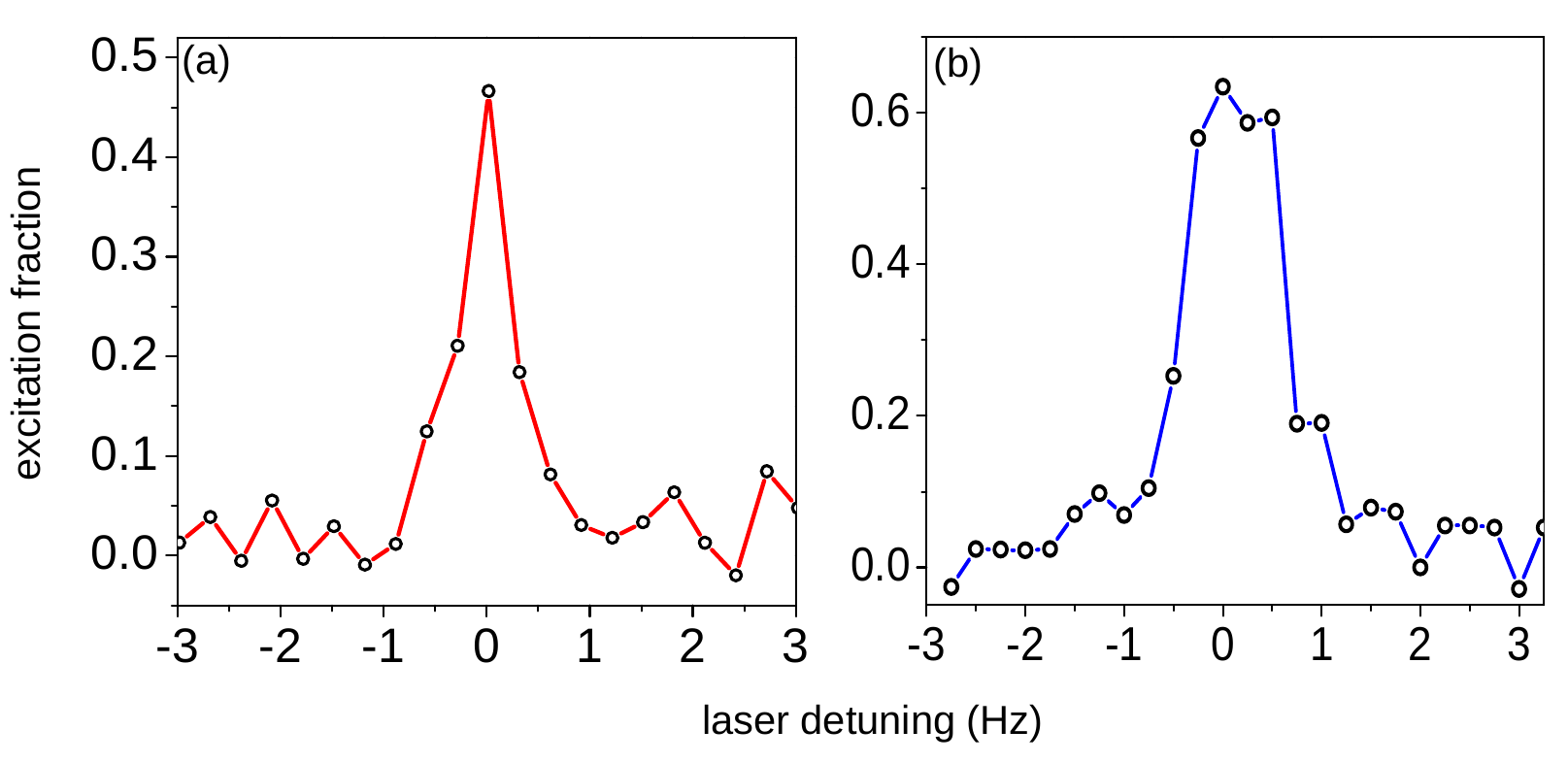}
    \caption[High resolution spectroscopy of the $^1S_0$-$^3P_0$ clock transition]{\label{Narrowsummary} (a) High resolution spectroscopy of $^{87}$Sr yielding a FWHM linewidth of 0.5 Hz (b) High resolution spectroscopy of $^{171}$Yb, yielding a FWHM linewidth of 1 Hz.}
\end{figure}

Since the narrowest resonances provide higher frequency resolution, lattice clocks need to operate with the best possible spectral linewidths for both stability and accuracy.  To date, the ability to observe the narrowest spectra has not been limited by the lattice trapped atoms, but rather by the stable lasers used to probe the transition.  The coherence time of these lasers typically limit the choice of probe time, which gives a minimum Fourier-resolvable linewidth.  Examples of the narrowest observed features are shown in Fig.~\ref{Narrowsummary}, for both Sr~\cite{Martin2013a} and Yb~\cite{jiang_making_2011}.  We note that the requirements on the laser coherence are quite stringent, since the laser frequency must be stable not only during the spectroscopic probing, but during many such probings to scan the laser frequency across the spectral lineshape. With a transition frequency of 429 THz for Sr and 518 THz for Yb, the observed spectral features correspond to a line quality factor approaching $10^{15}$, among the narrowest ever recorded for coherent spectroscopy.  We note that high resolution spectroscopy can also function as an optical spectrum analyzer to study the noise spectra of ultra-stable laser systems~\cite{Bishof2013a}.

Ramsey spectroscopy is also used for clock operation. The Ramsey interrogation scheme benefits from a slightly narrower spectral fringe compared to the one-pulse Rabi case, and can also be useful in some cases to reduce stability limitations from the Dick effect. It has also been a useful tool for cold collision studies in the optical lattice clock (e.g.~\cite{Lemke09a,Martin2013a}).  For the lattice bound atoms, there is no Doppler broadening of the carrier transition, so long as the Ramsey pulses are sufficiently gentle to avoid excitation to higher motional states.  In comparison to free space spectroscopy, this drastically reduces the number of fringes in the spectral pattern as well as any light shifts from the probe.

\subsection{Systematic effects in lattice clocks}\label{section:systematics}

With the obvious advantages in spectroscopic precision of the $^1S_0$-$^3P_0$ transition in an optical lattice, the sensitivity of
the clock transition to external fields and operational conditions becomes a central issue for the lattice clock as an accurate atomic frequency
standard. Here we consider many of the relevant effects influencing the uncertainty to which these standards can be operated.  The relative importance of these effects has and will continue to change, as the optical lattice clocks evolve both in their implementation and their performance levels.  We start by considering an effect central to the lattice clock: the Stark effect from the optical lattice.

\subsubsection{Optical Lattice Stark shifts}

A neutral atom becomes polarized in the presence of the electric field of a laser.  This effect leads to an ac-Stark shift of atomic states and enables the atom to be spatially confined by a laser beam.  For an electric field $E$, the Stark shift is simply $U=-1/2 E_i \alpha_{ij} E_j$, where $\alpha$ is the atomic polarizability tensor of a particular atomic state, for Cartesian dimensions $i$ and $j$.  This polarizability can be written as the sum of three irreducible spherical tensors of rank 0, 1, and 2, yielding the scalar, vector, and tensor polarizabilities (written here as $\alpha^S$, $\alpha^V$, and $\alpha^T$).  We first study each of these three contributions to the lattice Stark shift.

As each species of optical clocks employ $J=0$ `scalar' clock states, the Stark shift from the scalar polarizability $U=- \alpha^s E_0^2/4$ dominates.  While a typical scalar Stark shift may be of order $\delta f_S \simeq 1$ MHz, as described in Sec.~\ref{section:magic}, operation of the optical lattice at the magic wavelength (see Table~\ref{MagicTable}) constrains $\alpha^s$ for each clock state to be equal.  By so doing, this scalar Stark shift of the clock transition is nulled. As discussed in Sec.~\ref{section:magic}, part of the `magic' in a magic wavelength optical lattice is not simply that a zero crossing in the scalar Stark shift exists.  A critical detail is that the cancelation of the scalar Stark shift is fairly insensitive to the precise lattice laser frequency.  For example, by operating within 500 kHz of the magic wavelength, a typical lattice laser intensity in a Sr lattice clock enables cancelation of the scalar Stark shift at the $10^{-18}$ fractional frequency level.

If the clock states had identically zero total angular momentum, then the vector and tensor polarizability would also be zero.  However, for fermionic isotopes, state mixing from hyperfine interaction yields a non-zero vector and tensor polarizability in the excited clock state \cite{Boyd07c}.  The vector light shift is given by
\begin{equation}
 \Delta f_{\text{vector}}= -\alpha^{V}\frac{m_F}{2F}\xi \frac{E^2}{2h},\label{vectorpol}
\end{equation}
where $\xi$ is the degree of ellipticity of the light field. For pure circular (linear) polarization, $\xi$ = $\pm$1 ($\xi$=0).  Here, we have assumed that the lattice light propagation wave-vector is aligned along the atomic quantization axis.  The vector light shift can be viewed as a pseudo-magnetic field
$d\vec{B}$ applied along the light propagation axis, with $|d\vec{B}|$ given by the atomic properties, light polarization
state, and light intensity.  If the quantization axis, typically determined by a bias magnetic field, is not aligned with the light propagation, then the combined effect of the Zeeman shift from the bias B-field and the pseudo-B-field vector light shift is given by the appropriate vector sum of the two.

The vector light shift is nulled for linear polarization of light.  This is readily achieved for a 1-D optical lattice, although care must be taken because of stress-induced birefringence of the vacuum viewports through which the optical lattice passes.  For a 2- or 3-D optical lattice, the electric field in different dimensions can sum to yield unwanted elliptical polarization which can vary site-to-site in the optical lattice.  The magnitude of the vector polarizability has previously been estimated or calculated \cite{Porsev04a,Katori03a}.  Experimentally, an upper limit on the vector polarizability in Sr was determined by analyzing frequency measurements of $\sigma$- and $\pi$- transitions from different $m_F$ states in the presence of a bias magnetic field~\cite{Boyd07c}.  Since then, the vector polarizability has been directly measured, both in Yb~\cite{Lemke09a} and Sr~\cite{Westergaard11a}.  In both cases, circular polarization can lead to significant vector lights ($> 100$ Hz).  In practice, a high degree of linear polarization reduces this effect considerably.  Just as significant, the $m_F$ dependence of the vector Stark shift permits cancelation of the effect by averaged interrogation for equal but opposite $m_F$ magnetic substates.  In this way, the vector Stark shift does no presently contribute in a significant way to the measurement uncertainty of lattice clocks.

The tensor light shift for a given clock state is given by \cite{angel_hyperfine_1968,Ovsiannikov06a,Romalis99a}
\begin{equation}
\Delta f_{\text{tensor}}=-\alpha^{T}\frac{3m_F^2-F(F+1)}{F(2F-1)}\left(\frac{3\cos^2\phi'-1}{2}\right)\frac{E^2}{2h},
\label{tensorpol}
\end{equation}
where $\phi'$ is the angle between the light polarization axis and the quantization axis. As with the vector light shift, the tensor light shift induces a polarization sensitive effect to the lattice clock.  Notably, the geometric term in parentheses changes from 1 to -1/2 as $\phi'$ is varied from 0 to $\pi/2$. Unlike the vector Stark shift, the $m_F^2$ dependence of the tensor Stark shift precludes trivial cancelation of the effect through averaging of transitions from opposite signed magnetic substates.  Fortunately, the tensor polarizability is small.  In the case of $^{171}$Yb, the insufficient angular momentum ($F=1/2$) dictates that the tensor polarizability is zero \cite{angel_hyperfine_1968}.  It has been measured in the case of $^{87}$Sr \cite{Westergaard11a}.  There it was shown that the tensor shift, under some conditions, could be as large as the $10^{-16}$ level, but could be straightforwardly controlled to much better than the $10^{-17}$ level.

The scalar, vector, and tensor Stark shifts discussed above all scale with $E^2$, first order in the optical lattice intensity.  Another critical systematic stems from the hyperpolarizability, $\gamma$, contributing a shift which scales as $E^4$.  The atomic hyperpolarizability includes both one- and two-photon resonances \cite{Ovsiannikov06a}, and the differential hyperpolarizability between the clock states remains non-zero at the magic wavelength.  The primary contributions to the hyperpolarizability stem from two-photon resonances connecting to the $^3P_0$ state in the neighborhood of the magic wavelength, for both Sr \cite{Brusch06a} and Yb \cite{Porsev04a,Barber08a}.  In both cases, the differential hyperpolarizability leads to a Stark shift around the magic wavelength of approximately 0.5 $\mu$Hz $(U_0/E_r)^2$, where $U_0$ gives the lattice depth in units of photon recoil.  At $U_0 = 100 E_r$, this leads to a magnitude for the shift of $10^{-17}$, but the uncertainty of the shift has been determined to a fraction of the shift~\cite{Barber08a,Westergaard11a}.  In the case of Yb, warmer atomic temperatures~\cite{Lemke09a,Poli08a} (originating from the second stage of Doppler cooling) have historically required deeper lattice depths and thus have required somewhat more care in dealing with the hyperpolarizability shift.  This situation can be mitigated by optimized or additional cooling, perhaps utilizing quenched sideband cooling on the clock transition itself.  As with the non-scalar Stark shifts described above, in general the hyperpolarizability shift exhibits lattice polarization dependence.  This is additional motivation for good lattice polarization control, but has also led to a proposal to cancel the hyperpolarizability shift altogether~\cite{Taichenachev06a}.

The lattice Stark shifts considered above are due to electric-dipole allowed (E1) couplings.  Higher multipole couplings, via notably magnetic dipole (M1) and electric quadrupole (E2), can also lead to lattice Stark shifts.  Since these M1/E2 couplings are much weaker than their E1 counterparts, the resulting Stark effects are much smaller.  Nevertheless, they cannot be ignored when considering the smallest possible uncertainty for these lattice clocks.  The authors of~\cite{Taichenachev08a} discussed a subtle M1/E2 effect, stemming from quantized atomic motion in the optical lattice.  For the red-detuned lattice, atoms are trapped in the anti-nodes of the electric field of the optical potential.  The optical potential varies along its axis as $\cos^2 x \simeq 1 - x^2$, leading to two different sources of Stark shift.  The first is an E1 Stark shift common to all atoms and proportional to optical lattice intensity, $I$.  The second, given by the harmonic confinement of the atom, dictates an additional shift given by the particular motional state populated by the atom, and proportional to the lattice trap frequency, which scales as $\sqrt{I}$.  With only E1 couplings, at the magic wavelength the total Stark shift is equal for both clock states, resulting in the expected zero differential shift for the clock frequency.  However, the effect of M1/E2 couplings is to modify the second shift scaling as $\sqrt{I}$.  In general, the E1 and M1/E2 Stark shifts cannot be simultaneously canceled for the two clock states, frustrating the existence of a perfectly magic wavelength.  The residual shift is $ \Delta f_{M1/E2}\propto (n+1/2)\sqrt{I}$ where $n$ is the motional quantum number of the atom.  While the expectation is that weak M1/E2 couplings would keep this effect small, the authors of \cite{Taichenachev08a} made an alarming theoretical estimate that the effect could be as large as $10^{-16}$.  This effect was directly probed by~\cite{Westergaard11a} in a Sr lattice clock, by searching for a Stark shift with the appropriate $\sqrt{I}$ dependence.  Fortunately, no dependence was observed, constraining this effect to be below $10^{-17}$ for a lattice depth of $100 E_r$. Recent work characterizing the lattice Stark shifts for the JILA Sr clock has demonstrated that statistical analysis of extensive experimental data supports a purely linear model for the dependence of shift on intensity \cite{Bloom2014}.

\subsubsection{Zeeman shifts}

\begin{figure}[t!]
\centering
\includegraphics[width=\textwidth]{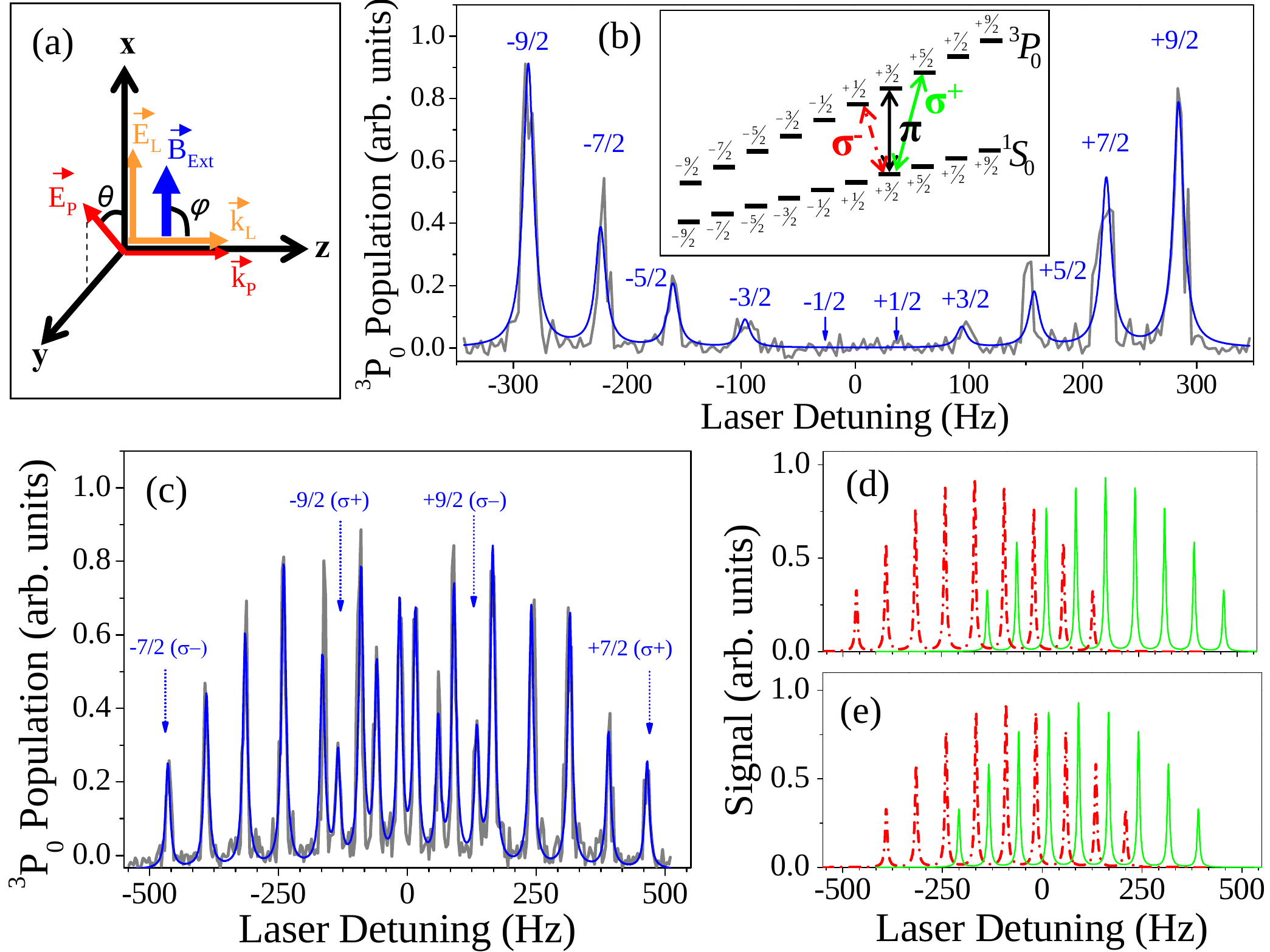}
\caption[Spectroscopy of resolved $\pi$ transitions]{\label{pifig}
(a) Relevant experimental field orientation for lattice
spectroscopy. The lattice laser propagates along the $z$-axis and is
linearly polarized along the $x$-axis, parallel to the bias magnetic
field such that $\varphi\approx\pi/2$.  The probe laser propagates
co-linearly with the lattice beam and the linear probe polarization
can be rotated relative to the quantization ($x$) axis by an angle
$\theta$. (b inset)  The large nuclear spin ($I=9/2$ for $^{87}$Sr) results in 28
total transitions, and the labels $\pi$, $\sigma^+$, and $\sigma^-$
represent transitions where $m_F$ changes by 0, +1, and $-$1
respectively.  The HFI state mixing modifies the $^3P_0$ $g$-factor,
making the magnitude about 60\% larger than that of $^1S_0$. (b)
Observation of the $^1S_0$-$^3P_0$ $\pi$-transitions when $\theta=0$
in the presence of a 58 $\mu$T magnetic field.  (c) Observation of the
18 $\sigma$ transitions when the probe laser polarization is
orthogonal to that of the lattice ($\theta=\frac{\pi}{2}$) when a
field of 69 $\mu$T is used.  In (b) and (c), data is shown in grey
and fits are shown as solid lines in blue. The peaks are labeled by the ground
state sublevel of the transition (and the relevant polarization in
(c)).  The relative transition amplitudes for the different
sublevels are strongly influenced by the Clebsch-Gordan
coefficients.  Here, Fourier-limited transition linewidths of 10 Hz
are used. (d and e) Calculations of the 18 $\sigma$-transition
frequencies in the presence of a 69 $\mu$T bias field, including the
influence of Clebsch-Gordan coefficients. The solid and dashed-dot
curves show the $\sigma^+$ and $\sigma^-$ transitions, respectively.
(d) Spectral pattern for $g$-factors {\small$g_I\mu_B/h=-1.85$ Hz/$\mu$T}
and {\small$\delta g\mu_B/h=-1.09$ Hz/$\mu$T}. (e) Same pattern as in (d)
but with {\small$\delta g\mu_B/h=+1.09$ Hz/$\mu$T}.  The qualitative
difference in the relative positions of the transitions allows
an absolute determination of the sign of $\delta g$ compared to that of $g_I$.}
\end{figure}

The sensitivity of a clock transition to magnetic fields has played a prominent role in nearly all types of atomic frequency standards.  In the case of the optical lattice clock, both first- and second-order Zeeman shifts can be relevant.  The nuclear spin, $I$, of the fermionic lattice clocks provides $2I+1$ magnetic sublevels for each $J=0$ clock state. A magnetic field $B$ gives a linear shift of the sublevels, which for $\pi$-transitions ($\Delta m_F$ = 0) shifts the clock transition frequency by
\begin{equation}
\Delta f_{\text{B1}}= - m_F \delta g \mu_B B/h \label{pidg}
\end{equation}
where $\mu_B/h$ $\cong$ 14 kHz/$\mu$T, and $\delta g$ is the difference in the $g$-factors of the $^3P_0$ and $^1S_0$ states. The ground state $g$ factor is determined by the nuclear $g$-factor, {\small$g_I=\frac{\mu_I(1-\sigma_d)}{\mu_B |I|}$} where $\mu_I$ is the nuclear magnetic moment, and $\sigma_d$ is the diamagnetic correction.  For $^{87}$Sr, $\mu_I=-1.0924(7)\mu_N$~\cite{Olshewski72a} and $\sigma_d=0.00345$~\cite{Kopfermann58a}, yielding a small Zeeman sensitivity of $g_I \mu_B/h=-1.850(1) $Hz/$\mu$T for the ground state. Lacking nuclear spin-induced state mixing, the $^3P_0$ $g$-factor would be essentially identical to the $^1S_0$ $g$-factor, such that $\delta g=0$.  Such is the case for bosonic isotopes.  However, since the hyperfine interaction modifies the $^3P_0$ wave function, a differential $g$-factor is introduced between the two states~\cite{Boyd06a}. This can be interpreted as a paramagnetic shift arising from the distortion of the electronic orbitals in the triplet state, and hence the magnetic moment \cite{Lahaye75a,becker_high-resolution_2001}.  If the state mixing in the system is known, then $\delta g$ is given by
\begin{equation}
\delta g =
-\left(\tilde{\alpha}_0\tilde{\alpha}-\tilde{\beta}_0\tilde{\beta}\right)\sqrt{\frac{8}{3I(I+1)}}.
\label{dgeq}
\end{equation}
Here $\tilde{\alpha}_0$, $\tilde{\beta}_0$, $\tilde{\alpha}$, and $\tilde{\beta}$ are state-mixing coefficients resulting from the
hyperfine and spin orbit interactions \cite{Boyd07c}.  The mixing increases the magnitude of the $^3P_0$ $g$-factor by $\sim$ 60\%.
The resulting first-order Zeeman sensitivity (shown schematically in Fig.~\ref{pifig}(b) inset) is an important systematic effect for the development of lattice clocks, as stray magnetic fields can deteriorate the spectroscopic accuracy of the system.

As seen in Eq.~\ref{pidg}, a $\pi$-transition ($\delta m_F = 0$) is only sensitive to $\delta g$, not $g_I$ which is common to both electronic states.  On the other hand, a $\sigma$-transition ($\delta m_F = \pm 1$) is sensitive to both $g_I$ and $\delta g$.  Measurement of the frequency splittings for both $\pi$- and $\sigma$- transitions can be used together to determine the value of $\delta g$.  The added value of the $\sigma$- transition measurements is that, since $g_I$ is already known well for the lattice clock species, the measured splittings can be used to self-calibrate the value of the B-field.  An example of this type of measurement is shown in Fig.~\ref{pifig}, for the case of $^{87}$Sr.  Here it can be seen that the hyperfine interaction \emph{increases} the magnitude of the $^3P_0$ $g$-factor (i.e.~$\delta g$ has the same sign as $g_I$).  Using data like this, $\delta g \mu_B /h $ has been determined experimentally to be -1.084(4) Hz/$\mu$T  \cite{Boyd07c}.  Similar measurements have been conducted for $^{171}$Yb, yielding a $\delta g \mu_B /h = -1.91(7)$ Hz/$\mu$T \cite{Lemke12a}.

The second order Zeeman shift must also be considered for high-accuracy clock operation. The two clock states are both $J=0$ so the shift arises from levels separated in energy by the fine-structure splitting, as opposed to the more traditional case of alkali(-like) atoms where the second order shift arises from nearby hyperfine levels.  As a result, the fractional frequency shift from second order Zeeman effect for optical lattice clock species is significantly smaller than that of clock transitions present in alkali(-like) atoms and ions.  The clock shift is dominated by the interaction of the $^3P_0$ and $^3P_1$ states since the ground state is separated from all other energy levels by optical frequencies. Therefore, the total shift can be approximated by the repulsion of the two triplet states (which are separated in energy by $h\Delta\nu_{10}$) as
\begin{equation}
  \Delta f_\mathrm{B2}^{(2)}\cong  -\frac{2\mu_B^2}{3(\Delta\nu_{10})h^2}B^2.
\label{secondorder}
\end{equation}
From Eq.~\ref{secondorder} the resulting second order Zeeman shift for Sr is {\small$\Delta f_\mathrm{B2}^{(2)}\cong-2.33 \times 10^{-5} B^2 $ Hz/ $\mu$T$^2$} \cite{Taichenachev06a,Baillard07a,Boyd07c,Ludlow08b}, and {\small$\Delta f_\mathrm{B2}^{(2)}\cong-6.2 \times 10^{-6} B^2 $ Hz/ $\mu$T$^2$} for Yb \cite{Taichenachev06a,Poli08a,Lemke09a}.

\subsubsection{Stark shift from Blackbody Radiation}

We now consider the Stark shift arising from blackbody radiation (BBR) bathing the lattice-trapped atoms. Because room-temperature BBR lies at frequencies below the detunings of intermediate states that contribute to the electric dipole polarizability of the clock states, as described in Section \ref{sec:systematics}, the BBR Stark shift can be written as:
\begin{equation}
h\Delta f_\mathrm{BBR}=-\frac{\Delta\alpha_s  \langle  E^2(T)\rangle}{2}(1+\eta(T^2)),
\end{equation}
Precise determination of the shift then requires accurate knowledge of the differential static polarizability, $\Delta\alpha_s$, the dynamic correction factor, $\eta$, and the BBR field given by the radiative temperature bathing the atoms, $T$.  The static polarizability can be computed semi-empirically, from known values of the relevant E1 transition frequencies, matrix elements, and lifetimes.  However, it is difficult to properly include contributions from intermediate states with poorly known matrix elements, including high lying states and the continuum.  In this regard, sophisticated ab-initio calculations have been implemented to compute these atomic properties with higher accuracy (e.\ g.\ \cite{porsev_multipolar_2006,Dzuba10a}).  At room temperature, the BBR shift for Sr and Yb is $5.5 \times 10^{-15}$ and $2.7 \times 10^{-15}$, respectively.  The BBR shift for Hg, like for its Group IIb counterparts Zn and Cd, is smaller by approximately one order of magnitude.  For both Sr and Yb, the room-temperature BBR shift represents the largest uncanceled systematic shift of the clock transition frequency.  Furthermore, the uncertainty in the BBR shift has previously limited the overall uncertainty of these clocks for several years at the $10^{-16}$ level~\cite{Ludlow08b,Lemke09a}.

A significant source of uncertainty originated from calculation of the polarizability of the clock states. These calculations approached the $1\%$ level for Sr \cite{porsev_multipolar_2006} and, complicated by the large number of electrons and core-excited states, at the $10\%$ level for Yb \cite{Dzuba10a,porsev_multipolar_2006}.  Improved measurements of the dipole matrix elements to low-lying intermediate states could provide useful constraints on the polarizability calculations, as could precise knowledge of the magic wavelengths \cite{Porsev08a}.  But more directly, the static polarizability can be measured  via the Stark shift under application of known static \cite{Simon98a} or even long-wavelength \cite{rosenband_blackbody_2006} electric field.  In the case of Yb, the differential static polarizability for the clock transition has recently been measured at the 20 ppm level using a static electric field \cite{Sherman12a}.  A high-accuracy measurement of the Sr differential static polarizability has also been performed~\cite{Middelmann12a}.  Furthermore, recent ab-initio calculations of the ytterbium and strontium polarizability using a coupled-cluster all-order approach have reduced the uncertainties in theoretical calculations and demonstrate very close agreement to the more precise experimental measurements \cite{SafronovaPorsev12,SafronovaPorsev13}.

The second piece contributing to the BBR shift uncertainty comes from the dynamic correction, $\eta$.  This correction term is computed for each clock state as a sum over intermediate states, and is most significant for the $^3P_0$ clock state. For both Yb and Sr, this sum is dominated by the lowest-lying coupled-state, $^3D_1$.  A recent measurement of this dipole matrix element allowed determination of $\eta$ in Yb at the $<3 \%$ level \cite{Beloy12a} and in Sr at the $<1 \%$ level \cite{Nicholson2014}. Calculations based on other atomic properties like the static polarizability and magic wavelength have been used to determine $\eta$ in Yb \cite{Beloy12a} and Sr \cite{Middelmann12a} at the few percent level.  Additionally, recent ab-initio calculations have provided improved theoretical values of $\eta$ for both Yb and Sr \cite{SafronovaPorsev12,SafronovaPorsev13}.

The third source contributing to the overall BBR shift uncertainty is knowledge of the BBR environment bathing the atoms.  This thermal radiation field is complicated by temperature inhomogeneities of the vacuum system enclosing the lattice-trapped atoms, by optical and infrared transparency of viewports typically used on the vacuum system, and by the complex geometries and non-unit emissivities of the vacuum apparatus.  At 1 K uncertainty in the BBR field, the room temperature BBR shift has a frequency uncertainty of 3.5 and 7 $\times 10^{-17}$ for Yb and Sr.  A cryogenically-cooled environment benefits from the strong $T^4$ dependence of the BBR shift, and can realize uncertainties at the $10^{-18}$ level \cite{Ushijima2014,Middelmann11a}. Alternatively, room temperature solutions also exist.  For example, efforts to maintain temperature uniformity of the vacuum enclosure around the lattice-trapped atoms has lead to a BBR uncertainty at the $10^{-17}$ level \cite{FalkeLemke13a}.  More recently,  calibrated in-situ thermal probes were used to monitor the radiative thermal environment illuminating the atoms, leading to a BBR shift uncertainty at the $<2 \times 10^{-18}$ level \cite{Nicholson2014,Bloom2014}.  Additionally, a room-temperature radiative thermal shield has been used in a Yb lattice clock to reduce BBR uncertainty to $1 \times 10^{-18}$ \cite{Beloy2014}.

\subsubsection{Cold Collision Shift}

Large ensembles of ultracold atoms offer atomic clocks a measurement of the atomic state with very high signal-to-noise ratio, allowing time and frequency measurements with unprecedented levels of precision and speed.  However, large atom density can give rise to significant atomic interactions.  These interactions can perturb the clock transition frequency, compromising the accuracy of the atomic frequency standard.  Density-dependent collisional shifts play an important role in the operation of the highest accuracy Cs fountain standards (e.\ g.\ \cite{Wynands05a,DosSantos02a,Szymaniec07a,Gibble93a,Leo01a}). In fact, the reduced collisional interaction in Rb fountain standards was a key motivation for their development \cite{Kokkelmans97a,Sortais00a}.  For the optical lattice clock, the density related frequency shift is the only source of error that plays a competing role between the clock stability and accuracy. Keeping this under control thus has a critical consequence.

In a 3-D optical lattice clock, the large number of lattice sites lead to an atom filling-factor of less than unity.  In the absence of tunneling, it is expected that atomic interactions can thus be minimized \cite{Katori03a}. As the clock accuracy continues improving, eventually one would need to consider the long-range dipolar interaction effects in a 3D lattice~\cite{Chang04a}.  For a 1-D optical lattice, the two dimensional lattice sites typically have multiple occupancy and the use of fermionic isotopes at ultracold temperatures seems advantageous.  Anti-symmetrization of the wavefunctions for identical fermions eliminates collisions from even partial-wave collision channels, including the lowest order $s$-wave.  At the same time, the lowest odd-wave ($p$-wave) collisions can be suppressed at sufficiently low temperatures.  This fermionic resistance to cold collisions (and collision shifts) makes fermions particularly good candidates for atomic frequency standards (e.g.\ \cite{Gibble95a}). As an example, suppression of collision shifts for a radio frequency transition has been experimentally observed in ultracold fermionic lithium atoms~\cite{Zwierlein03a,Gupta03a}.

Nevertheless, collision shifts were observed in fermionic optical lattice clocks, first in $^{87}$Sr \cite{Ludlow08b,Campbell09a} and later in $^{171}$Yb \cite{Lemke09a}.  The breakdown in collision suppression was considered to be likely due to one of two mechanisms.  One was that finite atomic temperatures prevented the $p$-wave collision channel from being completely suppressed.  The second was that, even though the fermionic atoms were prepared in identical quantum states, during spectroscopy the atoms evolved into non-identical superpositions of the clock states, becoming distinguishable and able to interact via the $s$-wave collision channel~\cite{Hazlett2013}.  Inhomogeneous evolution of the population is a residual Doppler effect due to weak atomic confinement in dimensions orthogonal to the lattice axis \cite{wineland_laser_1979,Campbell09a,Blatt09}.

A simple estimate can be made for the relative size of $s$- and $p$-wave collision shifts~\cite{Campbell09a,Lemke11a}.  The $p$-wave collision shift scales with $b^3k^2$, where $b^3$ is the $p$-wave scattering volume, and $k$ is the deBroglie wavenumber.  Conversely, the $s$-wave collision shift scales with the scattering length, $a$.  The ratio $b^3k^2/a$ estimates the relative contributions of $p$-wave to $s$-wave collisional shifts.  However, in the case where the atoms are largely indistinguishable, the $s$-wave shift is further suppressed by the degree of indistinguishability \cite{Campbell09a,Gibble09a,Rey09a,Lemke11a}.  As a result, either $s$- or $p$-wave interactions have the potential to contribute to cold collision shifts, depending on the experimental details and the case-specific values of $a$ and $b$.

Following observations of Hz-level cold collision shifts in Sr and Yb, a number of efforts explored these effects experimentally \cite{Campbell09a,Swallows11a,Lemke11a, Bishof11a, Ludlow11a,Nicholson12} and theoretically \cite{Gibble09a,Rey09a,Yu10a,Band06a}. In the case of $^{171}$Yb, it was found that the dominant interaction responsible for the cold collision shift was a $p$-wave one between ground state ($^1S_0$) and excited state ($^3P_0$) atoms \cite{Lemke11a}.  While the very existence of a cold collision shift serves as a potential stumbling block to reaching clock accuracy at the highest levels, it has been shown that the responsible interactions can be manipulated to realize cancelation~\cite{Ludlow11a} or suppression~\cite{Swallows11a} of the cold collision shift.  Together with the strategy of confining the atoms at lower number densities per lattice site \cite{Brusch06a,LeTargatLorini13}, the uncertainty of the collision shift for the lattice clock can be controlled below $10^{-18}$ \cite{Nicholson12,Bloom2014}.

\subsubsection{Stark shift from interrogation laser}

While the two clock states have identical polarizabilities at the magic wavelength, their polarizabilities differ at the actual clock transition frequency.  Off-resonant couplings to intermediate states other than the clock states, driven by the interrogation laser, introduce a dynamic Stark shift on the clock transition.  The resulting shift depends on the differential polarizability for the clock states at the clock transition frequency, as well as the interrogation laser intensity needed to drive the transition.  The required laser intensity must be sufficiently high to drive the transition for atoms confined in the optical lattice (i.e.\ $\langle n|e^{i\vec{k}\cdot\vec{x}}|n\rangle$ \cite{wineland_laser_1979}).  This Stark shift is present at the $10^{-17}$ level \cite{Ludlow08b,Lemke09a,Kohno09a,Falke11a}, and recent measurements have placed the uncertainty at the level of 10$^{-18}$~\cite{Bloom2014}. As laser coherence times continue to increase, the required laser intensity will be reduced, resulting in smaller Stark shifts.  Furthermore, techniques have been proposed to further reduce the sensitivity of the clock transition to the interrogation laser intensity \cite{yudin_hyper-ramsey_2010,Taichenachev10a}.

\subsubsection{Doppler effects}

A primary motivation for tightly confining the atoms in the optical lattice is to perform spectroscopy on the clock transition without the Doppler and recoil frequency shifts.  However, there are a number of effects that can introduce residual motional sensitivity.  One such effect is quantum tunneling between sites of the optical lattice, along the axis of interrogation.  This effect has been considered for a 1-D optical lattice \cite{Lemonde05a}, and is notably relevant for shallow lattices.  By aligning the lattice axis along gravity, gravity-induced non-degeneracy between lattice sites can further suppress tunneling.  In this case, it has been estimated that for even modest trap depths, tunneling related motional effects can be straightforwardly kept below the $10^{-17}$ level \cite{Lemonde05a}.

Relative vibration between the lattice field and the clock laser can also lead to residual Doppler shifts.  Any such motion that is synchronized to the experimental cycle time is notably problematic, as it does not average away statistically and leads to a systematic shift.  Such effects are a concern for other types of atomic frequency standards \cite{Wilpers07a,rosenband_frequency_2008}.  The problem is best minimized in an optical lattice clock by maintaining a passively-quiet opto-mechanical environment. The phase of the lattice and/or clock laser can also be actively stabilized~\cite{Ma94a}, so that no residual vibration occurs in the atomic reference frame.  Detection and cancelation of such residual Doppler shifts can be made by spectroscopically probing in two, counter-propagating directions.

One particularly pernicious residual Doppler effect is associated with switching rf power in an acousto-optic modulator.  The rf power is typically pulsed to switch on and off the interrogation of the clock transition, and is known to induce phase chirps of the clock laser from both rf ringing and thermal effects \cite{degenhardt_influence_2005}.  These effects must be carefully characterized and controlled, or can be compensated with active stabilization~\cite{Swallows12}.

The $2^{nd}$-order Doppler shift accounts for relativistic time dilation.  It is simply $\Delta \omega = \frac{1}{2} \beta^2 \omega_L$, where $\beta=v/c$ is proportional to the atomic velocity and $\omega_L$ is the laser (angular) frequency.  The velocity of the ultracold lattice-trapped atoms is characteristically given by the sample temperature.  For Sr, a temperature of $T=2.5$ $\mu$K corresponds to a velocity of $\sim 1.5$ cm/s.  This results in a $2^{nd}$-order Doppler shift below 1 $\mu$Hz ($< 10^{-20}$).  Shaking of the trap results in even smaller atomic velocities, with negligible $2^{nd}$-order Doppler contributions.

Finally, in the well-resolved sideband limit, the atomic motion occurs at a modulation frequency far-removed from the carrier.  Even at relatively low modulation frequencies corresponding to weak confinement in the transverse axes of a 1-D lattice, the effect of motional line-pulling of the transition frequency is negligible due to the negligible amplitude of the motional sidebands.

\subsubsection{DC Stark shifts}

Similar to the Stark shifts on the clock transition frequency produced by the blackbody and probe laser fields, static electric fields will induce static Stark shifts.  Optical lattice clocks benefit from the fact that the atomic sample is trapped in optical potentials, which are usually far-removed from physical surfaces of a vacuum chamber where stray charges may accumulate.  Metallic components of the vacuum system used in optical lattice clocks are electrically grounded, acting as a Faraday cage for the atoms.  However, charge can potentially accumulate on insulator surfaces, such as glass optical viewports with or without dielectric optical coatings.  From the typical geometry of lattice clock enclosures and the ability for stray charges to dissipate to ground, it has been estimated that that stray static electric fields will cause static Stark shifts below the $10^{-17}$ level.  However, in one case, it was shown that a stray charge buildup on an in-vacuum mirror can lead to a very large Stark shift at the $10^{-13}$ level \cite{Lodewyck11a}.  Here, electrical discharge was limited by very high electrical resistive paths to ground, with discharge times of hundreds of days.  Any such effects must be properly avoided or controlled, especially as lattice clocks are pushed to the $10^{-17}$ performance and better. The DC Stark shift can be directly measured \cite{Lodewyck11a}, and recent measurements have pushed this shift uncertainty to 2.1 $\times$ 10$^{-18}$ in Sr \cite{Bloom2014}.

\subsubsection{Other effects}

A number of other systematic effects have been considered for the optical lattice clock.  Among these are line pulling, servo error,
stray laser Stark shift, AC Zeeman shift, and others (e.g.\ \cite{Takamoto05a,Ludlow08b,Falke11a,Lemke09a,Baillard07a,Bloom2014}).  Fortunately, these effects are often small, and do not represent a fundamental limitation to lattice clocks today.

\subsection{Optical lattice clocks based on Fermions or Bosons}

In previous sections we have considered the electronic structure that makes alkaline earth (-like) atoms so attractive as optical frequency standards.  The two clock states ($^1S_0$ and $^3P_0$) have very weak coupling to each other, stemming from the forbidden dipole transition between these spin states.  From a practical standpoint, only the fermionic isotopes have a useful level of coupling, originating from the non-zero nuclear spin and the resulting hyperfine mixing in the $^3P_0$ state.  The bosonic isotopes, with no nuclear spin, lack this state mixing.  Yet even in the bosonic isotopes, the clock states themselves possess many ideal properties for an optical lattice clock.  By artificially inducing dipole coupling between these states, their utility can be realized in a clock, as with fermionic isotopes.  The first proposals to drive a weak transition between these states in a bosonic optical lattice clock exploited the rich dynamics of multi-level systems~\cite{Santra05a,Hong05a}.  Utilizing coherent population trapping (CPT) \cite{Arimondo96a}, as done in electromagnetically induced transparency, these schemes proposed two~\cite{Santra05a} or three~\cite{Hong05a} laser fields to resonantly drive population between the clock states.  The effect shares some characteristics with the CPT approach to Cs clocks, where in that case the clock states are separated by microwave frequencies but driven by two coherent optical fields \cite{Vanier05a}.  For the optical clock proposals, the strength and detuning of the laser fields can be chosen to yield a transition linewidth at the Hz to mHz level, or less.  The obvious tradeoff in these proposals is controlling the AC Stark shifts induced by the laser fields on the clock states.  Such control looked possible to reach accuracies of $10^{-17}$ level, and particularly for proposals involving pulsed, Ramsey-like interrogation fields~\cite{Zanon06a}.  However, careful control of multiple interrogation laser fields adds further experimental complexity.

Rather than using multiple laser fields to drive the clock resonance, another approach \cite{Taichenachev06a}, referred to as magnetic-induced spectroscopy, proposed using one laser field nearly resonant with the clock level energy spacing together with a bias DC magnetic field.  The DC magnetic field induces state mixing of $^1P_1$ and the upper clock state $^3P_0$, much like what the nuclear spin field does in the case of fermions.  The optical field then probes the weakly-allowed transition.  Here again, the strengths of the magnetic and laser field can be varied to set the resonance linewidth.  The effect was experimentally demonstrated shortly after the original conception \cite{Barber06a}.  A narrow 20 Hz wide resonance in the ground state was seen employing a bias magnetic field of about 1 mT.  The ability to avoid using extra lasers as in the CPT schemes makes this implementation more straightforward.  However, the presence of a large bias magnetic field and a strong laser drive requires careful control of the second order Zeeman shift and the Stark shift. Because the bosonic isotopes have $J=F=0$ for both clock states, the first order Zeeman shifts are zero. The potential of magnetic induced spectroscopy was demonstrated in several frequency evaluations of the bosonic isotopes of Yb and Sr, where total frequency uncertainties were controlled at the 30 Hz \cite{Baillard07a}, 1 Hz \cite{Akatsuka08a,Akatsuka10a}, and sub-Hz \cite{Poli08a} levels.

Another proposal for exciting the clock transition in bosonic isotopes utilized the lattice field itself to couple the clock states \cite{Ovsiannikov07a}.  More specifically, in addition to the lattice standing wave, an additional running wave introduced to induce state-mixing.  The state-mixing wave is a relatively intense, circularly polarized field which induces mixing between $^3P_1$ and $^3P_0$, making the dipole transition from $^1S_0$ possible.  This approach enjoys the convenience of using a single laser field at the magic wavelength to induce state mixing.  A major drawback of this approach is the high optical intensities required to create sufficient mixing.  At these high field amplitudes, higher order light shifts become important~\cite{Ovsiannikov07a}.

While all of the proposals discussed in this section have some differences, they share several basic features. Notably, their shared goal is to enable interrogation of the naturally forbidden clock transition in bosonic alkaline earth atoms.  There are several reasons why interrogation of the bosonic isotopes might be attractive.  From a practical perspective, the bosons often have higher isotopic abundance.  Combined with the simpler cooling scheme of bosons (e.g.\ \cite{Mukaiyama03a,Loftus04b}), it is usually easier experimentally to get a large sample of ultracold bosonic alkaline earth atoms compared to the fermionic case.  Perhaps more significantly, lacking nuclear spin the bare $^1S_0$ and $^3P_0$ bosonic clock states have no angular momenta, and intermediate states that are important for the lattice Stark shifts have no hyperfine structure \cite{Porsev04a}.  As a result, there is no polarization dependent light shifts on the clock transition, as the vector and tensor terms of the polarizability are basically zero.  Consequently, control of the lattice-induced Stark shifts are simplified.  The spinless clock states also have no Zeeman substructure, which means that there is no first order Zeeman sensitivity.  Furthermore, no substructure means that no optical pumping for state preparation is required, as often employed in fermionic isotopes.  The lack of substructure also gives the simplest possible absorption spectrum.

The primary disadvantage to probing the bosonic isotope is that, in all cases, at least one extra field is required to induce the clock transition.  More than introducing experimental complexity, it requires careful control of these fields and their respective field shifts to achieve high clock accuracy.  Techniques have been proposed to reduce the sensitivity of these fields on the resulting shifts \cite{Zanon06a,Taichenachev10a,yudin_hyper-ramsey_2010}, however, whether the bosonic species can compete with their fermionic counterpart in the clock accuracy is still unresolved.  While the boson's lack of first order Zeeman sensitivity is usually heralded as an advantage, it is in some ways a drawback.  In the fermionic case, this first-order sensitivity is easily canceled by averaging the equal but opposite first-order Zeeman shift for transitions from opposite spin states, $\pm m_F$.  At the same time, measurement of the transition splitting can be combined with precisely determined g-factors to directly read off the magnetic field magnitude in real time. This makes evaluation of the second-order Zeeman shift straightforward, without any additional measurement.  Finally, as a general rule, bosonic isotopes are expected to have larger collisional effects on the clock transition than their fermionic counterparts, due to the inability for two identical fermions to scatter with a $s$-wave interaction \cite{Campbell09a,Rey09a,Gibble09a}.  However, the cold collision physics of these different quantum particles is rich, and should be studied in detail for both bosons and fermions.

With these ideas in mind, the future role of bosonic isotopes in lattice clocks remains open.  Both fermionic and bosonic based lattice clocks have been developed, although fermionic systems are more commonly employed.  As laser coherence grows and enables longer probing times of the clock transition, the size of the extra field shifts in the boson case will shrink, making them more manageable.  As multi-dimensional lattice confinement effects are characterized more fully, the spin-free bosonic isotopes might offer simplicity.  The higher isotopic abundance and laser cooling simplicity of bosons may offer S/N benefits to improve clock stability. As both types of systems are refined, the pros and cons of each will become more pronounced.  In the meantime, both offer promise and together they provide greater variety in exploring optimal clock systems.  This variety will perhaps prove even more useful for exploring other interesting physics, including ultracold collisions, quantum degeneracy, many-body physics, and strongly coupled systems.

\subsection{Lattice clock performance}

At its core, the idea of trapping many quantum absorbers in an optical lattice is to realize an optical frequency standard with both very high stability and very low uncertainty.  Here we discuss both of these figures of merit, highlighting the performance that lattice clocks have so far demonstrated.  We turn our attention first to the frequency stability and then to the systematic evaluations of these systems, which provide insight into their potential accuracy in time and frequency measurement.  Finally, we discuss measurements of the absolute frequency of the clock transitions made by referencing to the caesium primary standard.

\subsubsection{Clock Stability}

In its simplest form, the fractional instability of an atomic frequency standard at averaging time $\tau$ can be written as:
\begin{equation}
    \label{eqn:stability1b}
    \sigma _y \left(\tau\right) = \frac{\delta f}{f_0}
    \frac{\eta}{S/N \sqrt{\tau}}.
\end{equation}
Here, we have assumed only that the frequency noise process dominating the instability is white.  $\eta$ is a parameter of order unity that depends
on the details of the spectroscopic lineshape.  We discussed in Sec.~\ref{highres} that the lattice clock can resolve very narrow spectral features, achieving a very small ratio $\frac{\delta f}{f_0}$.  This is the primary strength of optical frequency standards.  The quantity $S/N$ represents the signal-to-noise ratio at one second of measurement.  As $S/N$ increases, the resolution afforded by the narrow line $\frac{\delta f}{f}$ can be further enhanced.  A number of different
noise processes can play a role in limiting the achievable instability.  To highlight several of the most relevant, we can write the fractional instability as \cite{Lemonde00a}:
\begin{equation}
    \label{eqn:stability2}
    \sigma _y \left(\tau\right) = \frac{1}{\pi Q}\sqrt{\frac{T_c}{\tau}}
    \left(\frac{1}{N}+\frac{1}{N n_{ph}}+\frac{2
    \sigma^2_N}{N^2}+\gamma \right)^{\frac{1}{2}}
\end{equation}
Each term in parentheses gives the $S/N$ for different noise processes.  Here, $T_c$ is the experimental cycle time (of which a useful fraction is spent interrogating the clock transition), $N$ is the atom number, $n_{ph}$ is the number of signal photons detected for each atom, $\sigma_N$ is the uncorrelated rms (root mean squared) fluctuation of the atom number, $\gamma$ accounts for the frequency noise of the probe laser, and two pulse Ramsey spectroscopy is assumed. The first noise term in Eq.~\ref{eqn:stability2} is the most fundamental limit to instability, the quantum projection noise (see Section \ref{atomicnoise}).  The lattice clock is typically operated with $N=10^3$-$10^5$, meaning that $S/N_{QPN}$ is on the order of 100.  In terms of potential clock stability, it is this factor that sets the optical lattice clock apart from the trapped-single-ion standard.  The combination of narrow atomic resonances and measurement $S/N$ at this level give lattice clocks the potential to realize $10^{-17}$ fractional frequency instability or better in just one second.  QPN is a fundamental stability limitation, setting the standard quantum limit of measurement.  However, spin squeezing of the atomic sample, which trades fluctuations between atomic number and phase, can be utilized to improve upon QPN and beat the standard quantum limit (e.g.~\cite{Leroux10a,meyer_experimental_2001,appel_mesoscopic_2009,Gross10a,Riedel10a} and references therein).  In principle, such strategies could make measurements at the Heisenberg limit, with $S/N$ scaling as $1/N$.

The second noise term in Eq.~\ref{eqn:stability2} is the photon shot noise for the atomic state readout.  The long lifetime of $^3P_0$ and the very strong laser cooling transition $^1S_0$-$^1P_1$ from the ground state facilitate convenient implementation of shelving detection.  After atoms have been excited to $^3P_0$ on the clock transition, light resonant with $^1S_0$-$^1P_1$ illuminates the atom.  This transition can be driven many times, and the fluorescence collected, in order to measure the number of atoms remaining in $^1S_0$.  As a result, many photons can be collected per atom, so that the photon shot noise is typically much below the QPN (also termed atom shot noise).

The third noise term in Eq.~\ref{eqn:stability2} corresponds to technical fluctuations in the number of atoms probed during each experimental cycle.  The number of atoms loaded into the lattice fluctuates for each experimental cycle, contributing noise in the collected fluorescence signal.  This problem is typically overcome by measuring populations in both the ground and excited clock states, and computing the excitation fraction which is normalized against atom number fluctuations. Such an approach is readily compatible with the shelving detection scheme.  After detecting $^1S_0$-$^1P_1$ fluorescence from ground state atoms, atoms in $^3P_0$ can be optically pumped to a state with rapid decay to the ground state, at which point shelving detection can be repeated.  Several suitable intermediate states are available, including $^3S_1$ and $^3D_1$, which exploit cascaded decay to $^3P_1$ and then to $^1S_0$.  In order for the normalization to properly work, conditions must be held constant during both shelving detection pulses (e.g.~intensity of laser driving the $^1S_0$-$^1P_1$ transition) and the optical pumping and decay to the ground state must be efficient and stable.

The fourth noise term in Eq.~\ref{eqn:stability2} comes from frequency noise of the interrogation laser.  Of considerable concern is laser frequency noise which is periodic with experimental cycle time $T_c$, contributing to clock instability via the Dick effect (see Section \ref{laserstabatom}).  The problem is exacerbated by 'dead time' where no atomic frequency measurement is being made, but is rather spent in auxiliary processes such as atomic cooling, loading the optical lattice, state preparation, or state readout.  All standards that are not continuously interrogated are susceptible to this noise.  While it affects single-trapped-ion clocks, it is especially pernicious to lattice clocks since it can prevent them for reaching a much lower QPN instability.

\begin{figure}[t!]
    \centering
    \includegraphics[width=\textwidth]{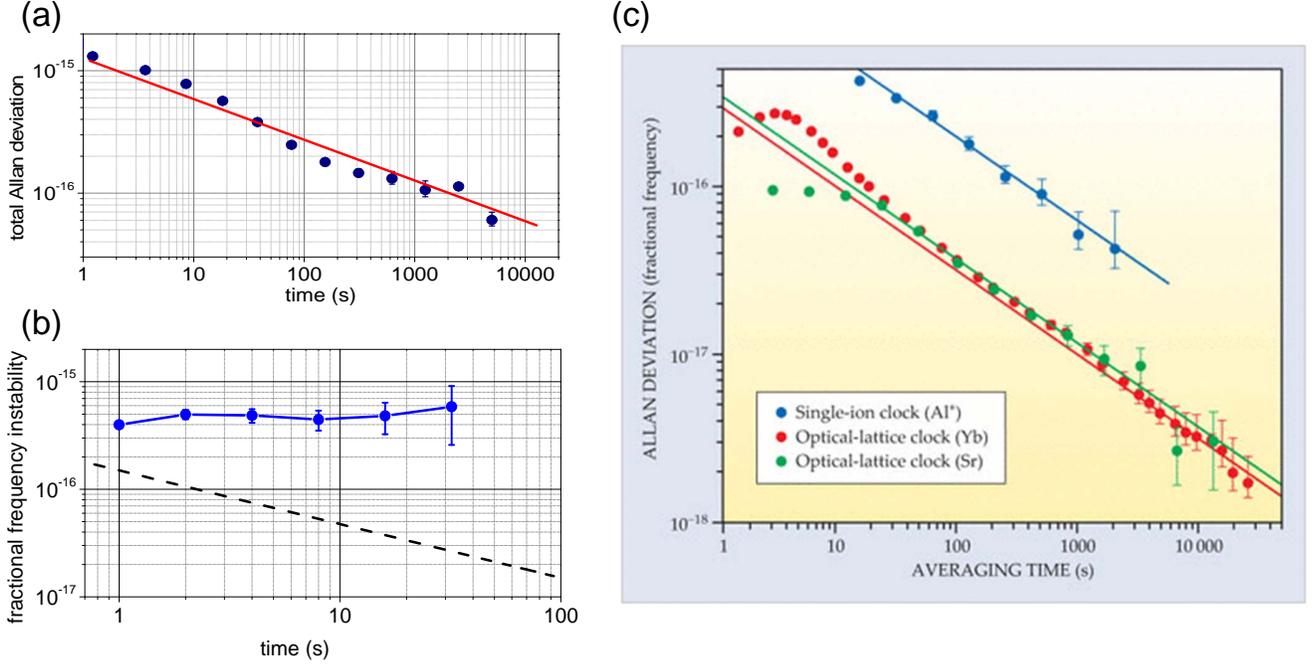}
    \caption{\label{stability} (a) Frequency stability of the JILA Sr versus NIST Yb lattice clocks, as measured in 2009. The total deviation confirms the predicted stability limit given by the clock laser (b) Measurements of the improved Yb clock instability at one to a few clock cycles. Blue circles are measured using atomic excitation as an (out-of-loop) frequency discriminator of the optical LO frequency, and includes contribution to the instability from the Dick effect, atomic detection noise, and the LO free-running instability.  The Dick-effect-limited instability (black dashed line) is $1.5 \times 10^{-16}/\sqrt{\tau}$. (c) Recently improved measurements of the fractional frequency instability for both the Yb (red) and Sr (blue) lattice clocks.  Reproduced with permission from \cite{Smart2014}. Copyright 2014, American Institute of Physics.}
\end{figure}

Figure~\ref{stability} shows the measured fractional instability between the JILA Sr lattice clock and the NIST Yb lattice clock (utilizing the optical fiber link~\cite{Foreman07b}) as measured in 2009.  The instability reached $1 \times 10^{-16}$ near 1000 s, demonstrate promise by crossing into the $10^{-17}$ decade.  Nevertheless, the measured instability is still nearly two orders of magnitude higher than their potential at $10^{-17}/\sqrt{\tau}$ or better.  As is the case for many optical clocks, this limitation was dominated by the Dick effect.  Consequently, recent efforts have targeted improving clock stability through reduction of the Dick effect \cite{Westergaard10a,lodewyck_frequency_2010,jiang_making_2011}.  In one case \cite{Westergaard10a}, rather than using a destructive measurement to readout the atomic population, a non-destructive measurement is employed to enable repeated spectroscopic probings before a required re-loading of the optical lattice due to finite trap time.  In this way, the dead time could be significantly reduced, reducing the Dick effect.  In another case \cite{jiang_making_2011,Nicholson12,Bishof2013a}, improved optical local oscillators were employed, aimed at reducing down-sampled frequency noise as well as enabling longer clock transition probe times to reduce the fractional dead time.  The results of \cite{jiang_making_2011} are highlighted in Fig.~\ref{stability}.  With the improvements described, the Dick-limited instability was calculated to be $1.5 \times 10^{-16}/\sqrt{\tau}$.  In lieu of a direct comparison between two clocks to measure the clock stability at all times, a maximum limit on the clock instability at short times could was made using the atomic transition as a discriminator of the stable laser used to probe the atoms.  The atomic response was measured using the same atomic detection utilized in clock operation.  The measurement indicated a clock instability of $<5 \times 10^{-16}$ at short times (seconds).

An approach that avoids the pernicious influence of the Dick effect is to use a synchronous interrogation method~\cite{Bize00}. Synchronous interrogation allows differential measurements between two atomic systems free laser noise.  Such measurements have yielded impressive measurement instability \cite{chou_quantum_2011,takamoto_frequency_2011}, although it should be noted that this approach does not measure independent clock stability.  With the recent implementation of an ultrastable optical local oscillator ($1 \times 10^{-16}$ at 1 - 1000 s~\cite{Swallows12,Nicholson12,Bishof2013a}), two Sr clocks were independently compared to demonstrate a frequency instability of a single clock at 3 $\times$10$^{-16}$/$\sqrt{\tau}$, approaching the QPN estimated for 1000 atoms with 160 ms coherent probe time.  A comparison between two Yb lattice clocks has demonstrated similar short-term stability, averaging to $1.6\times$10$^{-18}$ instability after 7 hours \cite{hinkley2013}.  Figure~\ref{stability}(c) shows record best frequency instability for both the Yb \cite{hinkley2013} and Sr \cite{Bloom2014} optical lattice clocks.  While lattice clocks are now demonstrating stability levels never before reached for any type of atomic clocks, they remain far from their potential.  As a result, further development in stable lasers will remain a high priority~\cite{kessler_sub-40-mhz-linewidth_2012, cole_tenfold_2013}.

\subsubsection{Systematic Evaluations}

In Sec~\ref{section:systematics}, we considered in detail the phenomena leading to systematic shifts of the clock transition frequency.  Any such shifts, if improperly controlled or compensated for, lead to frequency error of the standard.  To determine the overall uncertainty to which the natural transition frequency is being realized, systematic shifts and the ability to control these shifts must be characterized quantitatively.  When the mechanism yielding the shift is known precisely and described accurately with a sufficiently rigorous model, it is sometimes justified to measure the experimental parameters of the model and deduce the shift value and uncertainty.  For lattice clocks, such has previously been the case for the blackbody-radiation-induced Stark shift.  However, the optimal evaluation of a systemic shift consists of both a well-understood model explaining the shift, as well as a direct measurement of the frequency shift in the standard being evaluated.

\begin{figure}[t!]
\centering
    \includegraphics[width=\textwidth]{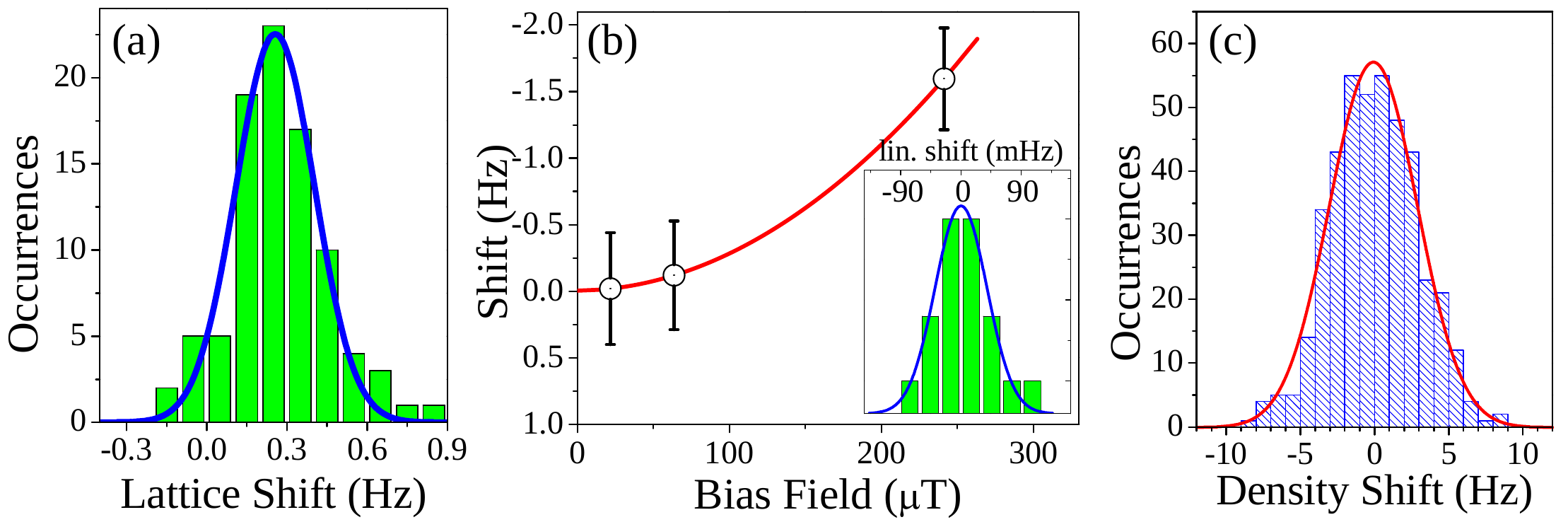}
\caption[]{\label{SrEval} Measurement of key systematic frequency shifts in a $^{87}$Sr optical lattice clock. The lattice Stark shifts (a) as well as second-order (b) and first-order (inset, b) magnetic field shifts are measured by direct optical comparison to a stable calcium optical standard.  On the other hand, (c) shows a measurement of the density shift using interleaved conditions of a single standard.  In this case, a large number of measurements facilitate precise determination of the shift. All histograms in the figure are accompanied by Gaussian fits to the data set.}
\end{figure}

A direct method to measure the shift is to compare the standard in question to a similar standard.  The standards are operated at different conditions, where the systematic effect yields different shifts, and the extrapolation to the zero shift case can be deduced.  One example of such an evaluation is \cite{Westergaard11a}, where lattice Stark shifts were carefully evaluated by comparison of two Sr lattice clocks.  We note that measurement between two systems can be synchronized to offer measurement stability below limitations of the local oscillator (e.g.~\cite{takamoto_frequency_2011}).

Often, only one system is being developed in a particular lab, in which case measurement may be made against another type of standard.  The conditions of the standard under test are varied quickly and controllably in time, and the resulting frequency variation is measured against the reference standard.  In this way, the reference standard serves predominantly as a stable frequency source.  Examples of such a measurement include \cite{Ludlow08b,Lemke09a}, where both a Sr and Yb lattice clock were evaluated by comparison with a calcium optical clock.  Figure~\ref{SrEval} (a) and (b) shows two such measurements of the lattice Stark and Zeeman shifts.

In practice, it is sometimes easy to measure many systematic effects by varying the standard's operating conditions, and looking for frequency shifts relative to the local oscillator used to probe the clock transition.  Such a technique is directly sensitive to local oscillator noise, and requires the local oscillator to be frequency-stable on the time scale over which the conditions are varied.  However, this technique requires no additional atomic standard, and thus simplifies the experimental process.  Fortunately for the optical lattice clock, stable lasers often exhibit sufficient frequency stability for convenient evaluation of many systematic effects in this manner.  The conditions being varied can frequently be changed on a relatively fast timescale, limiting the frequency wander of the local oscillator between measurements.  Many such measurements can be repeated to average down the measurement uncertainty.  Examples of such a measurement are~\cite{Boyd07b} and~\cite{Falke11a}, and Figure~\ref{SrEval} (c) shows a density shift measurement of this type.

\begin{table}
  \caption{A recent evaluation of systematic frequency shifts in an $^{87}$Sr lattice clock \cite{Bloom2014}.}
  \label{tab:systematics}
  \begin{tabular}{|l|cc|cc|}\hline
Systematic Effect & Corr. (10$^{-18}$) & Unc. (10$^{-18}$)\\ \hline
      Lattice Stark  & -461.5  & 3.7   \\
      Residual lattice vector shift & 0 & $<$0.1 \\
      Probe beam ac Stark   &0.8&1.3 \\
      BBR Stark (static)    &-4962.9 & 1.8 \\
      BBR Stark (dynamic) &-345.7 & 3.7 \\
      1st order Zeeman     &-0.2& 1.1  \\
      2nd order Zeeman     & -144.5 & 1.2 \\
      Density              &-4.7&0.6 \\
      Line pulling and tunnelling   &0&$<$0.1 \\
      DC Stark  &-3.5 & 2.1    \\
      Servo error       &0.4&0.6 \\
      AOM phase chirp   &0.6 & 0.4   \\
      2nd order Doppler  &0 &$<$0.1    \\
      Background gas collisions  &0 &0.6    \\ \hline
      Total Correction & -5921.2 & 6.4 \\ \hline

\hline
  \end{tabular}
\end{table}

To highlight recent progress on reducing the uncertainty of optical lattice clocks, the JILA $^{87}$Sr clock achieved an overall uncertainty of 6.4 $\times10^{-18}$ by the end of 2013 (see Table \ref{tab:systematics})~\cite{Bloom2014} and it was further reduced to 2.1 $\times10^{-18}$ by the end of 2014~\cite{Nicholson2014}. The excellent lattice clock stability has played an important role in facilitating the characterization of this level of low uncertainty for atomic clocks. Progress is being made in many other labs, and we anticipate that soon lattice clock uncertainty will be pushed to 1 $\times10^{-18}$ or below. Continued advancements in the clock stability will aid these efforts.  Efforts to measure and control the blackbody Stark shift and lattice Stark shifts will continue to play an important role.

\begin{figure}[t!]
\centering
\includegraphics[width=\textwidth]{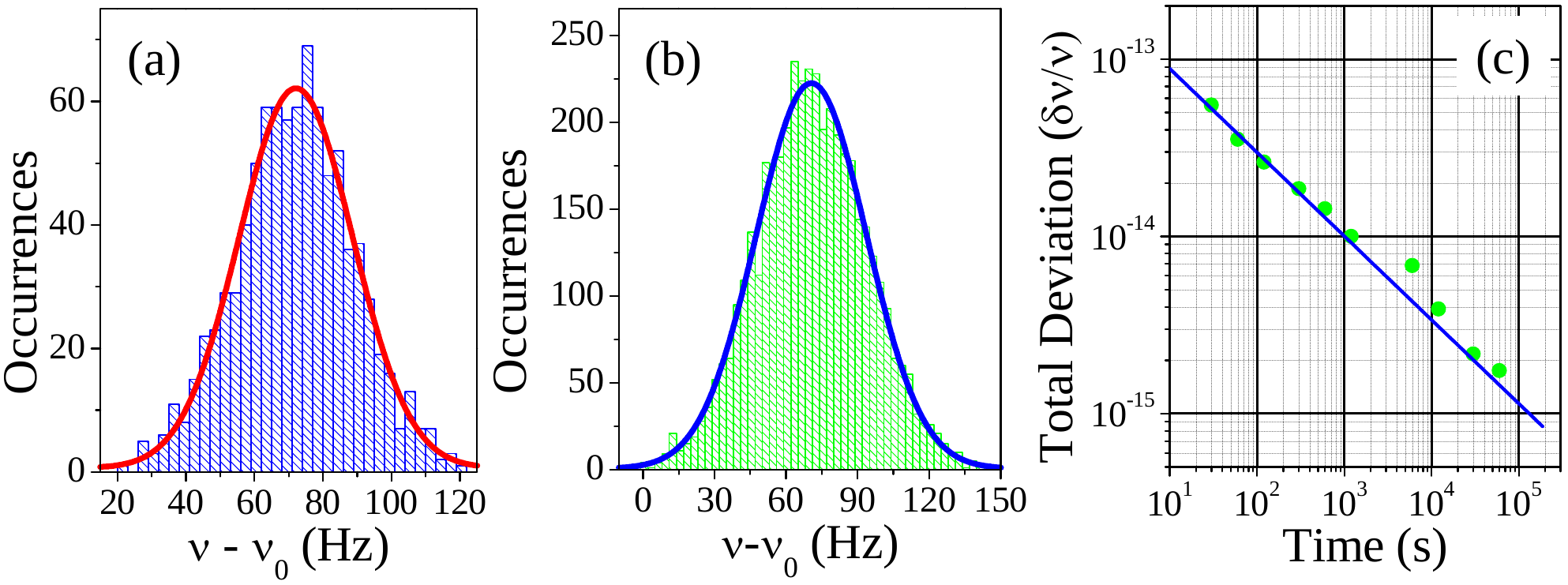}
\caption[Absolute frequency measurement of the $^1S_0$ - $^3P_0$ transition]{\label{fig:absfreq1} Absolute frequency measurement of the $^1S_0$ - $^3P_0$ transition in $^{87}$Sr. (a) Histogram of 880 measurements (without nuclear-spin-polarization) taken over a 24 hour period.  The corresponding Gaussian fit (solid curve) and data has a mean value of 71.8(6) Hz. (b) Histogram of the 50 hour absolute frequency measurement (with nuclear-spin polarization) using a hydrogen-maser calibrated in real-time to a caesium fountain. The resulting frequency is 70.88(35) Hz and the distribution is Gaussian as shown by the fit (solid curve).  In (a) and (b) the offset frequency, $f_0$, is 429 228 004 229 800 Hz and the data sets are corrected only for the maser offset.  When the Sr systematics are included the frequencies are in excellent agreement.  (c) Total deviation of the Sr-maser comparison for the data set in (b). The line is a fit to the data yielding a stability of 2.64(8)$\times10^{-13} \tau^{-0.48(1)}$ and extends out to the entire measurement time. }
\end{figure}

\subsubsection{Absolute Frequency Measurements}

Optical clocks have demonstrated systematic uncertainties which are fractionally smaller than that of the caesium primary standard.  Consequently, they are excellent candidates for primary time and frequency standards of the future.  However, at present, the SI second is defined relative to the caesium hyperfine clock transition.  By definition, any accurate frequency measurement must be traceable to a caesium primary reference.  Absolute frequency measurements of the clock transition frequency of optical lattice clocks are therefore usually made by referencing the highest performance caesium standards, the caesium fountain clock (e.g.~\cite{Wynands05a,Heavner05a,Bize04a,Bauch03a}).  For such a measurement, the systematic uncertainties of both the lattice clock and the caesium fountain clock often play an important role, as does the link between these standards.  An optical frequency comb is inevitably used to make the link between the optical and microwave frequency domains.  The standards are often spatially separated, requiring careful phase and/or frequency control of microwave and optical signals bridging the distance (e.g.~\cite{Foreman07a}).  The spatial separation between atomic standards is often accompanied by a change in gravitational potential, requiring the appropriate correction for the gravitational red shift (approximately $10^{-16}$ per meter of height change) \cite{vessot80}.  Because the caesium standard operates at microwave frequencies, its fractional stability can be 100 times lower than that of an optical lattice clock.  As a consequence, measurements must be made over longer timescales to reach sufficiently small statistical uncertainties (e.g.~see Fig.~\ref{fig:absfreq1}).  To reach an uncertainty level of $10^{-15}$ or below, absolute frequency measurements are typically made over the course of many hours or many days.  This long averaging times requires the standards to be operationally robust over these timescales.

\begin{table}
  \caption{Absolute frequency measurements of optical lattice clocks.}
  \label{tab:absfreq}
  \begin{tabular}{|l|l|c|}\hline
    & Absolute frequency and uncertainty (Hz) & reference \\ \hline
    $^{87}$Sr & 429 228 004 235 000 (20000) & SYRTE \cite{Courtillot03a} \\
            & 429 228 004 230 000 (15000)& SYRTE \cite{Courtillot05a} \\
            & 429 228 004 229 952 (15) & U.~Tokyo \cite{Takamoto05a} \\
            & 429 228 004 229 869 (19) & JILA \cite{Ludlow06a} \\
            & 429 228 004 229 879 (5) & SYRTE \cite{LeTargat06a} \\
            & 429 228 004 229 875 (4) & U.~Tokyo \cite{Takamoto06a} \\
            & 429 228 004 229 874 (1.1) & JILA \cite{Boyd07b} \\
            & 429 228 004 229 873.6 (1.1) & SYRTE \cite{Baillard08a} \\
            & 429 228 004 229 873.65 (0.37) & JILA \cite{Campbell08a} \\
            & 429 228 004 229 874.1 (2.4) & U.~Tokyo \cite{Hong09a} \\
            & 429 228 004 229 872.9 (0.5) & PTB \cite{Falke11a} \\
            & 429 228 004 229 873.9 (1.4) & NICT \cite{YamaguchiShiga2012} \\
            & 429 228 004 229 873.1 (0.132) & SYRTE \cite{LeTargatLorini13} \\
            & 429 228 004 229 873.13 (0.17) & PTB \cite{FalkeLemke13a} \\
            & 429 228 004 229 872.0 (1.6) & NMIJ \cite{AkamatsuInaba2014} \\
            & 429 228 004 229 873.60 (0.71) & NICT \cite{HachisuFujieda2014} \\ \hline
    $^{88}$Sr & 429 228 066 418 009 (32) & SYRTE \cite{Baillard07a} \\
            & 62 188 138.4 (1.3) & Tokyo 87-88 isotope shift \cite{Akatsuka08a} \\ \hline
    $^{171}$Yb & 518 295 836 591 600 (4400)& NIST \cite{Hoyt05a} \\
            & 518 295 836 590 865.2 (0.7) & NIST \cite{Lemke09a} \\
            & 518 295 836 590 864 (28) & NMIJ \cite{Kohno09a} \\
            & 518 295 836 590 865.7 (9.2) & KRISS \cite{Park2013} \\
            & 518 295 836 590 863.1 (2.0) & NMIJ \cite{YasudaInaba2012} \\
            & 518 295 836 590 863.5 (8.1) & KRISS \cite{Park2013} \\ \hline
    $^{174}$Yb & 518 294 025 309 217.8 (0.9) & NIST \cite{Poli08a} \\ \hline
    $^{199}$Hg & 1 128 575 290 808 162 (6.4) & SYRTE \cite{McFerran12a} \\ \hline

  \end{tabular}
\end{table}

\begin{figure}[b!]
\centering
    \includegraphics[width=\textwidth]{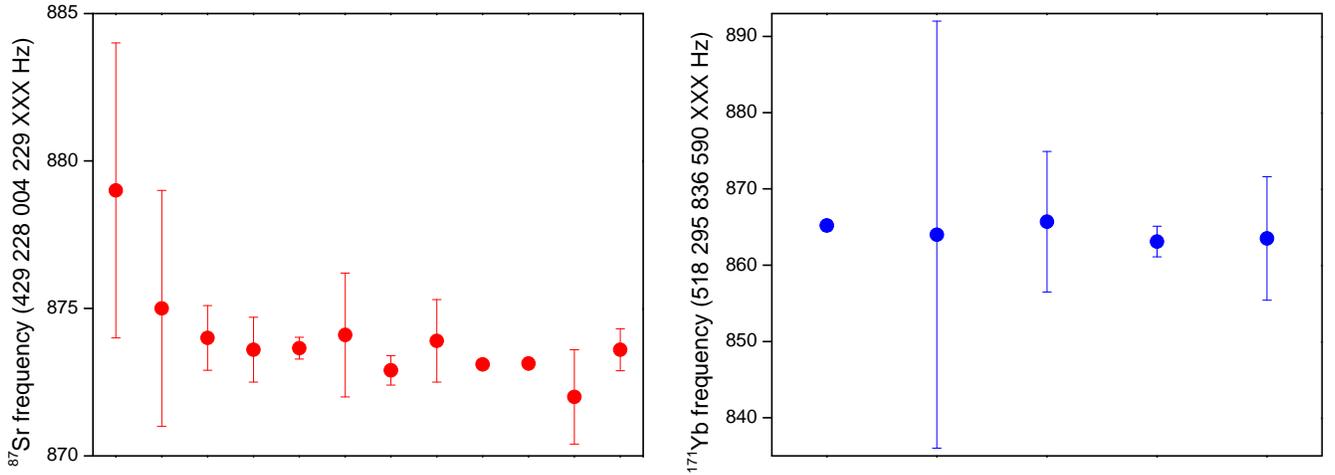}
\caption[Absolute frequency measurement of the $^1S_0$ - $^3P_0$
transition \label{fig:absfreq1}]{\label{fig:absfreq2} Absolute frequency measurements of the
$^1S_0$ - $^3P_0$ transition in $^{87}$Sr and $^{171}$Yb (see also Table~\ref{tab:absfreq}).}
\end{figure}

Absolute frequency measurements have been made for optical lattice clocks utilizing Sr, Yb, and Hg.  Table~\ref{tab:absfreq} lists many absolute frequency measurements that have been published in the literature. Figure~\ref{fig:absfreq2} plots recent measurements for the absolute frequencies of the $^{87}$Sr and $^{171}$Yb optical clocks.  We emphasize the excellent agreement between various measurements made in different laboratories around the world, indicating the ability of the lattice clock to serve as an accurate optical frequency standard.  In fact, as we will see in the next section, the agreement in these measurements has allowed the international comparison to provide a useful constraint on variation of fundamental constants.  Both the Sr and Yb standards have now been recommended as secondary representations of the SI second.

\section{APPLICATIONS AND FUTURE PROSPECTS}\label{section:future}

With the rapid progress and high performance levels of optical clocks, two natural questions arise.  What will the primary impacts of these advanced timekeepers be, and can they get even better? We explore these questions, beginning with the definition of time and frequency itself.  We consider the role of atomic clocks in the measurement of fundamental physics as an important field that these clocks have benefitted and will continue to benefit.  We also look at quantum control techniques which may ultimately benefit the optical clock, and how these clocks continue to benefit our study of quantum systems.  Finally, we conclude by considering atomic clocks which operate at the very highest optical frequencies and beyond.

\subsection{Primary standards and worldwide coordination of atomic time}

International Atomic Time (TAI) and Coordinated Universal Time (UTC) are maintained and disseminated by the Time, Frequency and Gravimetry Section of the Bureau International des Poids et Mesures (BIPM) in Paris \cite{guinot_atomic_2005}. They are the result of worldwide cooperation of about 70 national metrology laboratories and astronomical observatories that operate atomic clocks of different kinds. Each participating laboratory k realizes an approximation to UTC, denoted UTC(k), which is used as the reference for local clock comparisons and frequency distribution. Time transfer between the laboratories is performed by comparing local clocks to the time information received from the satellites of global navigation systems and via dedicated two-way links on geostationary telecommunication satellites. In this way, the differences between the UTC(k) time scales can be established with a transfer uncertainty that reaches less than 1~ns for well calibrated links.

A large ensemble of about 400 clocks, mainly commercial caesium clocks and hydrogen masers, are reported to BIPM and are averaged to obtain the time scale Echelle Atomique Libre (EAL).
The used algorithm is designed to provide a reliable scale with optimized frequency stability for a selected averaging time, assigning statistical weights to individual clocks based on their performance during the last 12 months.
The frequency instability of EAL reaches about $4\times10^{-16}$ over 30 to 40 days.
In a second step, measurements of primary caesium clocks made over the previous one year period are introduced to calculate the relative departure of the second of the free atomic time scale from the SI second as realized by the primary clocks. By application of a gentle frequency steering which should not compromise the intrinsic stability of EAL, the free scale is transformed into an accurate atomic time scale TAI (Temps Atomique International).
Presently, 11 atomic fountains in 8 laboratories contribute regularly to the calibration of TAI and a frequency uncertainty of about $3\times10^{-16}$ is obtained. Several more primary caesium clocks are under development worldwide. UTC is finally derived from TAI after the addition of leap seconds. These are introduced at irregular intervals, following the convention to maintain UTC in agreement to within 0.9~s with an astronomical time scale defined by earth's rotation \cite{nelson_leap_2001}
The dissemination of UTC by the BIPM takes the form of a time series of [UTC - UTC(k)] for selected dates in the past month.

With the rapid improvement in the development of optical frequency standards, it has been demonstrated that the accuracies of a number of systems now surpass those of primary caesium standards. The direct comparison of two optical frequency standards can be performed with lower uncertainty than the SI second is realized. This calls in the long term for a redefinition of the second in terms of an atomic transition frequency in the optical range. In order to approach this change and to introduce novel frequency standards into metrological use, the concept of ''secondary representations of the second'' has been defined \cite{gill_when_2011}. A formal procedure has been established of taking note of measurements of transition frequencies in atoms and ions relative to the caesium frequency standard and of the pertinent uncertainty evaluations.
As the result of an evaluation in 2012, seven values for optical transition frequencies in the atoms $^{87}$Sr and $^{171}$Yb and in the ions $^{27}$Al$^+$, $^{88}$Sr$^+$, $^{171}$Yb$^+$ (two transitions), and
$^{199}$Hg$^+$ have been recommended with uncertainties in the range $1\dots 4\times 10^{-15}$. Obviously, the uncertainty of the recommended SI frequency value can never be lower than those of the best available primary frequency standard.
In the microwave range, the ground-state hyperfine transition frequency of $^{87}$Rb has been recommended with an uncertainty of $1.3\times 10^{-15}$. Measurements with a rubidium fountain \cite{guena_improved_2012,Guena2014} are now reported in comparison to TAI regularly, and in principle data from optical frequency measurements could also be used, as has been demonstrated in a retrospective exercise already \cite{wolf_comparing_2006}. From such comparisons one can assess whether the reproducibilities of the new standards between successive periods of operation are in agreement with their stated uncertainties. If this is verified, they could constitute very valuable sources for the monitoring and steering of TAI, even if their full intrinsic uncertainty cannot be immediately used since they realize only secondary representations. To benefit fully from performance of optical frequency standards for  the realization of time scales requires significant improvements in time transfer and in technology for flywheel frequency standards that are needed to handle dead time \cite{parker_invited_2012}. Continuous operations of optical atomic clocks for periods of several days have been established in several laboratories. However, since these systems involve a number of lasers including optical frequency combs, and laser cooled and trapped atoms and ions, the reliability of their operations is still being gradually improved.

\subsection{Technological Applications}\label{sec:tecapps}

Advances in timekeeping impact a variety of applications.  Atomic clocks are a critical component in Global Navigation Satellite Systems (GNSS), mature advanced atomic clocks based on optical transitions and/or laser cooled atomic/ionic samples could be used to improve navigation and timekeeping capabilities.  At short timescales GNSS are not currently limited by the atomic clock stability, but by atmospheric disturbances which would need to be mitigated to take full advantage of the performance discussed here.  However, GNSS systems could already benefit from the superb long term stability provided by high performance clocks as the ultra-low drift would allow significantly extended operation between updates or re-synchronization, compared to currently deployed Rb vapor cell clocks.  The improved stability would translate to improved GNSS system integrity, enabling autonomous operation within given acceptable position ranging errors for timescales of days or weeks, instead of hours.  Applications which can benefit from improved system integrity include precision airplane approaches at airports~\cite{WeissPTTI}. Advanced clocks will also be needed onboard deep space missions to aid in navigation and timekeeping.  Deep space navigation is usually implemented by Doppler velocimetry and ranging in a two-way configuration. Stable clocks on board of spacecrafts would allow a down-link-only operation with significantly better accuracy and coverage of spacecraft observation \cite{prestage_progress_2009}. Other applications which are poised to benefit from next generation clocks include radar, where the improved short term stability results in ultra-low phase noise microwaves for high resolution and extended dwell times,  similarly radio-astronomy using synthetic aperture techniques, and communication networks.  Optical frequency synthesis using optical clock and comb architecture will enable on-demand coherent frequency generation for academic and industry applications.

In addition, the optical clock technology platform parallels that of emerging inertial sensor technology based on atom interferometry, and advances in one field can be incorporated into and benefit the other.  Common tools for these systems include frequency stabilized lasers, ultra high vacuum systems, and low noise electronics. Atomic sensors using this toolbox include absolute gravimeters and gravity gradiometers, which have applications in geophysical monitoring and research, as well as oil and mineral exploration and gravity aided navigation.  Gyroscope configurations show promise for inertial navigation systems, and may enable high performance navigation in sea and space environments where GNSS is not available.


\subsection{Optical clocks for geodetic applications}\label{sec:geodesy}
According to General Relativity, a clock ticks slower in a gravitational potential compared to a clock outside of it. The corresponding fractional frequency difference between the clocks is given by $\frac{\Delta f}{f}=-\frac{\Delta U}{c^2} $, where $\Delta U=U_1-U_2$ is the gravitational potential difference between the positions of the clocks and $\Delta f=f_1-f_2$ their frequency difference. On earth, the gradient of the gravitational potential results in a fractional frequency change of approximately $10^{-16}$ per meter height difference for a clock at rest. By combining the gravity potential provided by optical clocks and its derivative (the gravity field) as measured by gravimeters, one can estimate the size and location of a density anomaly~\cite{bondarescu_geophysical_2012}, an important application in earth exploration.

When comparing two clocks at different locations, relativistic time dilation from the rotation of the earth and higher-order general relativistic corrections need to be taken into account \cite{petit_relativistic_2005} and for contributions to international time scales by referencing the clocks to a well-defined reference geopotential  \cite{soffel_iau_2003}. The equipotential surface of this geopotential  (gravitational plus centrifugal components) closest to mean sea level is called the geoid and corresponds to a water surface at rest. The height above the geoid defines an orthometric height system in geodesy, closely approximating equipotential surfaces. Geopotential differences tell us in which direction water flows. This has important applications in coastal protection, engineering, and water resource management. Currently, heights within a country are determined through geometric leveling with theodolites supported by local gravimetry along leveling lines. This is performed in loops with a total length of more than 30,000~km for a country such as Germany with an area of 360,000~km$^2$. Establishing such a leveling network with typical single-setup distances of around 50~m is a time-consuming and costly task. Most importantly, errors in single measurements accumulate, compromising the overall height system to an accuracy of a few centimeters within a country. An alternative approach uses accurate GNSS (global navigation satellite system) data together with gravity field modelling from satellite gravimetry supported by terrestrial gravimetry, which in principle is capable to extend height systems across continents \cite{denker_regional_2013}. However, it should be noted that GNSS only provides geometrical heights above an ellipsoid. Different approaches to obtain the height above the geoid produce height deviations of several tens of centimeters and disagree with purely terrestrial measurements \cite{woodworth_towards_2012, gruber_intercontinental_2012}. A conceptually new and independent method to overcome these limitations and simplify the connection between height systems is ``relativistic geodesy'' or ``chronometric leveling'', which allows long-distance potential difference measurements  \cite{vermeer_chronometric_1983, bjerhammar_relativistic_1985, shen_determination_2011, delva_atomic_2013}. It is based on a frequency comparison between two remote optical clocks via optical fibers (see Sec.~\ref{sec:distribution} or free-space microwave \cite{levine_review_2008, piester_remote_2011,delva_time_2012, fujieda_carrier-phase_2014} or optical \cite{fujiwara_optical_2007, Djerroud10a, exertier_t2l2:_2013, giorgetta_optical_2013} satellite links to provide a direct height difference measurement between two remote locations. Alternatively, a mobile clock (operating during transport) together with careful modeling of its speed and geopotential trajectory can be used.
Geodesy and frequency metrology are inextricably linked: A remote frequency comparison probes the accuracy of clocks \emph{and} the geodetic model simultaneously, since the height difference between the clocks enters the systematic uncertainty evaluation of the frequency standard \cite{pavlis_relativistic_2003}. Therefore, relativistic geodesy should be performed using high performance transportable optical clocks. These can be calibrated through side-by-side measurements with the reference clock before being  transported to a remote site for a geopotential comparison.  Furthermore, this approach would relax the requirements of the uncertainty evaluation of the involved frequency standards and thus improve the height resolution. Instead of performing an evaluation of the \emph{accuracy} of the clock, one would evaluate its \emph{reproducibility}. It allows the clock to have a less precisely known but constant shift from its unperturbed transition frequency. The frequency uncertainty in terms of reproducibility is the uncertainty in keeping the shifts constant, without knowing their exact magnitude. An example is the black-body radiation shift discussed in Sec.~\ref{sec:ionBBR}.
The uncertainty of the shift has two contributions: i) the uncertainty in the differential atomic polarizability and ii) the uncertainty in the radiation field experienced by the atoms, usually characterized by an effective temperature.
If we assume the polarizability (as an atomic parameter) to have a well-defined and constant value, we can neglect its uncertainty in the uncertainty evaluation for a reproducible clock. This is qualitatively different from the uncertainty in the electric field determination which may fluctuate between frequency comparisons. The same argument relaxes the requirements on the evaluation of many other uncertainty contributions.
A world-encompassing network of optical clocks operating at a level of $10^{-18}$ with a suitable infrastructure for high-level frequency comparison would not only provide a more accurate time standard, but also form the basis for a unified, long-term stable geodetic height reference frame \cite{lehmann_altimetrygravimetry_2000, soffel_iau_2003}.

Ultimately, the accuracy of clocks on earth will be limited by the knowledge of the local gravity potential. A master clock in space on a sufficiently well-known orbit \cite{duchayne_orbit_2009, gill_optical_2008} would eliminated this issue and provide a gravitationally unperturbed signal. At the same time, such a ``master clock'' in space \cite{Schiller07a,gill_optical_2008} would enable high-stability time and frequency transfer between earth-bound clocks using microwave \cite{salomon_cold_2001, levine_review_2008, piester_remote_2011,delva_time_2012, fujieda_carrier-phase_2014} or optical links \cite{fujiwara_optical_2007, Djerroud10a, exertier_t2l2:_2013, giorgetta_optical_2013} to establish a unified world height system.

\subsection{Optical clocks in space}\label{sec:clocksinspace}
Optical clocks in space hold the promise of boosting the significance of tests of fundamental physics, in particular tests of Einstein's theory of relativity and applications such as positioning, time and frequency transfer, and directly related to the latter, accurate geoid determination and monitoring~\cite{Cacciapuoti2009}.
Most of these applications have been discussed in previous reviews~\cite{maleki_applications_2005, gill_optical_2008, dittus_lasers_2009} and in two space mission proposals involving optical clocks, namely the SAGAS (Search  for  Anomalous Gravitation using Atomic Sensors) \cite{wolf_quantum_2009} and the EGE (Einstein Gravity Explorer) \cite{Schiller09a} projects. Unfortunately, both missions have not been selected for implementation. However, they provide concrete mission scenarios and thus serve as baselines for space-borne tests with optical clocks. Most importantly, for such missions to be successful in the future, a continued effort into the development of space-qualified (trans)portable optical clocks is essential.

The unification of all fundamental forces including gravity is a formidable task. Such a quantum field theory of gravity should at some scale differ in its predictions  from general and special relativity as developed by Einstein. It is therefore important to devise experiments which probe relativity at different scales. The foundation of general relativity lies in the equivalence principle, comprising the weak equivalence principle (WEP), related to the universality of free fall, Local Lorentz Invariance (LLI), related to velocity-dependent effects, and Local Position Invariance (LPI), related to the universality of the gravitational red shift.
Except for the universality of free fall, optical clocks on satellites in space can outperform terrestrial tests of these principles with only modest requirements on the clock performance, owing to the long unperturbed integration time in a space environment and the strong modulation in gravitational potential and velocity achievable on an appropriately chosen orbit.

LPI tests come in two flavours: i) absolute redshift measurements in which a terrestrial clock is compared to a clock in a spacecraft, and ii) null redshift measurements or tests of the universality of the redshift in which two different types of clocks on board of the same spacecraft are compared. In both experiments the clock(s) in the spacecraft are subject to a strongly varying gravity potential. Any deviation from Einstein's theory of relativity should manifest itself in a modulation of the frequency ratio between the clocks. The SAGAS project proposes to use an optical clock on board of a spacecraft on a Solar System escape trajectory which is compared to a ground clock using an optical carrier link together with appropriate infrastructure to independently measure the spacecraft's velocity and acceleration \cite{wolf_quantum_2009}. It is expected that the much larger variation in gravitational potential and the long mission duration results in an improvement by 4 orders of magnitude over the previous best test by Gravity Probe A \cite{vessot_test_1980}. A similar improvement is expected from the EGE project in which a satellite hosting an optical and a microwave clock revolves around the earth on a highly elliptical orbit \cite{Schiller09a}. Frequency comparisons between the onboard clocks and between the onboard and ground clocks using a microwave link provide null and absolute redshift measurements, respectively.
These measurements can also be interpreted as a coupling of the fine-structure constant to the gravitational potential (see Sec.~\ref{sec:constants}).

LLI tests using clocks can be implemented by measuring the special relativistic time dilation effect scaling as $\Delta f/f\approx -(v_1^2-v_2^2)/2c^2$ (Ives-Stilwell test) for large velocities in the absence of strong gravitational potentials. Such a test could be performed within the SAGAS mission scenario when the spacecraft will leave the solar system at high speed. The expected uncertainty of this test is at $3\times 10^{-9}$ almost a factor of 30 smaller than the best terrestrial test to date \cite{reinhardt_test_2007}. Assuming a violation of LLI in the form of a preferred frame of reference (the cosmic microwave background) through which the solar system races with a speed $v_s$, the time dilation effect gets amplified to $\Delta f/f\approx -(v_1-v_2)v_s/2c^2$ \cite{reinhardt_test_2007}, which can be measured by SAGAS to a level of $5\times 10^{-11}$ relative uncertainty, an improvement by almost two orders of magnitude \cite{wolf_quantum_2009}. Another test of LLI is performed through Kennedy-Thorndike type experiments, in which the independence of the outcome of an experiment to the velocity with respect to a preferred frame is probed. Such experiments probe the relation between time dilation and spatial Lorentz contraction by comparing the frequency of an atomic standard with the resonance frequency of an optical cavity \cite{hils_improved_1990}. The EGE mission scenario predicts an improvement by a factor of 20 over the best terrestrial measurements owing to the large velocity changes during the highly elliptical orbit \cite{Schiller09a}.

Parametrized Post-Newtonian gravity (PPN) describes metric theories of gravitation in the weak field limit using a set of parameters, which are zero for the case of Newtonian gravity. One of the most important parameters is $\gamma$ and describes the amount of curvature produced by a unit rest mass. A non-zero $\gamma$ changes the delay suffered by light traversing a strong gravitational potential (Shapiro time delay) compared to Newtonian gravity and results in gravito-magnetic effects \cite{will_confrontation_2006}. This effect can be measured by spacecraft laser ranging during occultation. Within the SAGAS mission proposal a measurement uncertainty of $u(\gamma)\leq 10^{-8}$ is expected, limited by the onboard clock uncertainty. This corresponds to a two to four orders of magnitude improvement over previous results \cite{bertotti_test_2003}.

Besides these fundamental physics applications, optical clocks in space could act as stable time and frequency servers and provide links for time and frequency transfer between continents to establish improved time scales and a well-defined height system using relativistic geodesy (see Sec.~\ref{sec:geodesy}).

\subsection{Fundamental Physics Measurements}\label{sec:constants}

Understanding how systems evolve in time is a key goal of many scientific theories or models, whether that system be a single atom or the entire universe.  The ticking rate of an atomic clock is determined by the basic properties of sub-atomic particles and how they interact to form an atom.  It depends on the most basic parameters, the fundamental constants of nature. As their name suggests, these fundamental constants are typically assumed to be fixed in value throughout space and time.  However, if they varied, as some theories which seek to unify the fundamental forces predict, then so too does the ticking rate of an atomic clock.  As such, atomic clocks serve as one of several vital tools to explore this possible variation through time, space, or through coupling to gravitational fields \cite{Karshenboim05a,Lea07a}. Atomic clocks complement astronomical and other measurements which instead sample possible variation over a large fraction of the history of the universe \cite{Flambaum07a,Reinhold06a}. Atomic frequency standards, on the other hand, are locally operated on earth and are only useful for exploring fundamental constant variation during the time that they are operated for such measurements.  Presently, this is only on the timescale of years.  However, meaningful measurements can be made due to the unmatched measurement precision and accuracy of atomic clocks.

\begin{figure}[t!]
   \centering
     \includegraphics[width=5in]{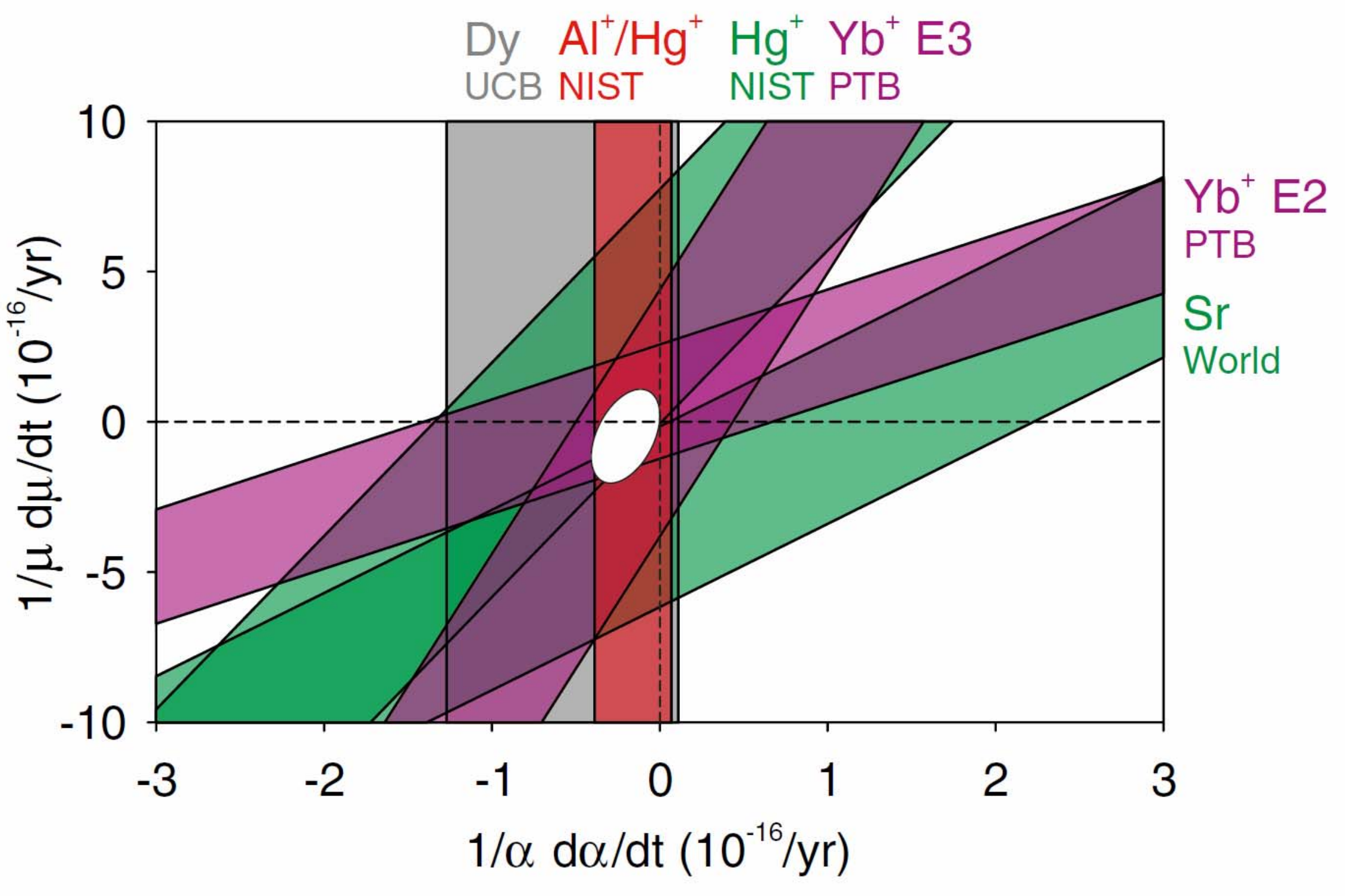}
    \caption[]{\label{alpha} Measurements between atomic clocks of different species can constrain possible variation of fundamental constants.  A number of comparisons between distinct atomic-clock-species are used here to constrain time variation of $\alpha$ and $\mu$.  Reproduced with permission from \cite{Huntemann2014}. Copyright 2014, American Physical Society.}
\end{figure}

The atomic and molecular transitions at the heart of these standards can depend on fundamental constants such as the fine structure constant ($\alpha$), the electron-proton mass ratio ($\mu$), and the light quark mass.  As these clocks advance in measurement precision, their ability to constrain the fluctuations of these constants improves. Optical clock transitions exhibit dependence on the fine structure constant through relativistic corrections to the transition frequency \cite{Karshenboim05a,Angstmann04a,Lea07a}.  The Cs microwave clock transition, based on hyperfine splitting, is additionally dependent on the electron-proton mass ratio $\mu=m_e/m_p$.  Thus, absolute frequency measurements of different species can be used to explore possible temporal variations of $\alpha$ and $\mu$.  For example, the fractional frequency drift rate of the Sr clock frequency measured against Cs constrains a linear combination of the variations $\frac{\delta\alpha}{\alpha}$ and $\frac{\delta \mu}{\mu}$ in atomic units as
\begin{equation}
    \label{alphaeq}
    \frac{\delta (f_{Sr}/f_{Cs})}{f_{Sr}/f_{Cs}}=(K_{Sr}-K_{Cs}-2)\frac{\delta
    \alpha}{\alpha}-\frac{\delta \mu}{\mu}.
\end{equation}
Sensitivity coefficients ($K$) for various species have been calculated (e.g.~\cite{Angstmann04a,flambaum_search_2009}). The sensitivity of the Cs clock to $\alpha$-variation is moderate, $K_{Cs}=0.83$.  On the other hand, the sensitivity for atomic frequency standards based on Sr or Al$^+$ is low, $K_{Sr}=0.06$ and $K_{Al^+}=0.008$.  Standards based on neutral mercury and ytterbium have larger values, $K_{Hg}=1.16$ and $K_{Yb}=0.45$.  Some atomic species exhibit quite large sensitivity, and thus are particularly well-suited to exploring $\alpha$-variation.  Notable among these are ion standards based on mercury or the octupole transition of ytterbium, $K_{Hg^+}=-3.19$ and $K_{Yb^+}=-5.95$ (octupole).  Measurements of the transition frequencies of these clocks can be measured at different times, ideally over an interval of many years, and require comparisons among different clock species, ideally between clocks with varying sensitivity to fundamental constant variation (e.g.~a clock with high sensitivity measured against one with low sensitivity, or two clocks with high sensitivity of opposite signs).  Observed drift rates can be extracted by linear fits to such data. From Eq.~\ref{alphaeq} it is seen that drift rates for more than two species are needed to constrain the $\alpha$ and $\mu$ dependence. Figure~\ref{alpha} combines the results of a variety of two-species comparisons made over time, in order to tightly constrain both $\alpha$ and $\mu$ variation.  Data is taken from measurements and analysis using Yb$^+$ \cite{Huntemann2014,Tamm2014}, Hg$^+$ and Al$^+$ \cite{fortier_precision_2007,rosenband_frequency_2008}, Sr \cite{LeTargatLorini13}, and Dy \cite{Leefer2013}, and often using Cs as the second system.  In this case, overall constraints of $\delta \alpha / \alpha = -2.0(2.0) \times 10^{-17}$/yr and $\delta \mu / \mu = -0.5(1.6) \times 10^{-16}$/yr resulted. We note some other measurements helping to constrain fundamental constant variation, including hydrogen \cite{Fischer04a}, Sr$^+$ \cite{Barwood2014,Madej2012}, Sr \cite{Falke2014,Blatt08a}, Rb \cite{Guena2014}, and an additional high accuracy measurement and analysis using Yb$^+$ \cite{Godun2014}, among many others.  As more species are
compared with increasing accuracy, an improved sensitivity to temporal variations can be expected.

Because optical standards have achieved lower measurement instability and systematic uncertainty than Cs standards, direct optical clock comparisons can be very useful for studying $\alpha$ variation.  In this case, $\alpha$ variation can be measured directly with only two different standards, as the $\mu$ dependence vanishes.  A notable example of such a measurement is the comparison of the Hg$^+$ and Al$^+$ ion clocks at NIST over approximately a one year interval \cite{rosenband_frequency_2008}.  In addition to independently constraining $\alpha$ variation to $\leq 2.3 \times 10^{-17}$ per year, as shown above this measurement could be combined with others versus caesium to aid $\mu$-variation constraint.  Further improvements in the obtained result can be realized by simply making additional Hg$^+$-Al$^+$ ratio measurements, as many years have now elapsed since those results were published.  Furthermore, another exciting possible measurement involves determining the ratio of two different optical transitions in Yb$^+$, one a quadrupole transition and the other an octupole transition.  As mentioned above, the octupole transition has large negative sensitivity to $\alpha$-variation, while the quadrupole transition possesses reasonably-sized positive sensitivity.  Furthermore, because the effects of some systematic shifts common to both transitions are suppressed, such a ratio measurement has significant potential to explore $\alpha$-variation \cite{Lea07a}.

Frequency measurements can also be analyzed to search for couplings of the $\alpha$ and $\mu$ values to the gravitational
potential, as the earth's elliptical orbit brings the atomic frequency standards through the annually varying solar gravitational
potential. For example, assuming the coupling of these constants is given by dimensionless parameters, $k_{\alpha}$ and $k_{\mu}$, the Sr
frequency can vary sinusoidally over the course of a year by the relation
\begin{equation}
    \label{gravity}
    \frac{\delta (f_{Sr}/f_{Cs})}{f_{Sr}/f_{Cs}}=-\left[
    (K_{Sr}-K_{Cs}-2)k_{\alpha} - k_{\mu} \right] \frac{G m_{sun}}{a
    c^2} \epsilon \cos(\Omega t),
\end{equation}
where $G$ is the gravitational constant, $m_{sun}$ is the solar mass, $a\simeq 1$ au is the semi-major orbital axis, $c$ is the
speed of light, $\epsilon \simeq 0.0167$ is the orbital ellipticity, $\Omega\simeq\sqrt{\frac{G m_{sun}}{a^3}}$ is the earth's angular
velocity around the sun. The orbit of the earth is well known so the frequency data can be fit to Eq.~\ref{gravity} using only a single
free parameter, which is the total amplitude of the cosine variation.  From a recent analysis fitting Sr/Cs measurements made over years at diverse locations around the world, the amplitude of annual variation is determined to be $1.2(4.4) \times 10^{-16}$ \cite{LeTargatLorini13}.  As in the case of linear drift analysis, data from at least one other species is needed to solve for $k_{\alpha}$ and $k_{\mu}$. Hg$^+$ results have also been tested for gravitational variance \cite{fortier_precision_2007} and the combined Sr-Cs and Hg$^{+}$-Cs data can be used to place independent constraints on $k_{\alpha}$ and $k_{\mu}$ (e.g.~\cite{Blatt08a}).  In addition, H-maser frequency measurements have also been tested for gravitational variation~\cite{Ashby07a} and can be used in the same analysis.  The H-maser introduces a possible gravitational sensitivity to the light quark mass $k_q$ that can be extracted when combined with the Sr and Hg$^+$ data~\cite{Blatt08a}.  Together, these results give among the most stringent limits to date for the gravitational coupling of fundamental constants.

In addition to measurements of the possible variation of fundamental constants, optical clocks are useful tools for other fundamental physics measurements. The ability to measure time very precisely allows one to explore how physical systems can influence time evolution.  A prominent example of this is the gravitational redshift: Einstein's theory of general relativity predicts that clocks tick slower in a gravitational potential.  This phenomena is regularly taken into account when comparing atomic clocks or frequency standards at different elevations on the earth's surface.  The gravitational redshift is roughly a $10^{-16}$ effect per meter of elevation gain, and has been measured most precisely by atomic clocks. Alternative gravitational theories predict corrections to the redshift beyond that given by general relativity, and as the performance of clocks continues to improve, so does their ability to characterize the redshift and explore any possible deviations.

\subsection{Quantum correlations to improve clock stability} \label{sec:squeezing}
Most applications of optical clocks demand a high stability, to reach a given frequency uncertainty in the shortest time possible. As outlined above, fluctuations in the number of atoms in a lattice clock and quantum projection noise for a fixed number of trapped ions poses a limit to the measurement accuracy \cite{itano_quantum_1993}. The maximum phase sensitivity in a Ramsey measurement with uncorrelated input states containing on average $N$ particles is given by $\Delta\phi\geq 1/\sqrt{N}$, also known as the standard quantum limit (SQL). Quantum mechanically higher resolution is allowed. The ultimate limit is given by Heisenberg's uncertainty relation which puts a lower bound on the measurement uncertainty of two conjugate variables such as phase and number of particles or energy and time, leading to the Heisenberg limit (HL) $\Delta\phi\geq 1/N$. In the limit of large $N$, this limit can not be further improved by any measurement strategy or specially designed input states \cite{giovannetti_quantum_2012, giovannetti_sub-heisenberg_2012, zwierz_ultimate_2012, hall_universality_2012, zwierz_erratum:_2011, zwierz_general_2010,ou_fundamental_1997}. Identification of measurement strategies and quantum correlated states that minimize the uncertainty of a given observable in a measurement, ideally under realistic noise models, is being pursued in the emerging field of quantum metrology \cite{giovannetti_quantum-enhanced_2004, giovannetti_advances_2011, gross_spin_2012, dorner_quantum_2012, escher_general_2011, luis_quantum-limited_2010}.

To be more specific, consider the frequency uncertainty from a single measurement of $N$ 2-level atoms with collective spin vector $\vec{J}=\sum_n^N \vec{J}_n$ \cite{arecchi_atomic_1972}.  This uncertainty can, in general, be described by
\begin{equation}\label{eq:errorprop}
\Delta\omega=\frac{(\Delta \hat{J}_z)_f}{\left|\frac{\partial \langle \hat{J}_z\rangle_f}{\partial\omega}\right|}
\end{equation}
where $(\Delta \hat{J}_z)^2_f$ denotes the variance of operator $\hat{J}_z$ with respect to the final detected state. Evaluated for uncorrelated atoms using Ramsey spectroscopy, this relation reproduces the standard quantum limit in Ramsey spectroscopy \cite{wineland_squeezed_1994} $\Delta\omega_{SQL}\sim 1/\sqrt{N}$. However, this limit can be overcome by correlating quantum states to reach Heisenberg limited frequency uncertainty which scales as $\Delta\omega_{SQL}\sim 1/N$. Equation~\ref{eq:errorprop} identifies two pathways to achieve this improved frequency resolution: (i) Reducing the projection noise $(\Delta \hat{J}_z)_f$ or (ii) increasing the signal slope $\left|\frac{\partial \langle \hat{J}_z\rangle_f}{\partial\omega}\right|$. Strategy (i) can be implemented by preparing spin-squeezed atomic states \cite{wineland_squeezed_1994, wineland_spin_1992, kitagawa_squeezed_1993} that exhibit reduced quantum projection noise along the measurement direction at the expense of increased noise orthogonal to it.  For example, imagine that all atoms are initially prepared in a superposition of two states labeled spin up and spin down.  The number of atoms in each state is traditionally inferred by measuring the scattering rate of photons out of a probe laser and into a detector.  Due to the multi-mode nature of the scattering process (i.e. photons scatter in all directions), it is in-principle possible to not only determine how many atoms are in each state, but also to determine the state of each individual atom.  This additional information leads to collapse of each atom into spin up or down, resulting in complete decoherence of the sample.  Decoherence can be evaded by building a detection system in which the N atoms uniformly couple more strongly to a single optical mode than the combined coupling to all other modes. The uniform coupling ensures that only collective information is gathered from the detection mode-i.e. how many total atoms are in spin up, but not which atoms are in spin up.  The collective measurement collapses the collective atomic wave function into an entangled state - a spin-squeezed state. The quantum-driven fluctuations of atoms between spin up and down is reduced, while the noise in an unused measurement basis (spin pointing left versus right) is increased.  The conditionally prepared entangled state can be used as an input for clock measurements whose precision increases faster than the standard quantum limit.

Radio-frequency neutral atom clocks below the SQL have been demonstrated using quantum non-demolition (QND) measurements and deterministic, light-mediated interactions to generate squeezed atomic states \cite{appel_mesoscopic_2009, louchet-chauvet_entanglement-assisted_2010, schleier-smith_squeezing_2010,Leroux10a} with a reduction in averaging time of up to a factor of $2.8(3)$ \cite{Leroux10a}.
Squeezed spin states and sub-SQL phase estimation have been experimentally observed for two trapped ions \cite{meyer_experimental_2001} in the radio-frequency regime. However, so far squeezing has not been realized on an optical transition. Schemes for squeezing the collective spin of atoms in a neutral atom optical lattice clock through a cavity-based QND measurement have been proposed \cite{meiser_spin_2008}.

Strategy (ii) can be implemented through maximally entangled states of the form $\psi_\mathrm{GHZ}=(\downN+e^{i\phi}\upN)/\sqrt{2}$  \cite{sanders_optimal_1995, bollinger_optimal_1996}, known as GHZ, Schrödinger-cat, NOON, or N-particle EPR states \cite{kafatos_going_1989, greenberger_bells_1990, bollinger_optimal_1996, monroe_schrodinger_1996, lee_quantum_2002}. They can be generated by implementing a nonlinear Ramsey interferometer using generalized $N$-atom $\pi/2$ Ramsey pulses, implementing a nonlinear rotation of the collective spin.
The atoms in these states are quantum mechanically correlated in such a way that they act as a single, macroscopic quantum system with a phase evolution between the two components which is $N$ times faster compared to uncorrelated atoms, allowing Heisenberg-limited resolution.
The largest GHZ states with high fidelity have been created in trapped ions systems using quantum phase gates \cite{milburn_ion_2000, molmer_multiparticle_1999, solano_deterministic_1999, roos_ion_2008}. 
This way, two \cite{home_deterministic_2006, haljan_entanglement_2005}, three \cite{leibfried_toward_2004}, six \cite{leibfried_creation_2005} and 14 \cite{monz_14-qubit_2011} entangled atoms and improved phase estimation have been demonstrated. The latest experiment by the Innsbruck group is particularly relevant here, since it is implemented on the optical clock transition of the \Ca ion. Scaling the system to hundreds of ions in a Penning trap has been proposed \cite{uys_toward_2011}. Implementation of GHZ Ramsey spectroscopy has also been proposed for a neutral atom optical lattice clock, where the GHZ state is created through the on-site interaction of an atom movable across the lattice \cite{weinstein_entangling_2010}.
A disadvantage of the larger signal slope using GHZ states is the concomitant increased sensitivity to laser phase noise \cite{huelga_improvement_1997,wineland_experimental_1998}, eliminating any stability enhancement of these maximally entangled states.
By engineering more symmetric states with reduced $\Delta \hat{J}_z^2$, such as Gaussian states, \textcite{andre_stability_2004} could show that a stability improvement by a factor of $1/N^{1/6}$ is achieved. \textcite{buzek_optimal_1999} analytically optimized measurement basis and input states to obtain Heisenberg-limited scaling in the limit of large $N$ for similarly symmetric correlated states.  In a numerical optimization approach, \textcite{rosenband_numerical_2012} has shown that for realistic $1/f$ local oscillator noise with a flat Allan deviation of 1~Hz, for up to 15 ions the protocol by \textcite{andre_stability_2004} and for more ions the \textcite{buzek_optimal_1999} approach provides the best improvement over the SQL, whereas GHZ states perform even slightly worse than a standard Ramsey experiment.

Short of better clock lasers, improved clock interrogation schemes can realize sub-SQL instability. Recently, such optimized measurement strategies based on a hierarchy of ensembles of clock atoms with increasing interrogation time have been proposed \cite{borregaard_efficient_2013, kessler_heisenberg-limited_2013, rosenband_exponential_2013}. In these schemes the phase noise of the laser is tracked and stabilized on time scales approaching the excited state lifetime of the clock atom through interrogation of several ensembles with successively longer probe times. This ensures a well-defined laser phase for the ensemble with the longest interrogation time and results in an exponential scaling of the instability with the number of atoms. Quantum correlated spin states can either be used to reduce the required number of atoms in each ensemble or to further improve the instability, approaching Heisenberg-limited scaling for an infinite number of atoms \cite{kessler_heisenberg-limited_2013}.
Current implementations of single ion clocks with their limited instability would benefit the most from these new schemes, requiring multi-ion traps tailored for metrological purposes \cite{pyka_high-precision_2013-1}.

In summary, Heisenberg scaling for improved clock stability remains an experimental challenge. Phase noise of the LO prevents the clock stability from scaling as the HL. However, in scenarios with realistic noise models and taking into account the reduced performance of uncorrelated states in the presence of noise, partially entangled states can still lead to a significant improvement in stability \cite{kessler_heisenberg-limited_2013,Komar2013}. The full potential of entanglement-enhanced metrology can only be realized if analytical models for identifying the optimum states and measurement basis for complex noise models are developed. This goes hand in hand with the development of efficient protocols for creating these metrologically relevant states. From an experimental point of view entanglement-enhanced metrology is only worth the effort if either a simple scaling of the number of (uncorrelated) particles is technologically challenging as is the case for trapped ion systems \cite{pyka_high-precision_2013-1}, compromises the accuracy of the clock as is the case for density-related shifts in neutral atom lattice clocks, or where entanglement offers other added values, such as reduced systematic shifts.

\subsection{Designer atoms}
Entanglement as a resource for spectroscopy and optical clocks is not limited to improved stability as outlined in the previous section. Efficient schemes for creating entangled atomic states and protect them against environmental decoherence have been developed in the context of quantum information processing \cite{wineland_experimental_1998, haffner_quantum_2008, blatt_entangled_2008}. These techniques allow the creation of ``Designer atoms'' consisting of two or more entangled atoms with engineered properties for spectroscopy \cite{roos_designer_2006}. In a ground-breaking experiment, \citeauthor{roos_designer_2006} have entangled two \Ca ions (wavefunction indices $1$ and $2$) in a linear Paul trap in first-order magnetic field insensitive states of the form
\begin{equation}
\ket{\Psi}=\frac{1}{\sqrt{2}}(\ket{m_1}_1\ket{m_2}_2+\ket{m_3}_1\ket{m_4}_2)
\end{equation}
with $m_1+m_2=m_3+m_4$, where $m_i$ indicates the magnetic quantum number of the $D_{5/2}$ state. This construction ensures that each part of the wavefunction shifts the same way in a magnetic field. This idea can be further extended to engineer first-order magnetic-field- and electric-quadrupole-insensitive entangled states relevant for ion clocks. Currently this goal is achieved by averaging six transitions to obtain a virtual clock transition free of these shifts. However, changes in the magnetic or electric fields between the interrogation of the different transitions would cause imperfect cancellation. Using six entangled ions, the equivalent to averaging over these six  transitions could be achieved in a single experiment. In addition to the insensitivity against external fields, the differential phase shift between the two parts of the wavefunction exhibits GHZ-type scaling with an energy difference of $\Delta E=6\hbar\omega_0$. However, efficient schemes to produce such complex maximally entangled states are yet to be developed.

\subsection{Active Optical Clocks and Superradiant Lasers}

The lasers with best frequency stability currently have linewidths of $<$100 mHz~\cite{Bishof2013a}, and are limited by the thermal noise in the optical cavities that provide frequency stabilization. An alternative solution is to make a narrow laser using the same high-$Q$ transitions used in optical lattice clocks. The atoms become spontaneously correlated, creating a collective atomic dipole that emits light whose phase stability directly reflects the phase stability of the atomic dipole~\cite{meiser2009}. We note also that passive schemes using ultranarrow atomic resonances enhanced with an optical cavity can take a similar advantage of atom correlations to realize excellent laser frequency stabilization \cite{Martin2011}. The continuous superradiant light source has never been demonstrated, and has the potential to produce laser-like light with linewidths approaching 1 mHz. The impact of mHz linewidth frequency references has the same potential to revolutionize the precision of clocks as has the development of optical frequency standards during the past decade. Recently, a proof-of-principle experiment has been carried out using a Raman transition in Rb \cite{Bohnet2012}.

Instead of relying on the coherence of the photons, the continuous superradiant sources relies on the atomic coherence.  In this approach, $N$ atoms trapped in an optical cavity spontaneously form a collective 1D polarization grating leading to collective and directional emission of photons into the cavity mode.  The superradiant emission grows as $N^2$, and occurs without the macroscopic buildup of photons within the cavity~\cite{kuppens_quantum-limited_1994}.  A key insight is that the system can be continuously repumped, an advance akin to moving from pulsed to continuous lasing.  The second key insight is that the emitted light reflects the phase stability of the atomic polarization grating and that the coherence of the grating surpasses the single particle decoherence rate.  The predicted linewidth of the light can be even less than the atomic linewidth, and the scaling is fundamentally different from the Schawlow-Townes laser linewidth.  Lastly, the requirements on the optical cavity are relaxed since the key parameter that must be made large is again the cavity-QED collective cooperativity parameter $NC\gg$1, while the cooperativity parameter is preferably small $C\ll$1.

The effort using $^{87}$Sr atoms has the potential to produce unprecedentedly narrow light approaching 1 mHz.  The atoms would be trapped in a magic-wavelength optical lattice inside of a high finesse optical cavity (finesse $F \sim 10^6$) resonant with the clock transition $^3$P$_0$ to $^1$S$_0$.  For a 1 mm cavity length, and $N$ = $10^5$, the collective cooperativity is very large $NC \sim 10^4$  so that the superradiance threshold can be easily achieved.  However, the small cooperativity $C\sim 0.1$, yields a predicted linewidth of the emitted light smaller than the 1 mHz transition linewidth.

Following superradiant decay to $^1S_0$ the atoms would be continuously repumped back to $^3P_0$ via the intermediate states $^3P_{1,2}$  and $^3S_1$ that also serves to provide Raman sideband cooling.  The continuous nature of the light emitted from the cavity would last for several seconds and would be limited only by losses from the optical lattice. Emission could be made truly continuous by continuously reloading atoms into the optical lattice from the side.  The 0.1 pW of generated optical power would be sufficient to stabilize the current 1 Hz linewidth laser with a precision of 1mHz in a feedback bandwidth of $\sim$1 Hz. The stability of current Sr optical lattice clocks would be improved by a factor $>$10 should this proposed scheme succeed.

\subsection{Many-body quantum systems}

One of the most exciting research directions for ultracold matter lies in the exploration of strongly correlated quantum many-body systems. Optical lattice clocks have emerged as an surprising new platform for this endeavor. Motivated by the desire to reduce the clock systematic uncertainty arising from atomic interactions, high resolution spectroscopy performed in optical lattice clocks has provided new insights and understandings of these effects.  This effort in turn makes the lattice clocks well-suited for the study of many-body spin interactions. When the spectral resolving power of the clock laser advances to be better than atomic interactions in the clock, the seemingly weakly interacting spin system actually demonstrates strong correlations with complex excitation spectra~\cite{Martin2013a,Rey2014a}, and even SU(N) symmetry~\cite{Gorshkov10a} can now be directly explored to study complex quantum systems with high degeneracy~\cite{Zhangx_2014,Scazza_2014,Cappellini_2014}. This is an exciting new research direction that builds on the advanced optical clock and will provide important guidelines for future advances of optical lattice clocks~\cite{Chang04a}.

Endowed with a number of attractive properties, ultracold group II atoms provide new opportunities for quantum simulation and quantum information science~\cite{Reichenbach07a,Daley08a,Gorshkov09a,Daley11a,Gorshkov10a}, leveraging on the efforts on optical manipulation, quantum engineering, clock-type precision measurement, and optical control of interactions~\cite{Blatt11a,Yamazaki10a}. The clean separation of internal and external degrees of freedom in an optical lattice clock system rivals that of ion-trap systems and is ideal for retaining quantum coherence for many trapped atoms, and for precise quantum measurement and manipulations~\cite{Ye08a}. In fact, even at the early stage of the lattice clock development, laser-atom interaction coherence time has been extended to hundreds of milliseconds~\cite{Boyd06a}, and has been further improved with more stable lasers~\cite{Martin2013a,Bishof2013a}. This high spectral resolution allows us to precisely control the electronic and nuclear spin configurations and to probe their interactions.  Specifically, we can use both the nuclear spin and long-lived electronic states ($^1S_0$ and $^3P_0$ states) to represent spins and orbitals in a quantum system. The two key features are the presence of a metastable excited state $^3P_0$ and the almost perfectly decoupled nuclear spin $I$ from the electronic angular momentum $J$ in these two states, because $J$ = 0.

The advantage of using pure nuclear spin states is that their coherence is largely insensitive to \emph{stray} magnetic or electric fields in the laboratory, and yet they can still be effectively manipulated via strong and deliberately applied laser fields so that state-specific resonances can be controlled, even in a spatially resolved manner.  In addition, by using the metastable electronic states to represent orbitals, one gains exceptional spectral selectivity to impose state-dependent optical forces on atoms in the lattice.  As such, schemes for generating spin-dependent interactions, similar to those relevant for trapped ions or in the bilayer lattice, can be implemented~\cite{Daley08a,Daley11a}.  In addition, it should also be possible to develop individual quantum bit addressability and readout using tomographic and site-resolved imaging techniques under applied inhomogeneous
magnetic fields. This spatial addressability and control are useful in several ways: 1) characteristics of the lattice may only be uniform in a small portion of the system, and this spatial addressing would allow the simulation to take place specifically and exclusively in that portion; 2) the non-uniform lattice parameters can be compensated for with the spatial addressing; 3) we can simulate non-uniform material systems, a capability of clear technological importance if aim to simulate materials for real world devices and that are thus deliberately shaped and crafted to specific tasks.

A major advantage of the optical lattice clock with many atoms is the enhanced signal to noise ratio for spectroscopy and hence the improved clock stability.  However, with atom–light coherence times reaching beyond one second, even very weak atomic interactions can give rise to undesired clock frequency shifts. This systematic uncertainty connects to many-body physics and is thus different from all other single-atom based effects.  An interesting discovery in our push for ever increasing accuracy of the Sr and Yb lattice clocks is the interaction-induced frequency shift on the clock transition even with spin-polarized fermionic atoms prepared under ultralow temperatures, where atoms collide with a single or very few partial waves. For identical fermionic atoms, antisymmetrization of the two-particle wave function forbids the $s$-wave interaction, and the $p$-wave is suppressed owing to the centrifugal potential arising from an angular momentum of $\hbar$. After intensive research efforts focusing on the atomic density-related frequency shifts in both Sr and Yb systems, we have come to very good understandings of these effects and have since suppressed the density-dependent frequency shift below 1$\times 10^{-18}$. The theory model developed by Rey \textrm{et al}. has also become capable of describing full many-body spin-spin interaction dynamics well beyond a simple mean-field treatment~\cite{Rey2014a,Martin2013a}.

The powerful spectroscopy resolution allows us to very effectively remove single-particle dephasing effects and reveal the underlying correlated spin dynamics. The decoupling between the electronic and nuclear spins implies that atomic scattering lengths involving states $^1S_0$ and $^3P_0$ are independent of the nuclear spin to very high precision. Of course the nuclear spin wavefunction can be engineered to dictate how the two atoms interact electronically via antisymmetrization of the overall wavefunction for fermions.  The resulting SU($N$) spin symmetry (where $N$ = 2$I$ + 1 can be as large as 10) together with the possibility of combining (nuclear) spin physics with (electronic) orbital physics opens up an exciting research direction for rich many-body systems with alkaline-earth atoms~\cite{Gorshkov10a}.

\subsection{Atomic clocks with even higher transition frequencies}

Going to higher and higher operating frequencies has been a recurrent trend in the development of precise clocks and the present status of optical clocks clearly shows the benefits in terms of stability and accuracy that can be obtained in comparison to atomic clocks operating in the microwave range. It is therefore evident to consider to carry this development forward and to enter the domain of vacuum ultraviolet and soft x-ray radiation. Candidates for suitable reference transitions may be sought in highly charged ions where the remaining electrons are tightly bound, and potentially in heavy nuclei where a few $\gamma$-transitions are known at energies of 1~keV and below, that are untypically low on the nuclear energy scale. It has been pointed out that both types of transitions may offer considerable advantages in terms of field-induced systematic frequency shifts. Comprehensive proposals have been developed for clocks based  on nuclear transitions \cite{peik_nuclear_2003,campbell_single-ion_2012} and on electronic transitions in highly charged ions \cite{derevianko_highly_2012}.

Nuclear transition frequencies are generally several orders of magnitude higher than those of transitions in the electron shell and are also less sensitive to shifts induced by external electric or magnetic fields because the characteristic nuclear dimensions and nuclear moments are small compared to those of the shell. If the interrogation is not performed with a bare nucleus, one has to consider the coupling of the nuclear and electronic energy level systems through the hyperfine interactions. Since the primary interest is in the nuclear transition, the choice of a suitable electronic configuration can be adapted to the experimental requirements. From general considerations it can be seen that for every radiative nuclear transition,
an electronic state can be selected based on angular momentum quantum numbers such that the hyperfine coupled nuclear transition frequency becomes immune against field-induced shift to a degree that can not be obtained for an electronic transition.

In an $LS$ coupling scheme the eigenstates of the coupled electronic and nuclear system are characterised by sets of quantum numbers $|\alpha,I;\beta,L,S,J;F,m_F\rangle$, where $I$ denotes the nuclear spin, $L,S,J$ the orbital, spin and total electronic angular momenta, $F$ and $m_F$ the total atomic angular momentum and its orientation.
$\alpha$ and $\beta$ label the involved nuclear and electronic configurations.
The choice of an integer total angular momentum $F$ (so that a Zeeman sublevel $m_F=0 \rightarrow 0$ is available) together with $J<1$ leads to
vanishing of the linear Zeeman effect, quadratic Stark effect and quadrupole shift \cite{peik_nuclear_2003}. In this scheme, the optimal electronic states for the interrogation of the nuclear transition are those with $J=1/2$ in the case of a half integer nuclear spin, and $J=0$ if the nuclear spin is integer. Alternatively, and more generally applicable also for higher values of $J$, a pair of transitions between stretched hyperfine states $|F=\pm(J+I), m_F=F\rangle \rightarrow |F'=\pm(J+I'), m_F'=F'\rangle$ can be used to realize a nuclear transition that is largely uncoupled from shifts in the electron shell  \cite{campbell_single-ion_2012}.

Because of its favorably low transition energy of about 7.8~eV, the transition between the nuclear gound state and an isomeric state in $^{229}$Th
\cite{beck_energy_2007} is considered as the experimentally most accessible system for a nuclear clock and a number of experimental projects have been started to investigate this potential.
For a high-precision nuclear clock, the case of trapped $^{229}$Th$^{3+}$-ions seems to be especially promising because its electronic level structure is suitable for laser cooling \cite{peik_nuclear_2003,campbell_multiply_2009}. The sensitive detection of excitation to the isomeric state will be possible using a double resonance scheme that probes the hyperfine structure of a resonance transition in the electron shell, in analogy to electron shelving as applied in single-ion optical clocks on electronic transitions.

An alternative option for a nuclear optical optical clock with $^{229}$Th  is based on the idea of performing laser M\"o{\ss}bauer spectroscopy with $^{229}$Th embedded in a crystal \cite {peik_nuclear_2003,rellergert_constraining_2010,kazakov_performance_2012}. While the systematic uncertainty of such a solid-state nuclear clock may not reach that of a realization with trapped and laser cooled ions, the potentially much larger number of nuclei may provide a frequency reference of high stability. The crystal field shifts of the nuclear resonance frequency will be dominantly due to electric fields and field gradients. A diamagnetic host with a lattice of high symmetry should be used.
Thermal motion will lead to a temperature-dependent broadening and shift of the nuclear line, where the line shape will depend on phonon frequencies and correlation times. For a solid state nuclear clock of high accuracy (beyond $10^{-15}$) the temperature dependence may be eliminated if the crystal is cryogenically cooled to well below the Debye temperature, so that the influence of phonons is effectively frozen out.

At higher nuclear transition energies, the methods envisaged here for $^{229}$Th will not be viable if radiative nuclear decay competes with the emission of conversion electrons, leading to changes of the charge state of the ion. In the case of trapped ions, internal conversion can be suppressed by using a sufficiently high charge state with an ionization potential that lies above the nuclear excitation energy. Laser cooling and state detection will then be performed using the methods developed for the trapped ion quantum logic clock.

Electronic transitions in highly charged ions also possess favorable properties as a reference for a highly accurate clock.
In a positive ion of net charge $Ze$ the binding energy of a valence electron is proportional to $Z^2$.
Within an isoelectronic sequence, transition energies between bound states can be expected to follow a similar scaling, modified by contributions from QED and finite nuclear size. Since the size of the electron cloud contracts with $1/Z$, size-dependent quantities like polarisabilities or electric quadrupole moments, that determine the sensitivities to external perturbations from electric fields, scale down rapidly with increasing $Z$. From this point of view, it may be advantageous to study highly-charged ions that show the same types of forbidden transitions like the neutral atoms or singly-charged ions that are used in optical clocks today. For $Z\approx 20$, the hyperfine-induced $^1S_0 \rightarrow ^3P_0$ transition in Be-like ions appears at a transition energy of about 30~eV with a natural linewidth on the order of 1~Hz
\cite{cheng_hyperfine_2008}. At still higher $Z$, the ratio of transition frequency to the natural linewidth decreases for this type of transition.

So far, proposals for clocks with highly charged ions have identified suitable transitions within the ground state configuration that provide a low sensitivity to field-induced frequency shifts at a transition frequency in the infrared or visible spectral range: hyperfine transitions in the electronic ground state of hydrogen-like ions \cite{schiller_hydrogenlike_2007} and electric quadrupole transitions within the $4f^{12}$ configuration of the Re$^{17+}$ sequence \cite{derevianko_highly_2012}.
Given the wide choice of positive charge states in different isoelectronic sequences, it is foreseeable that more opportunities may be discovered.

An important consideration in the pursuit of higher frequencies is that the development of low-noise coherent sources of radiation and of the required clockwork for the counting of periods
seems to pose major challenges because materials for amplifiers or mixers that provide a similar efficiency as it is now available in the visible spectral range are not known.
Promising results have been obtained with harmonic generation from near-infrared femtosecond frequency combs in gas jets
\cite{Jones05a,witte_deep-ultraviolet_2005,gohle_frequency_2005}. In this approach, the frequency of the comb modes can be stabilized, controlled and measured in the infrared spectral region, while the conversion of the original frequency comb into a sequence of odd harmonics makes the ensuing measurement precision available in the vacuum-ultraviolet.
This method has now permitted to perfom precision spectroscopy and frequency measurements of transitions in rare gases at XUV wavelengths around 50 nm
\cite{kandula_extreme_2010,cingoz_direct_2012,Benko_2014}.

\section*{Acknowledgments}
We thank Daniel Kleppner for his initial invitation of writing this review article in 2008. We are grateful to David Wineland for his suggestions and contributions to the introduction of this review. We are indebted to many of our coworkers who have made critical contributions to the work done at JILA, NIST, and PTB over many years. A large portion of this review is based on their work.  The research at JILA and NIST is supported by NIST, the Physics Frontier Center of NSF at JILA, DARPA, and NASA. The work at PTB is supported by the DFG through the Centre for Quantum Engineering and Space-Time Research (QUEST), ESA, and by the European Metrology Research Program (EMRP) in project SIB04. The EMRP is jointly funded by the EMRP participating countries within EURAMET and the European Union.
We thank Christian Tamm, Nils Huntemann, Ian Leroux, Heiner Denker, Tanja Mehlstäubler, Xibo Zhang, and Kyle Beloy for critical reading of the manuscript.

\bibliographystyle{apsrmp}

\end{document}